\documentclass[a4paper,11pt]{article}
\pdfoutput=1 % if you are submitting a pdflatex (i.e. if you have
\usepackage{jheppub} % for details on the use of the package, please
\usepackage[T1]{fontenc} % if needed
\usepackage{orcidlink} %To show orchid link
\usepackage{float}

\title{\boldmath Effect of invisible neutrino decay on neutrino oscillation at long baselines}
 
\usepackage{multirow}
 
\author[a,b]{Animesh Chatterjee \orcidlink{0000-0002-2935-0958},}
\author[b,1]{Srubabati Goswami \orcidlink{0000-0002-5614-4092},}
\author[b,1]{Supriya Pan \orcidlink{0000-0003-3556-8619},\note{Corresponding author.}}
\author[b]{Paras Thacker}

\affiliation[a]{European Organization for Nuclear Research (CERN), 1211 Geneva 23, Switzerland}
\affiliation[b]{Physical Research Laboratory, Ahmedabad, Gujarat, 380009, India}
\emailAdd{animesh.chatterjee@cern.ch}
\emailAdd{sruba@prl.res.in}
\emailAdd{supriyapan@prl.res.in }
\emailAdd{parasthacker9878@gmail.com}

\abstract{In this article, we study the effect of invisible neutrino decay of the third neutrino state for accelerator neutrino experiments at two different baselines, 1300 km with a liquid argon time projection chamber (LArTPC) detector (similar to DUNE) and 2588 km with a water Cherenkov detector (similar to P2O). For such baselines, the matter effect starts to become important. Our aim is to ascertain the sensitivity to mass hierarchy and octant of $\theta_{23}$ in these two experiments in the presence of a decaying neutrino state. We also compare and contrast the results of the two experimental setups. We find that, in general, hierarchy sensitivity decreases in the presence of decay. However, if we consider decay only in the opposite hierarchy (test scenario), in the 2588 km setup, the hierarchy sensitivity with the true hierarchy as IH is larger than the no decay case. We also study the dependence of hierarchy sensitivity with true $\theta_{23}$. We find that the dominant muon background in P2O plays an important role in how the hierarchy sensitivity depends on $\theta_{23}$. The octant sensitivity for both setups increases in the presence of decay except for the LArTPC setup in case true $\theta_{23}=49^\circ$. To understand the octant sensitivity results in the two setups, we check the synergy between electron and muon channels as a function of test $\theta_{23}$. We also study the degeneracies in the test $\theta_{23}-\delta_{CP}$ plane and find that combined analysis of the two setups removes all the degeneracies at $5\sigma$ significance.}

\begin{document} 
\maketitle

\section{Introduction}
\label{sec:intro}
Neutrino oscillations, which involve the conversion of one neutrino flavor into another, have been detected across various neutrino sources, including the sun, the atmosphere, reactors, and accelerators, utilizing a diverse range of detection techniques in terrestrial experiments\cite{McDonald:2016ixn, Kajita:2016cak, T2K:2011qtm, MINOS:2008kxu, DayaBay:2007fgu}. These findings conclusively establish the three-neutrino oscillation paradigm, with a substantial portion of the parameter space governing the phenomenon accurately established. 
However, there are still open challenges and unanswered questions,  namely, the ordering of neutrino masses (mass hierarchy), the exact value (i.e., octant) of the mixing angle $\theta_{23}$, and the value of the CP phase. Many ongoing, upcoming, and proposed experiments are expected to address these issues.  Although neutrino oscillation is established as a leading solution to the solar and atmospheric neutrino anomalies corroborated by reactor and accelerator experiments, imprints of physics beyond the standard model can appear as sub-leading effects. Many studies have been performed to explore a multitude of new physics scenarios. Neutrino decay is one such solution.
 
Massive neutrinos can decay in general. Neutrino decay was first proposed as a solution to the solar neutrino problem\cite{Bahcall:1972my, Pakvasa:1972gz} and has since been thoroughly explored in scientific research. Neutrino decay in the Standard Model, whether radiative or non-radiative, is constrained by cosmological observations due to precise measurements of the cosmic microwave background.\cite{Mirizzi:2007jd, Agashe:2013eba}. \\
In contrast, processes involving neutrino decay in physics beyond the Standard Model (BSM) such as  $\nu_{i}\rightarrow \nu_{f} + X_{f} $ are much less constrained. Here $\nu_{i}$ denotes the mass states (i=1,2,3), and $\nu_{f}$ and $X_{f}$ are the final state particles. From a phenomenological perspective, there are two types of neutrino decay: visible decay and invisible decay. In visible decay, the neutrinos in the final state are active and can be detected. This type of decay would result in the observable event rates at the detector being impacted by the decay products, which would have lower energies than the parent particle. A typical signature of visible neutrino decay would be a concentration of events at low energies\cite{Kim:1990km, Acker:1992eh, Lindner:2001fx, Coloma:2017zpg, Gago:2017zzy}. In invisible decay, the neutrino in the final state is not observable in the detector. This could occur if the neutrino in the final state is a light sterile neutrino (lighter than the active ones, allowing for the decay kinematically). Another possibility is that the decay product involves an active neutrino, which, due to its low energy, is unobservable. We are interested in discussing the phenomenon of invisible neutrino decay. In this particular scenario, a beam of relativistic neutrinos ($\nu_{i}$) with a lifetime of $\tau_{i}$ and mass $m_i$ is subject to depletion as a consequence of invisible decay by a factor of
\begin{equation}
    \xi (L,E)= exp(-\alpha_{i} \frac{L}{E}) = exp(-\kappa_{i})
\end{equation}
where E is the neutrino energy, L is the distance
between the source and the detector and
\begin{equation}
    \alpha_{i}= \frac{m_i}{\tau_{i}}; \, \, \kappa_i = \alpha_{i} \frac{L}{E}
\end{equation}
is the decay parameter. It is evident that, for a given ratio of L/E, neutrino decay is only sensitive when $\alpha_{i}\geq L/E$, and the no decay limit is $\alpha_i \rightarrow 0$.\\
Constraints on $\tau_{i}$ have been established using various neutrino sources. Neutrino decay as a possible explanation for the decrease in solar neutrinos has been explored early on\cite{Berezhiani:1991vk, Berezhiani:1992xg, Acker:1992eh}. Subsequently, many studies have investigated the combination of neutrino oscillation and decay as a solution to the solar neutrino problem, setting limits on the lifetime of the unstable state\cite{Choubey:2000an, Beacom:2002cb, Joshipura:2002fb, Bandyopadhyay:2002qg, Picoreti:2015ika}. These studies assumed that $\nu_{2}$ is the unstable state and derived constraints on $\tau_{2}$. The possibility of decay of Dirac/Majorana neutrino to a pseudo scaler has been studied from solar neutrino appearance in \cite{Funcke:2019grs}. A recent study\cite{Berryman:2014qha} has established constraints on both $\tau_{1}$ and $\tau_{2}$ using low-energy solar neutrino data. Additional bounds have been obtained from the observation of SN1987A data \cite{Frieman:1987as, Bionta:1987qt}. On the other hand, atmospheric neutrinos or long-baseline neutrino data have been used so far to bound the $\tau_{3}/m_{3}$ \cite{LoSecco:1998cd, Barger:1998xk, Lipari:1999vh, Fogli:1999qt, Choubey:1999ir, Barger:1999bg, Ashie:2004mr}. The combined analysis of Super-Kamiokande data together with long-baseline data from K2K and MINOS provides a lower bound on the decay parameter $\tau_{3}/m_{3} \geq 9.3\times 10^{-11}$ s/eV at $99\%$ C.L\cite{Gonzalez-Garcia:2008mgl}. Recent analysis\cite{Gomes:2014yua} of the  MINOS (using the full charged-current and neutral-current data sample) and T2K (charged-current data) imply  $\tau_{3}/m_{3} > 2.8\times 10^{-12}$ s/eV at $90\%$ C.L. bound on $\tau_{3}/m_{3} \geq 1.5 \times 10^{-12}$ s/eV was obtained from combined analysis of NO$\nu$A and T2K data \cite{Choubey:2018cfz}. However, adding MINOS data with T2K and NO$\nu$A leads to $\tau_{3}/m_{3} \geq 2.4 \times 10^{-11}$ s/eV \cite{Ternes:2024qui} at $90\%$ C.L. for both mass hierarchy.

The decay of ultrahigh energy astrophysical neutrinos has been considered by many authors\cite{Beacom:2002vi, Maltoni:2008jr, Pakvasa:2012db}. The IceCube experiment, as shown in \cite{Pagliaroli:2015rca}, can reach a sensitivity of $\tau/m > 10$ s/eV for both hierarchies for 100 TeV neutrinos coming from a source at a distance of 1 Gpc. Moreover, in \cite{Denton:2018aml}, the invisible neutrino decay with $\tau/m >10^2$ s/eV has been shown to explain the discrepancy seen in the track and cascade data set of high-energy neutrinos. The implications of invisible neutrino decay for future experiments like JUNO\cite{He:2014zwa}, RENO-50\cite{Kim:2014rfa}, DUNE \cite{DUNE:2015lol}, T2HK, and T2HKK \cite{Hyper-Kamiokande:2016dsw,Hyper-Kamiokande:2016srs} have been investigated. In DUNE, the bound of $5.1 \times 10^{-11}$ s/eV is shown to be obtained \cite{Ghoshal:2020hyo} in 90$\%$ C.L. The sensitivity to lifetimes of all neutrino mass states has been discussed in \cite{Martinez-Mirave:2024hfd} using solar, supernova and DSNB neutrinos in upcoming neutrino observatories.

In this paper, our focus is on investigating invisible neutrino decay of the third neutrino state for long baseline experiments. Toward this goal, we choose two setups: (i) A water Cerenkov detector at a distance of 2588 km from an accelerator beam. This baseline is proposed to be realized by the Protvino to ORCA (P2O) experiment; (ii) A liquid Argon time projection chamber (LArTPC) detector at a distance of 1300 km from a beam source of neutrinos, similar to the upcoming DUNE experimental setup. 
For these baselines, the matter effect starts becoming important. We include this in our study. Note that the 2588 km baseline of the P2O experiment closely corresponds to the bi-magic baseline of 2540 km \cite{Raut:2009jj, Dighe:2010js, Dighe:2011pa}. We examine how the bi-magic condition \cite{Dighe:2010js}, which implies near-independence of the probabilities on $\delta_{CP}$ for a particular hierarchy at specific energies, gets affected in the presence of neutrino decay. 
Furthermore, we seek to investigate how the three-neutrino oscillation phenomenology is modified in the presence of neutrino decay. Our particular interest lies in understanding the impact of the decay of the third mass state ($\nu_{3}$) on the determination of both neutrino mass hierarchy and the octant of $\theta_{23}$. In addition to studying the mass hierarchy, octant of $\theta_{23}$, the study also probes the degeneracies in $\theta_{23}-\delta_{CP}$ plane to determine if combining the results from the two baselines can resolve these ambiguities.

The paper is organized as follows. To start with, in section \ref{sec:decay-osc}, we have discussed the analytical framework for neutrino oscillation in the presence of decay and the oscillation probabilities. Then, in section \ref{sec:prob}, we study the effects of decay in $P_{\mu e}, P_{\mu\mu}$ in the context of determining mass hierarchy and octant of $\theta_{23}$. The details of experimental setups have been written in section \ref{sec:expr}. Next, in section \ref{sec:results}, we discuss the results of sensitivity to mass hierarchy and octant of $\theta_{23}$. Finally, we summarize our findings in section \ref{sec:conclusion}.

\section{Neutrino oscillations in the presence of neutrino decay}\label{sec:decay-osc}
In the presence of decay of neutrino states, the total Hamiltonian is non-Hermitian due to the contribution of decay,
\begin{equation}
    \mathcal{H}_m = H^{std}_m - \frac{i}{2} \Gamma_m,
\end{equation}
In the presence of neutrino decay of the third mass state $\nu_3$, the total Hamiltonian of the system is given by,
\begin{equation}
    H = \frac{1}{2E}  U \left[ \begin{pmatrix}
        0 & 0 & 0\\
        0 & \Delta_{21} & 0\\
        0 & 0 & \Delta_{31}
    \end{pmatrix} + \begin{pmatrix}
        0 & 0 & 0\\
        0 & 0 & 0\\
        0 & 0 & -i \alpha_3
    \end{pmatrix}
    \right]U^\dagger + \frac{1}{2E}\begin{pmatrix}
        A & 0 & 0\\
        0 & 0 & 0\\
        0 & 0 & 0
    \end{pmatrix},
\end{equation}
where $U$ is the PMNS matrix, $\Delta_{ij}=m_i^2-m_j^2$ are mass-squared differences and $A=2\sqrt{2}G_F N_e E$ is the matter potential of neutrino. Because of the non-Hermitian nature of the Hamiltonian, the calculation of the probabilities in the presence of decay in matter is not straightforward. It has been discussed in \cite{Chattopadhyay:2021eba} that in matter, the decay and mass eigenstates are necessarily different, and even if one starts with one decaying state in a vacuum, other states eventually also develop decaying components. Three generation probabilities in the presence of decay have been discussed in \cite{Chattopadhyay:2022ftv}. The probability in the presence of decay can be written as \cite{Chattopadhyay:2022ftv},

\begin{align} \label{eq:pme-dec}
    P_{\mu e}=&s_{13}^2 s_{23}^2 \left(1+ e^{-4\kappa_3}- 2 e^{-2\kappa_3}+ 4 e^{-2\kappa_3} \sin^2[(\hat{A}-1)\Delta] \right)\frac{\alpha_3^2 + \Delta_{31}^2}{\Delta_{31}^2(\hat{A}-1)^2 + \alpha_3^2} \nonumber\\
    &+\beta s_{13} \sin{2\theta_{12}} \sin{2\theta_{23}} \frac{\sin{\hat{A}\Delta}}{\hat{A}} \times \nonumber\\ & \Bigg[ \Big(\sin[(\hat{A}-1)\Delta+\delta_{CP}-\Delta] e^{-2\kappa_3} + \sin[\hat{A}\Delta -\delta_{CP}] \Big)
    \frac{\Delta_{31}^2(\hat{A}-1)-\alpha_3^2}{\Delta_{31}^2(\hat{A}-1)^2 + \alpha_3^2} \nonumber\\
    &+ \Big( \cos[\hat{A}\Delta -\delta_{CP}] - \cos[(\hat{A}-1)\Delta+\delta_{CP}-\Delta]e^{-2\kappa_3} \Big) \frac{A \alpha_3}{\Delta_{31}^2(\hat{A}-1)^2 + \alpha_3^2} \Bigg]
\end{align}

\begin{align}\label{eq:pmm-dec}
    P_{\mu\mu}=&1 - \frac{1}{2}\sin^2{2\theta_{23}} [ 1+ 2\sin^2{\Delta}e^{-2\kappa_3} -e^{-2\kappa_3} ] - s_{23}^2(1- e^{-4\kappa_3}) -\kappa_3 (4s_{23}^4 + \sin^2{2\theta_{23}}\cos{2\Delta}) \nonumber\\
    & - \frac{2}{\hat{A}-1}s_{13}^2 \sin^2{2\theta_{23}} \Big( \sin{\Delta} \cos[\hat{A}\Delta] \frac{\sin[(\hat{A}-1)\Delta]}{\hat{A}} - \frac{\hat{A}}{2}\Delta \sin[2\Delta] \Big) \nonumber\\
    &-4 s_{13}^2 s_{23}^2 \frac{\sin^2[(\hat{A}-1)\Delta]}{(\hat{A}-1)^2} + \beta c_{12}^2 \sin^2{2\theta_{23}} \Delta \sin[2\Delta] + \kappa_3^2 (8s_{23}^4 + \sin^2{2\theta_{23}}\cos{2\Delta})\rm{,}
\end{align}
where $\hat{A}=A/\Delta_{31}$, $\beta=\Delta_{21}/\Delta_{31}$, and $\Delta=\Delta_{31}L/E$.
Now, for the bi-magic baseline, putting $\sin{[(A-1)\Delta]}=0$ in the above equations will lead to,
\begin{align}\label{eq:pme-bm}
    P_{\mu e}^{\rm BM}=&s_{13}^2 s_{23}^2 \left(1+ e^{-4\kappa_3}- 2 e^{-2\kappa_3}\right)\frac{\alpha_3^2 + \Delta_{31}^2}{\Delta_{31}^2(\hat{A}-1)^2 + \alpha_3^2} +\beta s_{13} \sin{2\theta_{12}} \sin{2\theta_{13}} \frac{\sin{\hat{A}\Delta}}{\hat{A}} \times \nonumber\\ & \Bigg[ \Big(\sin[\delta_{CP}-\Delta] e^{-2\kappa_3} + \sin[\hat{A}\Delta -\delta_{CP}] \Big)
    \frac{\Delta_{31}^2(\hat{A}-1)-\alpha_3^2}{\Delta_{31}^2(\hat{A}-1)^2 + \alpha_3^2} + \nonumber\\ 
    &\Big( \cos[\hat{A}\Delta -\delta_{CP}] - \cos[\delta_{CP}-\Delta]e^{-2\kappa_3} \Big) \frac{A \alpha_3}{\Delta_{31}^2(\hat{A}-1)^2 + \alpha_3^2} \Bigg]
\end{align}

\begin{align}\label{eq:pmm-bm}
    P_{\mu\mu}^{\rm BM}=&1 - \frac{1}{2}\sin^2{2\theta_{23}} [ 1+ 2\sin^2{\Delta}e^{-2\kappa_3} -e^{-2\kappa_3} ] - s_{23}^2(1- e^{-4\kappa_3}) -\kappa_3 (4s_{23}^4 + \sin^2{2\theta_{23}}\cos{2\Delta}) \nonumber\\
    & + \frac{\hat{A}}{\hat{A}-1}s_{13}^2 \sin^2{2\theta_{23}} \Big(\Delta \sin[2\Delta] \Big) + \beta c_{12}^2 \sin^2{2\theta_{23}} \Delta \sin[2\Delta] +\kappa_3^2 (8s_{23}^4 + \sin^2{2\theta_{23}}\cos{2\Delta})
\end{align}
To retrieve the no decay limit, we put $\alpha_3 \rightarrow 0$ in the above equations to get,
\begin{align}\label{eq:pme-sm}
    P_{\mu e}^{\rm SM}=& 4s_{13}^2 s_{23}^2 \frac{ \sin^2[(\hat{A}-1)\Delta] }{(\hat{A}-1)^2} +2 \beta s_{13} \sin{2\theta_{12}} \sin{2\theta_{23}}  \cos[\Delta -\delta_{CP}] \frac{\sin{\hat{A}\Delta}}{\hat{A}}
    \frac{\sin[(\hat{A}-1)\Delta]}{(\hat{A}-1)} 
\end{align}
\begin{align}\label{eq:pmm-sm}
    P_{\mu\mu}^{\rm SM}=&1 - \sin^2{2\theta_{23}} \sin^2{\Delta} + \frac{\hat{A}}{\hat{A}-1}s_{13}^2 \sin^2{2\theta_{23}} \Delta \sin[2\Delta] -4 s_{13}^2 s_{23}^2 \frac{\sin^2[(\hat{A}-1)\Delta]}{(\hat{A}-1)^2}  \nonumber\\
    & - \frac{2}{\hat{A}-1}s_{13}^2 \sin^2{2\theta_{23}} \sin{\Delta} \cos[\hat{A}\Delta] \frac{\sin[(\hat{A}-1)\Delta]}{\hat{A}} + \beta c_{12}^2 \sin^2{2\theta_{23}} \Delta \sin[2\Delta]
\end{align}

\section{Probability level Analysis}\label{sec:prob}
\begin{figure}[H]
    \centering
    \includegraphics[width=0.4\linewidth]{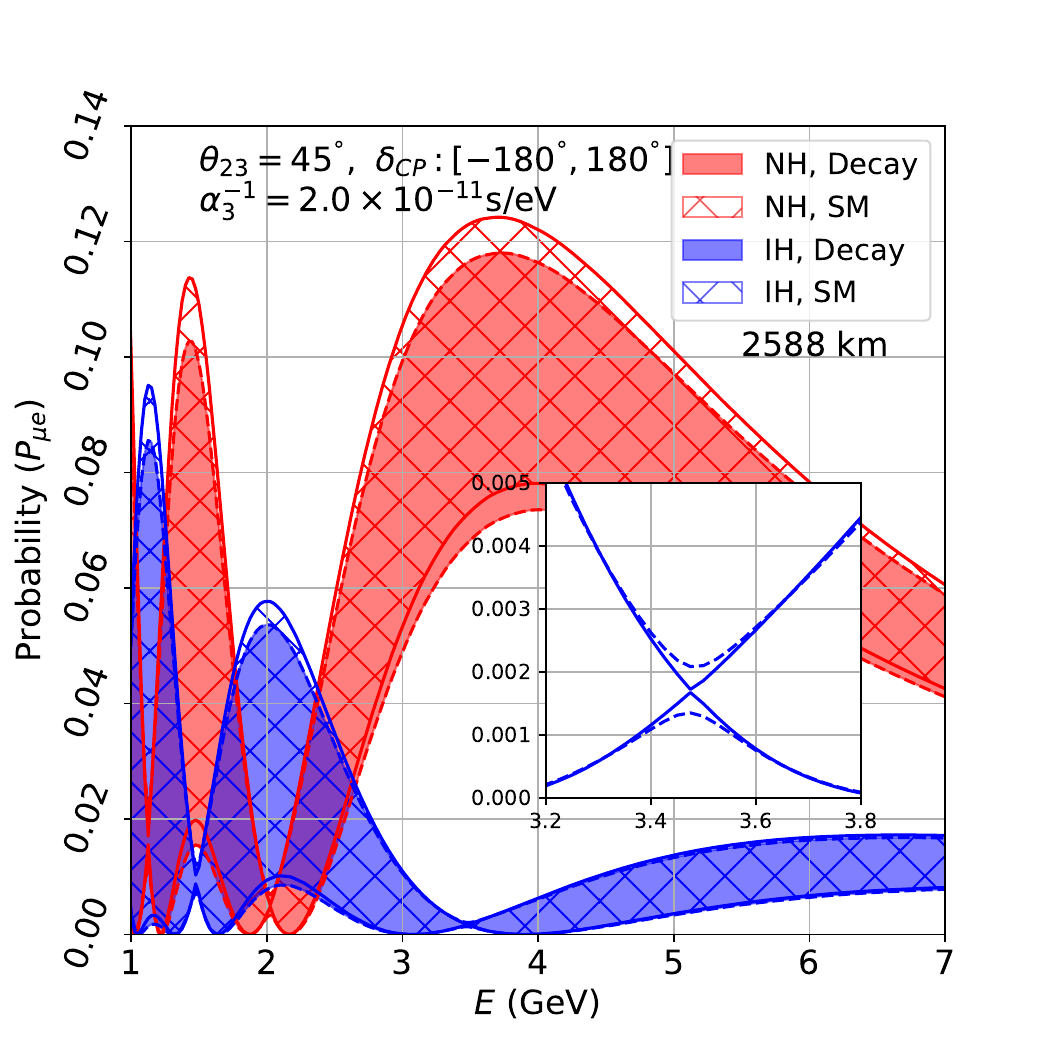}
    \includegraphics[width=.4\textwidth]{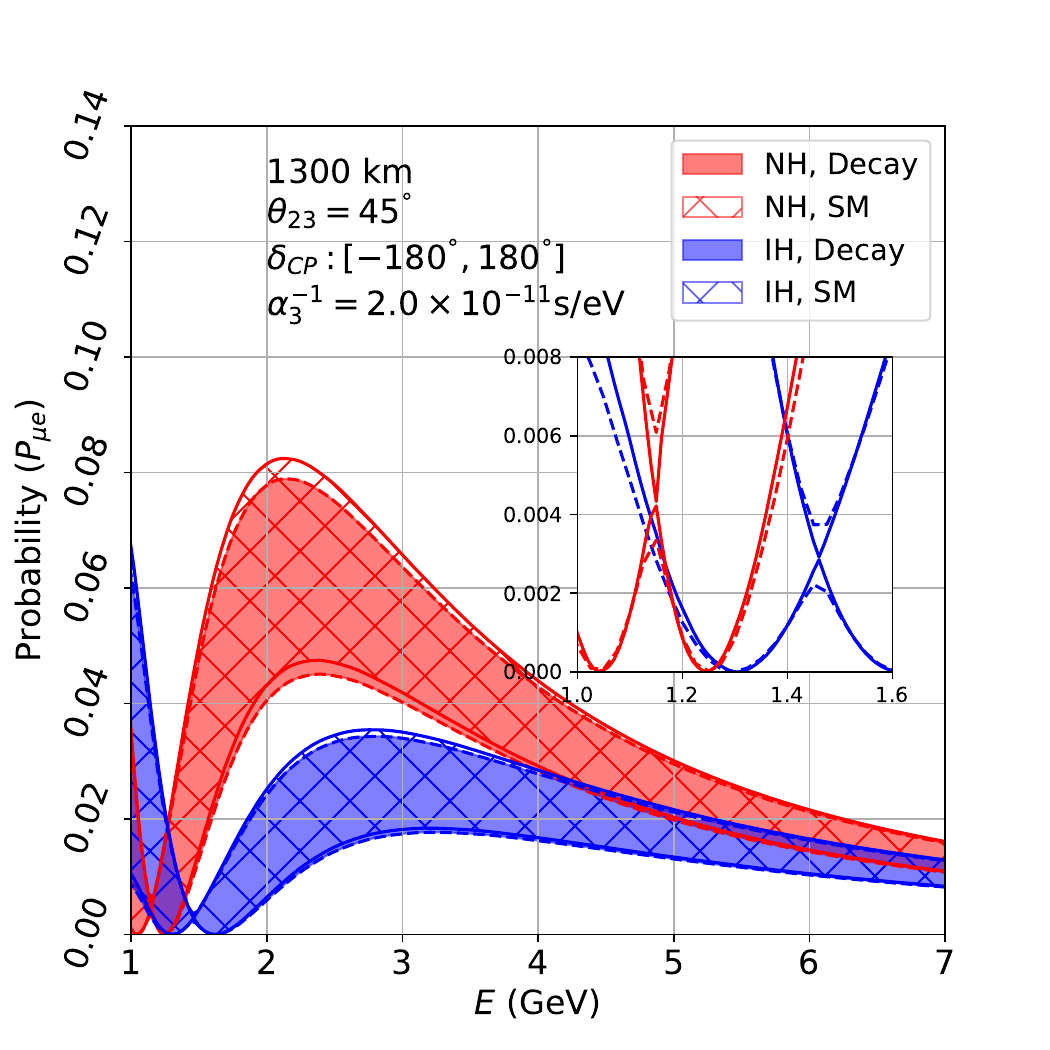}
    \includegraphics[width=0.4\linewidth]{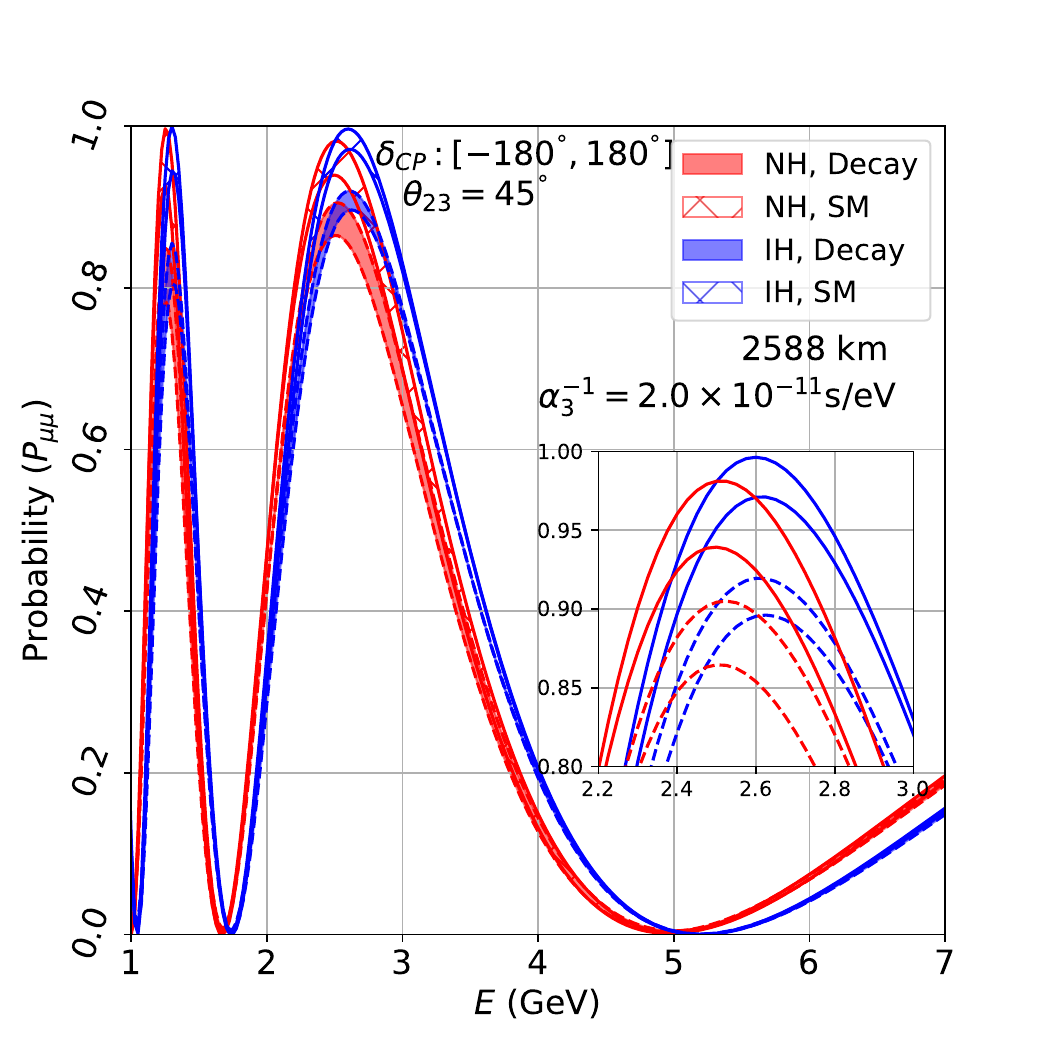}
    \includegraphics[width=.4\textwidth]{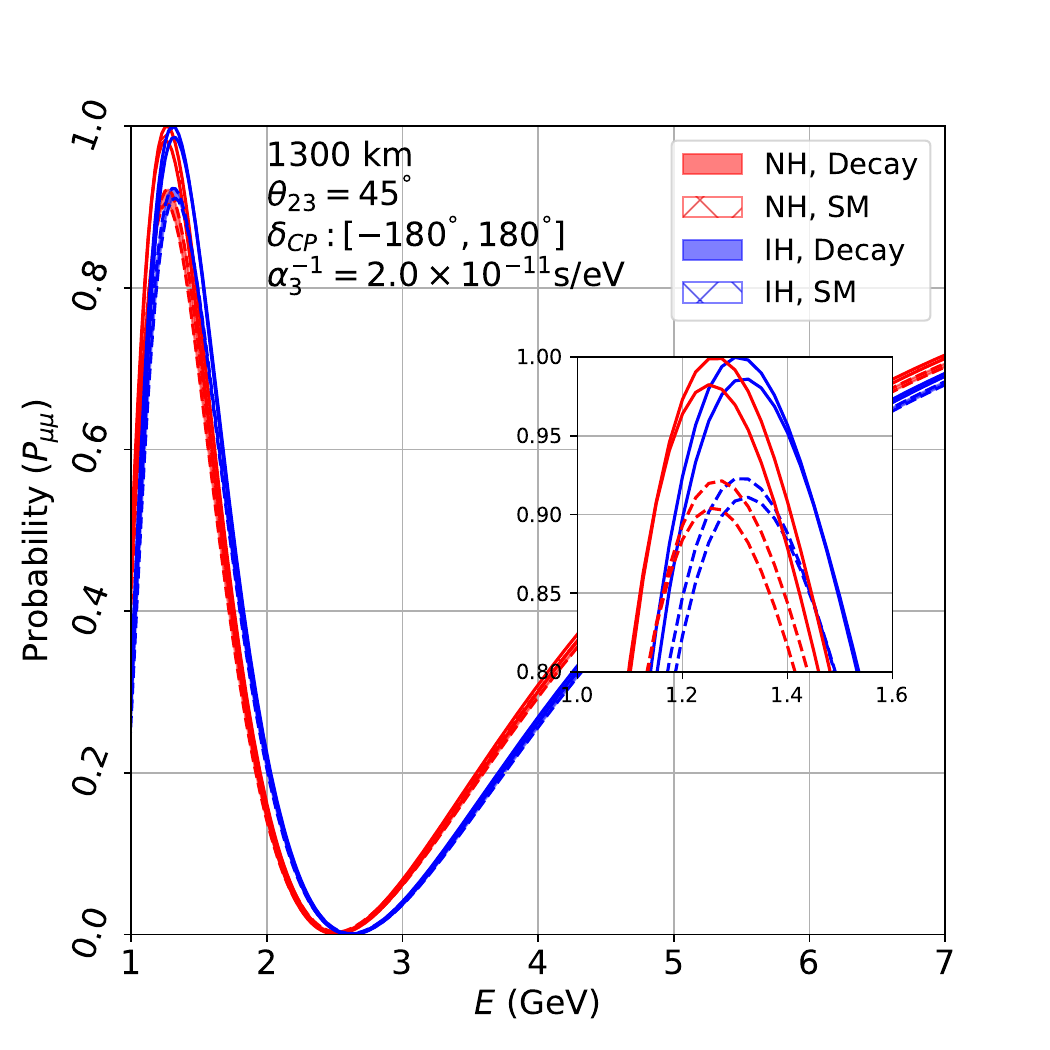}
    \caption{$P_{\mu e}$ (top) and $P_{\mu\mu}$ (bottom) as a function of energy $E$ at 2588 km (left) and 1300 km (right) without decay (hatched band) and with decay (solid band) for NH (red) and IH (blue).}
    \label{fig:pme-pmm-dec-nodec-p2o}
\end{figure}

In this section, we present the probabilities  $P_{\mu e}$  (top panel) and  $P_{\mu\mu}$ (bottom panel) for both the baselines of 2588 km (left) and 1300 km (right) in Figure \ref{fig:pme-pmm-dec-nodec-p2o}. The red (NH) and blue (IH) colored solid shaded bands correspond to decay, while the hatched area is for no decay. The width of these curves is due to the variation of $\delta_{CP}$ in the range $-180^\circ:180^\circ$. %The solid (dashed) lines are used for probabilities in case of no decay (decay) in the inset. 

The separation between NH and IH probability bands can give an idea about the hierarchy sensitivity. We observe the following points from these figures
\begin{itemize}
    \item The shaded probability bands for different hierarchies are lower in the presence of decay than the standard scenario with no decay. Comparing the dominant first terms in eq. \ref{eq:pme-dec} and eq. \ref{eq:pme-sm}, the $e^{-2\kappa_3}$ term leads to the lowering of $P_{\mu e}$ in the presence of decay. Similarly, $P_{\mu\mu}$ is affected due to the presence of $e^{-2\kappa_3}$ in the second term in eq.\ref{eq:pmm-dec} relative to eq. \ref{eq:pmm-sm}.

    \item In the appearance channel, decay affects NH more than  IH as the shift in bands of NH is greater than that of the bands of IH. This can lead to lower hierarchy sensitivity in the presence of decay.

    \item The width of the bands is narrower in the disappearance channel than the appearance channel as there is no dependence on $\delta_{CP}$ in leading terms as seen from eq.\ref{eq:pmm-dec}, but higher order terms show $\delta_{CP}$ dependence \cite{Gronroos:2024jbs}. The probability bands in the disappearance channel get lower when decay is present. However, there is no noticeable change in separation between the NH and IH bands near maxima in the decay case as compared to no decay. This can be understood more clearly from the insets in the lower panels.

    \item Near the minima of appearance probability curves, there are points where $\delta_{CP}$ dependence vanishes in the standard scenario. However, the inclusion of decay introduces dependence to $\delta_{CP}$ in these energies, e.g., 2 GeV, 3.5 GeV at 2588 km baseline, as can be seen from the inset of the plots in the upper panel. Thus, decay induces a slight dependence on $\delta_{CP}$ at the energies corresponding to the bimagic condition for 2588 km. Similar signatures can also be observed in 1300 km at 1.1 GeV and 1.5 GeV. The broadening of bands at minima can be explained by the dependence of the second term in eq. \ref{eq:pme-bm} on $\delta_{CP}$ only in the presence of decay.
\end{itemize}

\begin{figure}
    \centering
    \includegraphics[width=0.40\linewidth]{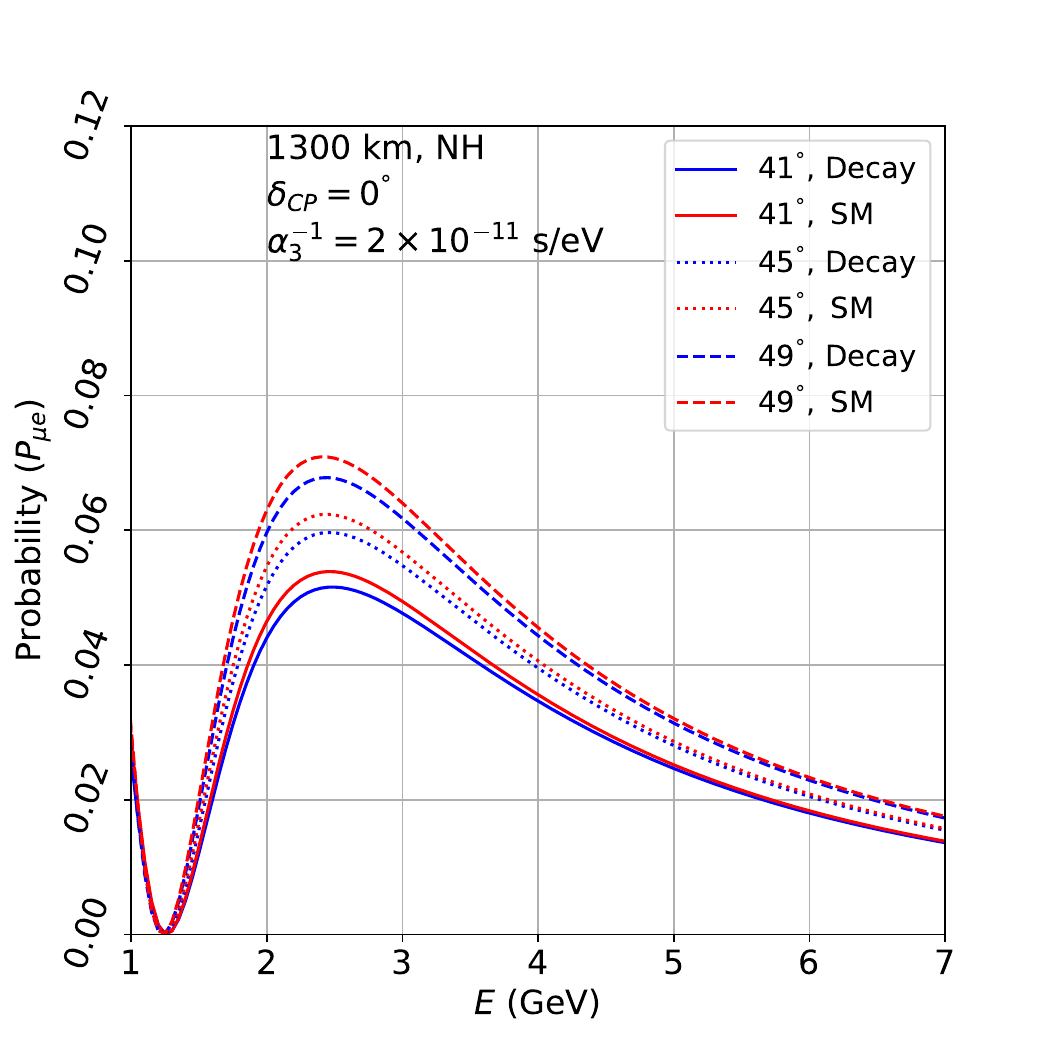}
    \includegraphics[width=0.40\linewidth]{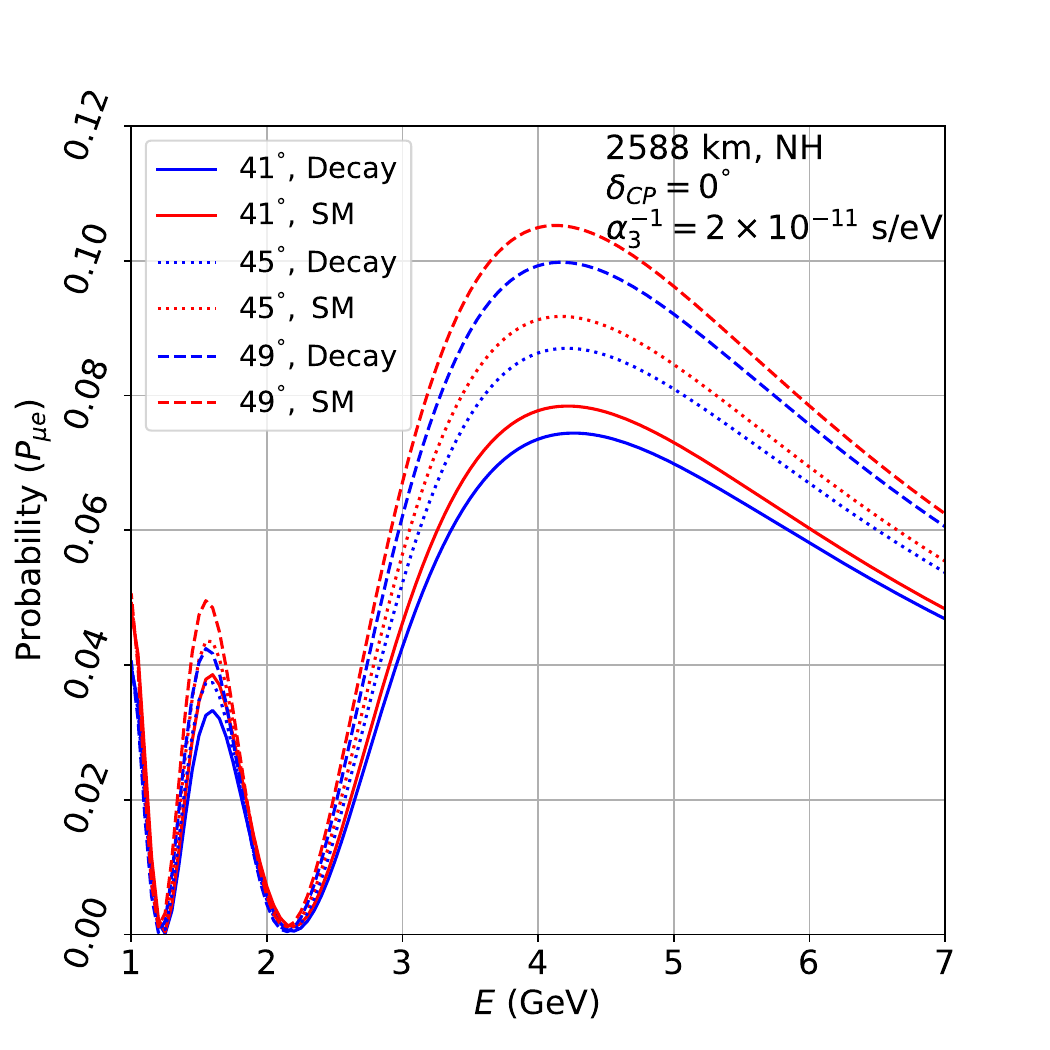}
    \includegraphics[width=0.40\linewidth]{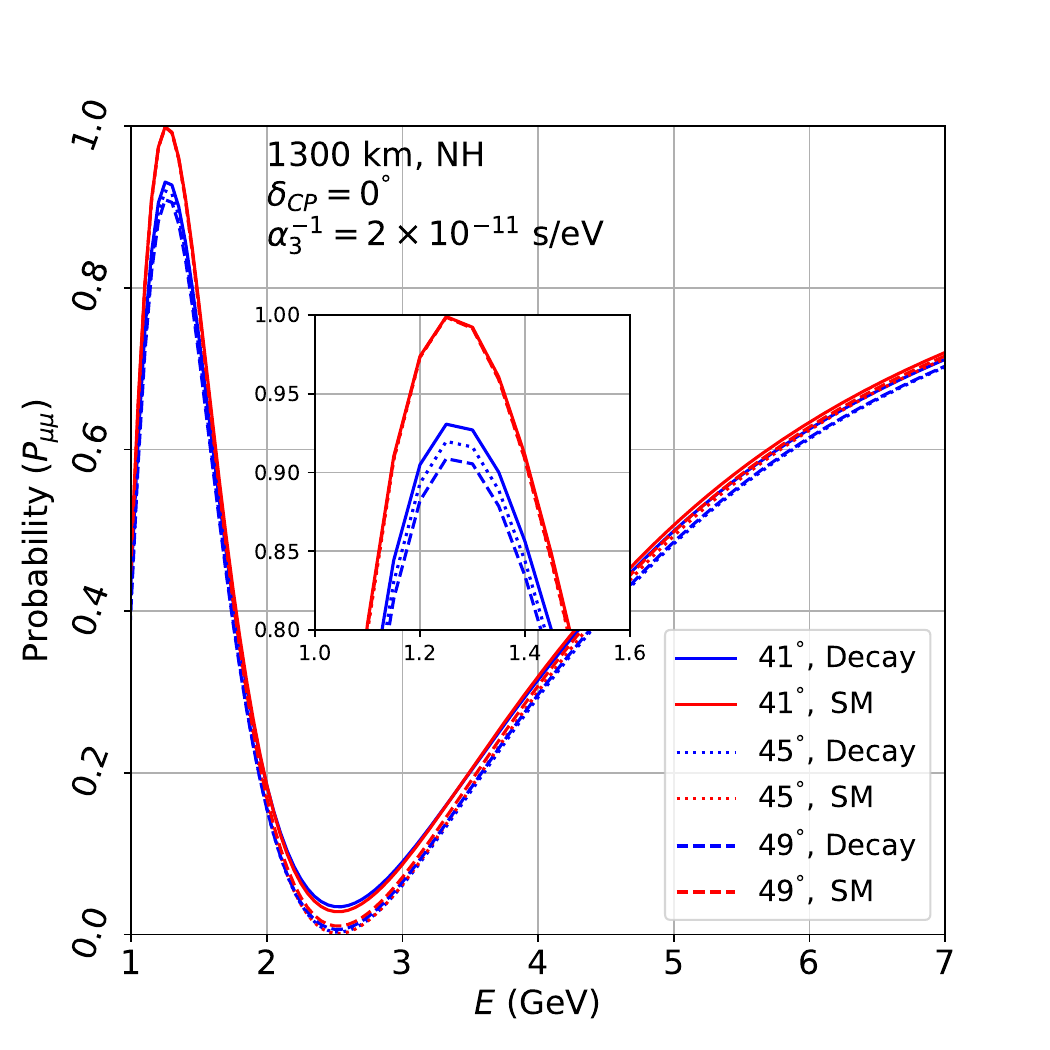}
    \includegraphics[width=0.40\linewidth]{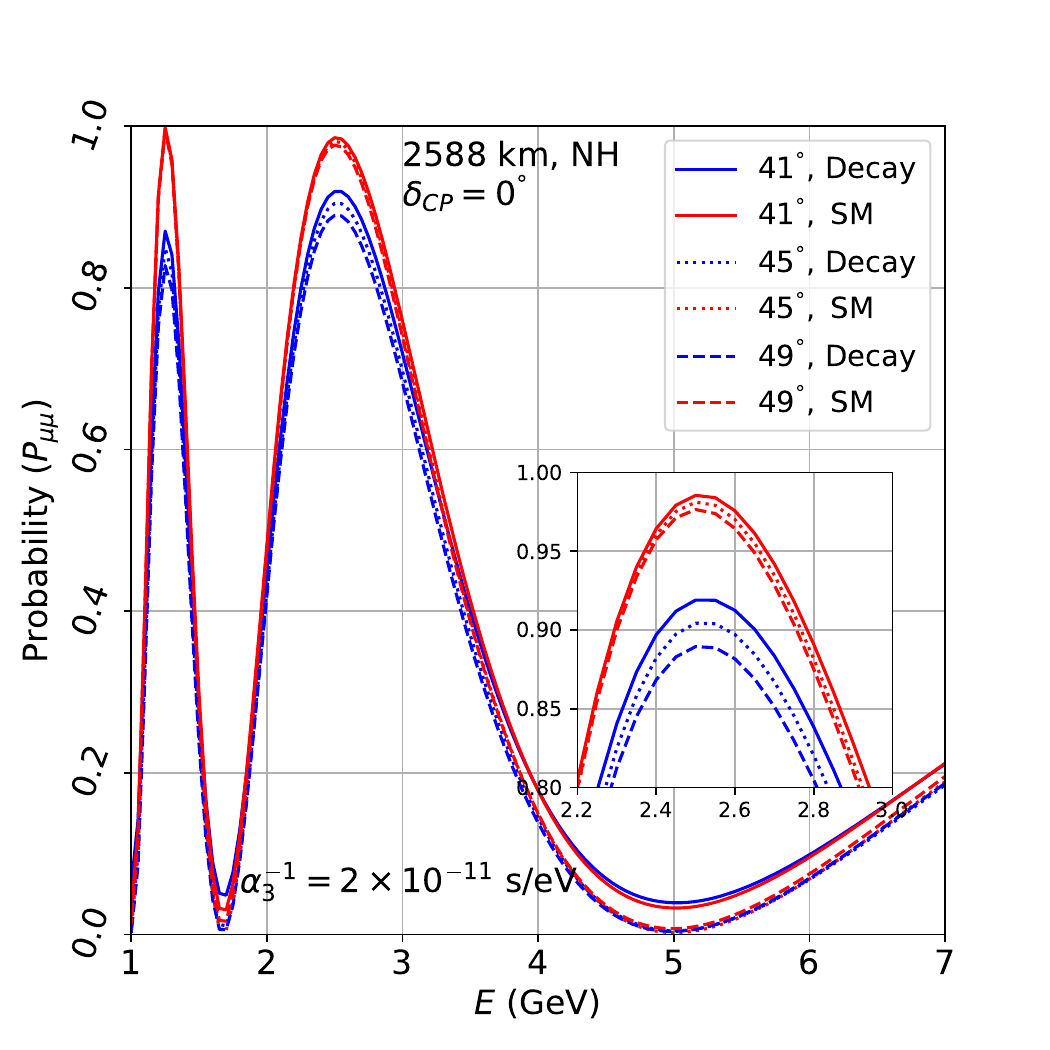}
    \caption{$P_{\mu e}$ (first row), $P_{\mu\mu}$ (second row) as a function of energy for $\gamma_3=2\times 10^{-11}$ s/eV (blue), no decay (red) in 2588 km (left) and 1300 km (right) baseline for NH. The solid, dotted, and dashed lines refer to values of $\theta_{23}$ of $41^\circ$ (LO), $45^\circ$, and $49^\circ$(HO).}
    \label{fig:pr-E-th23-gamma}
\end{figure}

Now, we discuss the change in the appearance and disappearance probabilities for different values of $\theta_{23}$ for two different setups. In figure \ref{fig:pr-E-th23-gamma}, the probability curves are depicted for no decay (red) and with decay using $\alpha_3 = 2\times 10^{-11}$ s/eV (red) with different values of  $\theta_{23}=41^\circ (\rm{solid}), 45^\circ (\rm{dotted}), 49^\circ (\rm{dashed})$ while considering both NH and IH. We also show an enlarged view of the first oscillation maximum of $P_{\mu\mu}$ in the bottom panel. We can infer the following from the above plots:

\begin{itemize}
    \item Both decay and $\theta_{23}$ can lower the probability amplitude near oscillation maxima. Thus, the effect of decay can mimic the effect of a lower $\theta_{23}$ and no decay near oscillation maxima.
    
    \item In the appearance channel ($P_{\mu e}$), the differences between probabilities corresponding to different $\theta_{23}$ are almost similar for decay and no decay scenarios near maxima. 

    \item In the disappearance channel ($P_{\mu \mu}$), the blue curves (decay) for different values of $\theta_{23}$ are more dispersed than the red curves in no decay case. This effect is relevant near both the first and second maxima of $P_{\mu\mu}$. This indicates that including decay may lead to higher octant sensitivity with a bigger contribution coming from $P_{\mu\mu}$.
\end{itemize}

\section{Experiment and Simulation details}\label{sec:expr}
To facilitate the simulation of the neutrino beams in LArTPC and P2O setups at 1300 km and 2588 km, we have employed the GLoBES \cite{Huber:2004ka} software package. We have described two different experimental setups used for this analysis below.

\subsection{A water Cherenkov detector configuration (2588 km)}\label{subsec:p2o}
%P2O: Protvino to ORCA(Oscillation Research with Cosmics in the Abyss) 
The Protvino accelerator facility, situated in the Moscow region, Russia, can be utilized as a robust platform for conducting extensive experimental research in the realm of neutrino physics. At its core is the U-70 synchrotron, with a circumference of 1.5 km, capable of accelerating protons up to 70 GeV and delivering a substantial annual proton-on-target (POT) yield of $4 \times 10^{20}$. This proton beam serves as the foundation for generating a neutrino beam with a power of 90 kW \cite{Choubey:2018rnl}.

The directed neutrino beam can be aimed at the ORCA detector, a component of the KM3NeT collaboration, strategically positioned 2588 km away from the Protvino accelerator facility. The ORCA detector, with a fiducial target mass of 4 Mton, is assumed to be operating in an environment characterized by an average matter density of 3.4 g/cc.

The experiment is envisioned to span three years each in the neutrino and antineutrino modes, leading to a total exposure of 2160 MW.Kton.yr. The energy resolution is modeled as a Gaussian distribution with a width of 30$\%$. This setup enables an exploration of neutrino oscillation phenomena in the presence of matter effects. The rules representing different (anti)neutrino channels along with the corresponding signal and background coefficients are listed in table \ref{tab:globes-rules-p2o}.

\begin{table}
\centering
\begin{tabular}{|l|c|l|}
\hline
\textbf{Rule}                          & \textbf{Signal Coefficient} & \textbf{Background Coefficients}                                              \\ \hline
$\nu_e$ Appearance          & 0.9                         & $\nu_{\tau}$ Appearance: 0.78 \\ 
                                     &                              & $\nu_{\mu}$ Disappearance: 0.45 \\
                                     &                              & Neutrino Beam: 0.9 \\
                                     &                              & Neutrino Misidentification: 0.2 \\
                                     &                              & Neutral-Current Background: 0.9 \\ \hline
$\nu_\mu$ Disappearance           & 0.7                          & $\nu_{\tau}$ Appearance: 0.22 \\
                                     &                              & Neutral-Current Background: 0.18 \\
                                     &                              & $\nu_{e}$ Appearance: 0.15 \\ \hline
$\bar{\nu}_e$ Appearance     & 0.9                          & $\nu_{\tau}$ Appearance: 0.78 \\
                                     &                              & $\nu_{\mu}$ Disappearance: 0.45 \\
                                     &                              & Antineutrino Beam: 0.9 \\
                                     &                              & Antineutrino Misidentification: 0.2 \\
                                     &                              & Neutral-Current Background: 0.9 \\ \hline
$\bar{\nu}_\mu$ Disappearance     & 0.7                          & Neutral-Current Background: 0.18 \\
                                     &                              & $\nu_{\tau}$ Appearance: 0.22 \\
                                     &                              & $\nu_{e}$ Appearance: 0.15 \\ \hline
\end{tabular}
\caption{Signal and Background coefficient for different rules in GLoBES file for P2O.}
\label{tab:globes-rules-p2o}
\end{table}

\subsection{A liquid Argon detector setup (1300 km)}\label{subsec:dune}

The experimental setup under consideration consists of a megawatt-scale muon neutrino beam source, a near detector (ND), and a far detector(FD). The ND will be placed close to the beam's source, while the FD, comprising a 40 Kton LArTPC detector, will be placed at a distance of 1300 km away from the neutrino source. In our study, we have considered events from FD only. The large LArTPC at an underground observatory can also observe atmospheric neutrinos. The proposed DUNE experiment has a similar experimental configuration\cite{DUNE:2016ymp}. In this analysis, a beam-power of 1.2 MW leading to a total exposure of $1\times 10^{21}$ POT/yr has been implemented for the numerical analysis. The energy resolutions, efficiency, and systematic errors are taken as per the .glb file in \cite{DUNE:2016ymp}. We assume the experiment will run for five years each in the neutrino and antineutrino modes, corresponding to a total exposure of 480 MW.Kton.yr. We have considered a constant matter density of 2.85 g/cc. The various background channels for different rules have been mentioned in table \ref{tab:globes-rules-dune}, and each of these backgrounds is listed in files as a function of energy. 

\begin{table}[htbp]
\centering
\begin{tabular}{|l|c|l|}
\hline
\textbf{Rule}  & \textbf{Background Channels} \\
\hline
$\nu_e$ Appearance  & $\nu_e$, $\bar{\nu}_e$, $\nu_\mu$, $\bar{\nu}_\mu$, $\nu_\tau$, $\bar{\nu}_\tau$, $\nu$ NC, $\bar{\nu}$ NC\\
\hline
$\nu_\mu$ Disappearance  & $\nu_\tau$, $\bar{\nu}_\tau$, $\nu$ NC, $\bar{\nu}$ NC\\
\hline
$\bar{\nu}_e$ Appearance & $\nu_e$, $\bar{\nu}_e$, $\nu_\mu$, $\bar{\nu}_\mu$, $\nu_\tau$, $\bar{\nu}_\tau$, $\nu$ NC, $\bar{\nu}$ NC\\
\hline
$\bar{\nu}_\mu$ Disappearance   & $\nu_\tau$, $\bar{\nu}_\tau$, $\nu$ NC, $\bar{\nu}$ NC\\
\hline
\end{tabular}
\caption{Background channels for different GLoBES rules in LArTPC detector.}
\label{tab:globes-rules-dune}
\end{table}

The table \ref{tab:tr-ts-param} provides a summary of the true and test values of various oscillation parameters used in the simulation. The true and test values of $\theta_{12}$ and $\theta_{13}$ are fixed at $33.4^\circ$ and $8.42^\circ$ respectively along with $\Delta_{21}$ at $7.53\times10^{-5}$ eV$^2$.
In this analysis, both the true and test values of $\delta_{CP}$ have been varied in a range from $-180^{\circ}$ to $180^{\circ}$ with a step size of $15^{\circ}$. The exploration of $\alpha_{3}^{-1}$ extends from $1.0\times10^{-11}$ s/eV to $3.0\times10^{-11}$ s/eV, with the true value set at $2.0\times10^{-11}$ s/eV. For hierarchy sensitivity, the sign of $\Delta_{31}$ is reversed, whereas, for octant sensitivity analyses, the test value is varied in the opposite octant in the range as specified in table \ref{tab:tr-ts-param} for a true value of $\theta_{23}$. 

\begin{table}[H]
    \centering
    \begin{tabular}{|l|r|c|}
    \hline
    \textbf{Parameters} & \textbf{True values} & \textbf{Test values}\\
    \hline
    $\theta_{12}$ & $33.4^{\circ}$ & $33.4^{\circ}$ \\
    $\theta_{13}$ & $8.42^{\circ}$ & $8.42^{\circ}$ \\
    $\theta_{23}$ (Hierarchy)& $41^{\circ}, 45^{\circ}, 49^{\circ}$ & $39^{\circ}$ : $51^{\circ}$ \\
    $\theta_{23}$ (Octant)& $41^{\circ}$ ($49^{\circ}$) & $45^{\circ}$ : $51^{\circ}$ ($39^{\circ}$ : $45^{\circ}$) \\
    $\delta_{CP}$ & $-180^{\circ}$ : $180^{\circ}$  & $-180^{\circ}$ : $180^{\circ}$ \\
    $\delta_{CP}$(CP discovery)  & $-90^{\circ}, 0^\circ, 90^{\circ}$  & $-180^{\circ}$, 0, $180^{\circ}$ \\
    $\Delta_{21}$ & $7.53\times10^{-5}$ eV$^2$ & $7.53\times 10^{-5}$ eV$^2$\\
    $\Delta_{31}$(Hierarchy) & $\pm2.45\times10^{-3}$ eV$^2$ & $\mp[2.35$ : $2.6]\times10^{-3}$ eV$^2$ \\
    $\Delta_{31}$(Octant) & $\pm2.45\times10^{-3}$ eV$^2$ & $\pm[2.35$ : $2.6]\times10^{-3}$ eV$^2$ \\
    $\alpha_{3}^{-1}$ & $2.0\times10^{-11}$ s/eV& $[1.0$ : $3.0]\times10^{-11}$ s/eV \\
    \hline
    \end{tabular}
    \caption{True and test values of oscillation parameters.}
    \label{tab:tr-ts-param}
\end{table}

\section{Results}\label{sec:results}
In this section, we evaluate the sensitivity of the two above-mentioned experiments regarding the determination of the mass hierarchy, the octant of $\theta_{23}$, and CP violation within scenarios involving potential invisible decay of the $\nu_{3}$ state. Initially, we examine the ability of these experiments to detect such decay by assessing its sensitivity under the assumption of no decay. 

\subsection{Decay sensitivity of the two experimental setups}\label{subsec:dune}
First, we have evaluated the sensitivity of the detectors to decay. The simulated true data is generated assuming no decay scenario, while decay is introduced in the test. All the oscillation parameters are kept fixed in true and test as in table \ref{tab:tr-ts-param} while $\theta_{23}=45^\circ, \delta_{CP}=0^\circ$. Our findings indicate that a $3 \sigma$ sensitivity level can be achieved at a decay rate of approximately $2.9\times 10^{-11}$ s/eV ($3.3\times 10^{-11}$ s/eV) and a $5 \sigma$ sensitivity level at around $1.7\times10^{-11}$ s/eV ($1.9\times10^{-11}$ s/eV) for LArTPC (P2O) setup Sensitivity for NH and IH, as given by red and blue curves, are almost identical.
\begin{figure}[H]
\centering 
\includegraphics[width=0.4\textwidth]{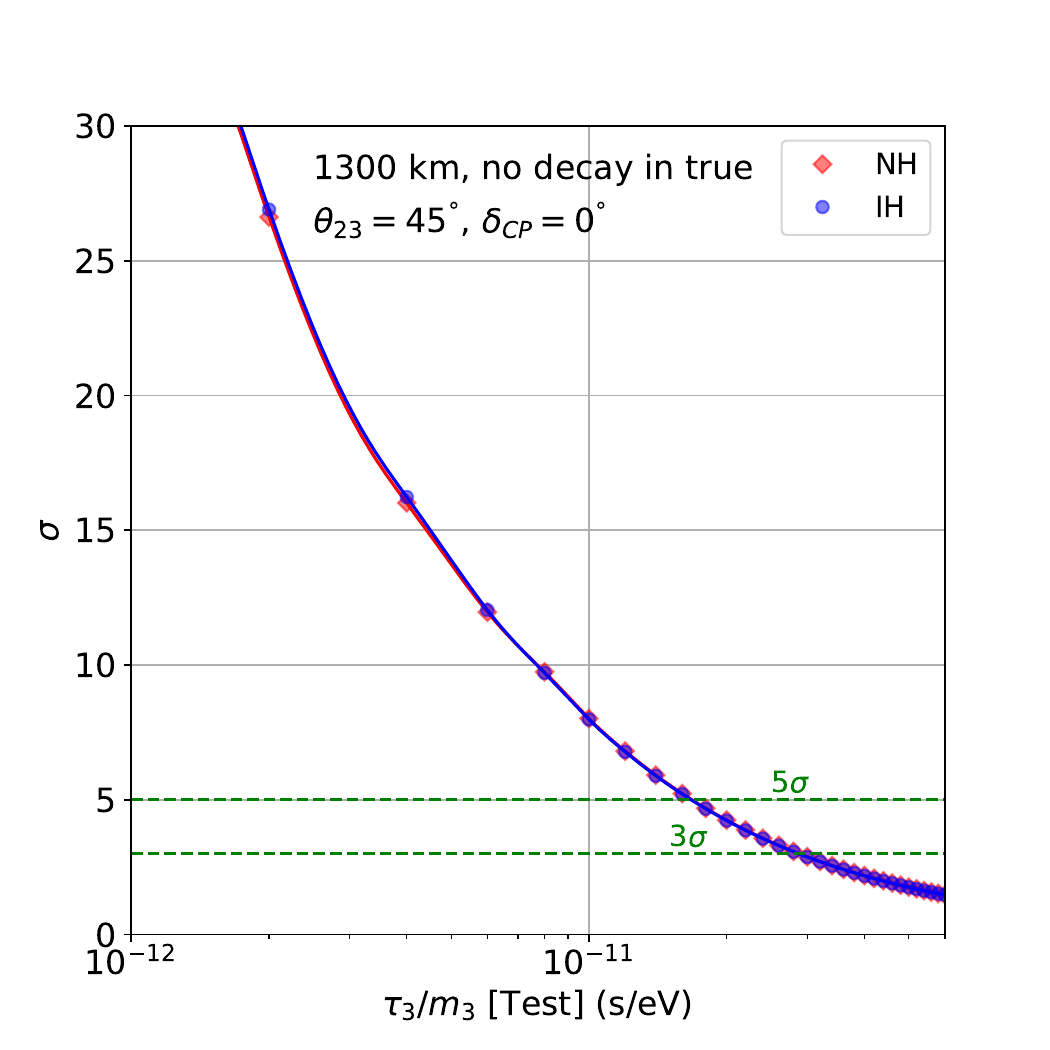}
\includegraphics[width=.4\textwidth]{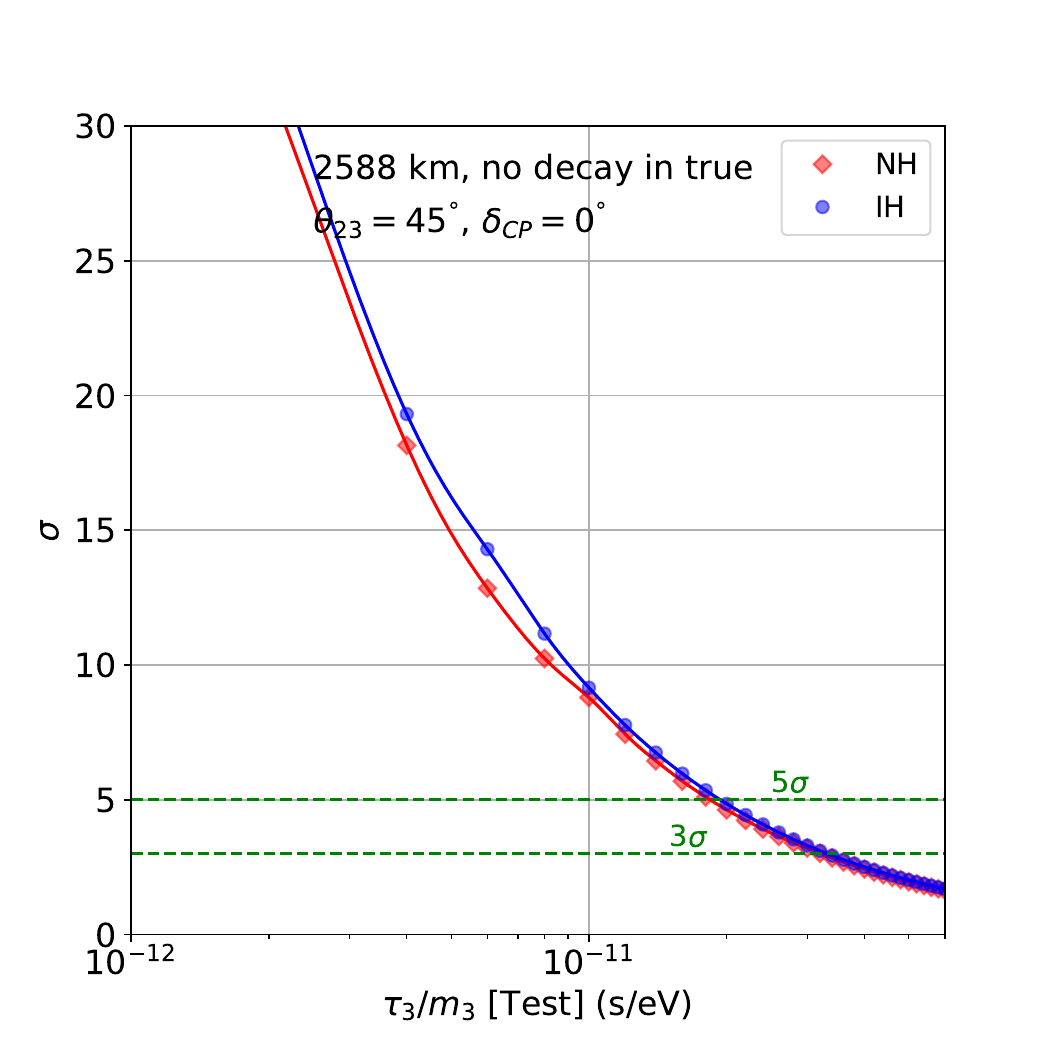}
\caption{Sensitivity of LArTPC (left), P2O (right) setups to detect decay as a function of decay parameter $\alpha_3^{-1}=\tau_3/m_3$ in test for $\theta_{23}=45^\circ, \delta_{CP}=0^\circ$ for NH (red) and IH (blue).}
\label{fig:chi-tau-dune-p2o} 
\end{figure}

\subsection{Mass Hierarchy Sensitivity}\label{subsec:hierarchy}
\begin{figure}
    \centering 
    \includegraphics[width=.32\textwidth]{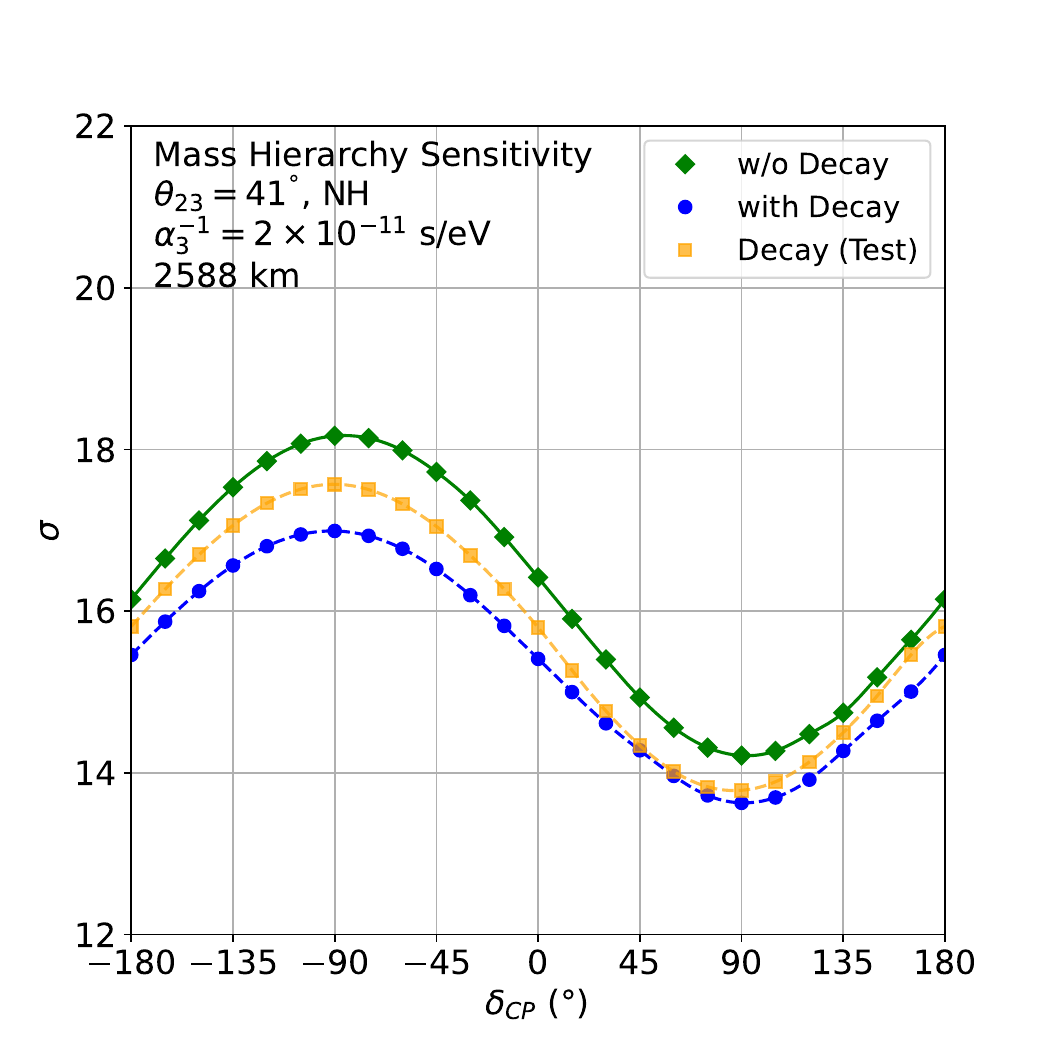}
    \includegraphics[width=.32\textwidth]{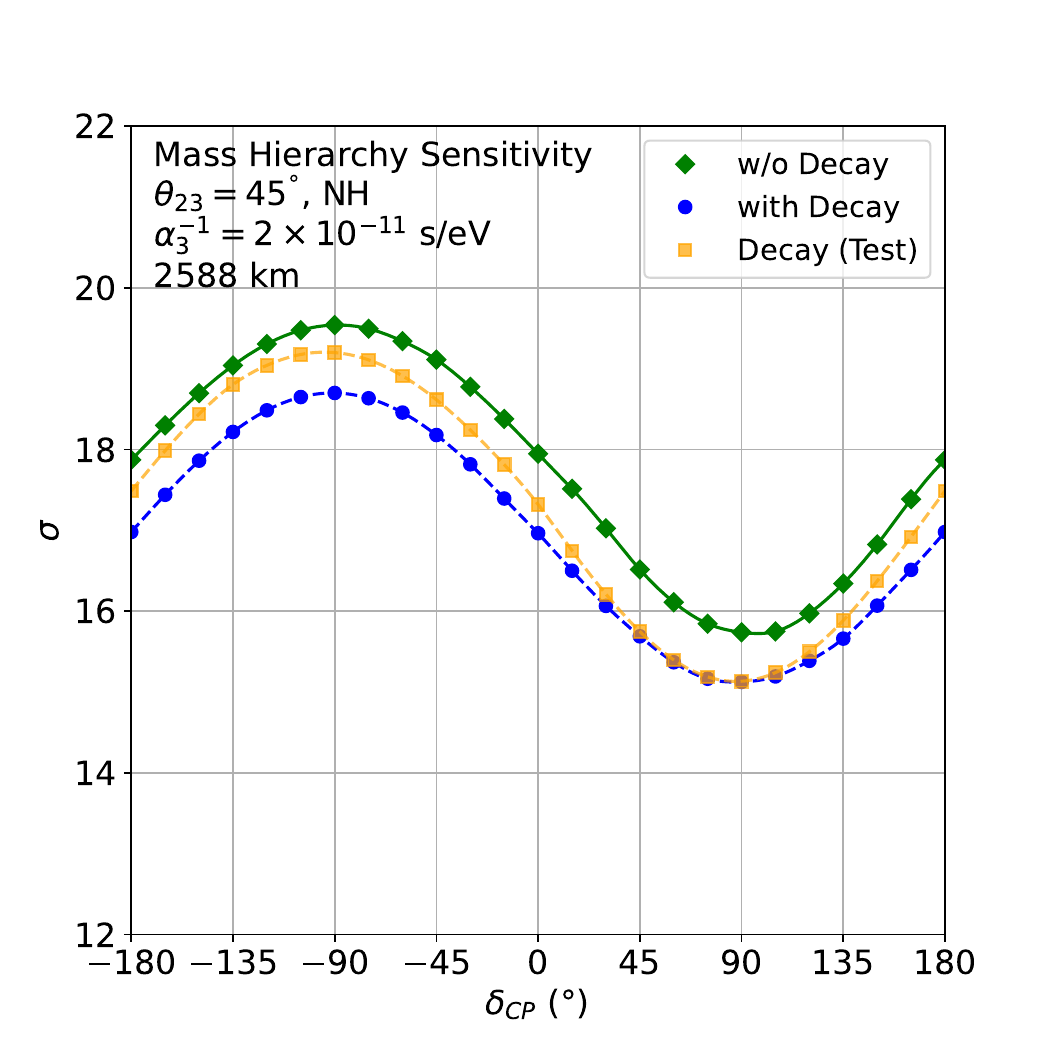}
    \includegraphics[width=.32\textwidth]{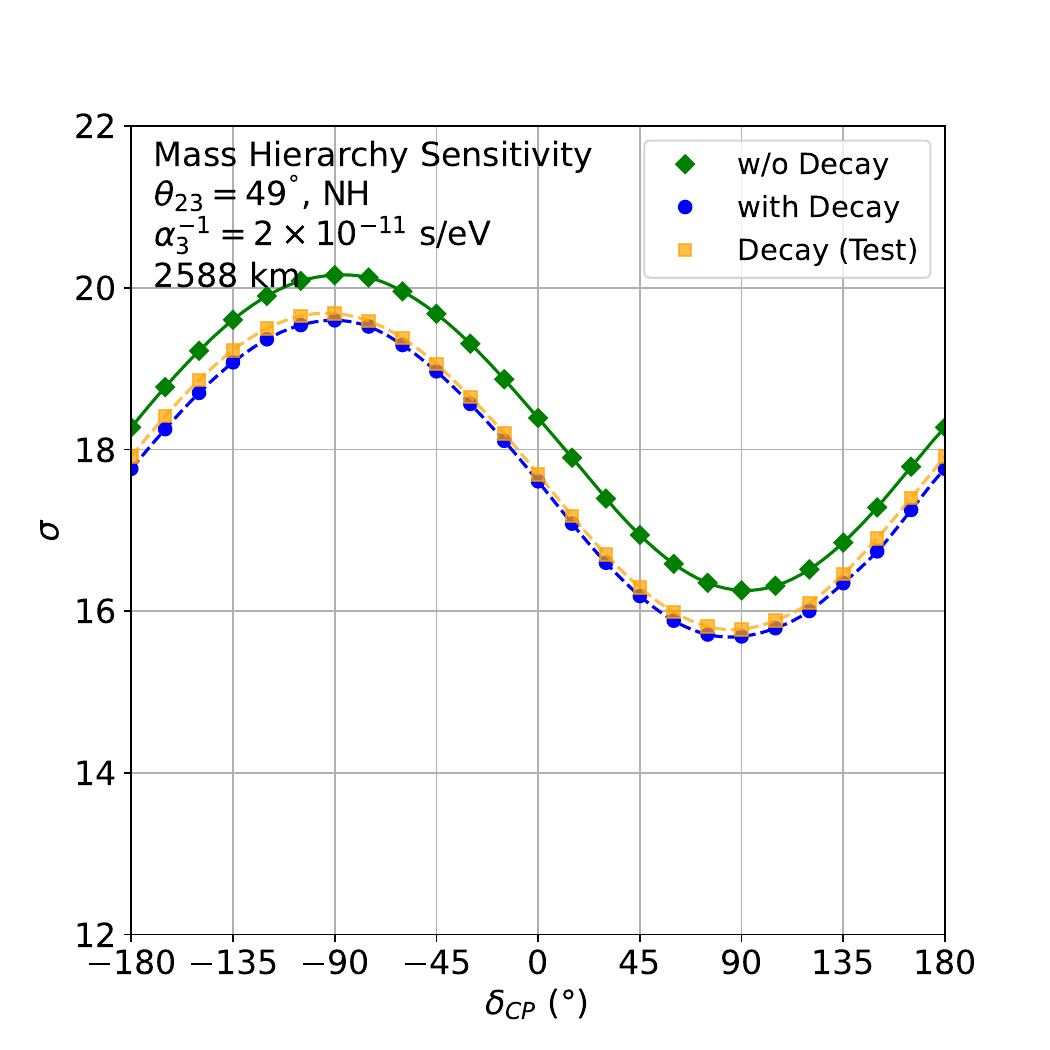}
    \includegraphics[width=.32\textwidth]{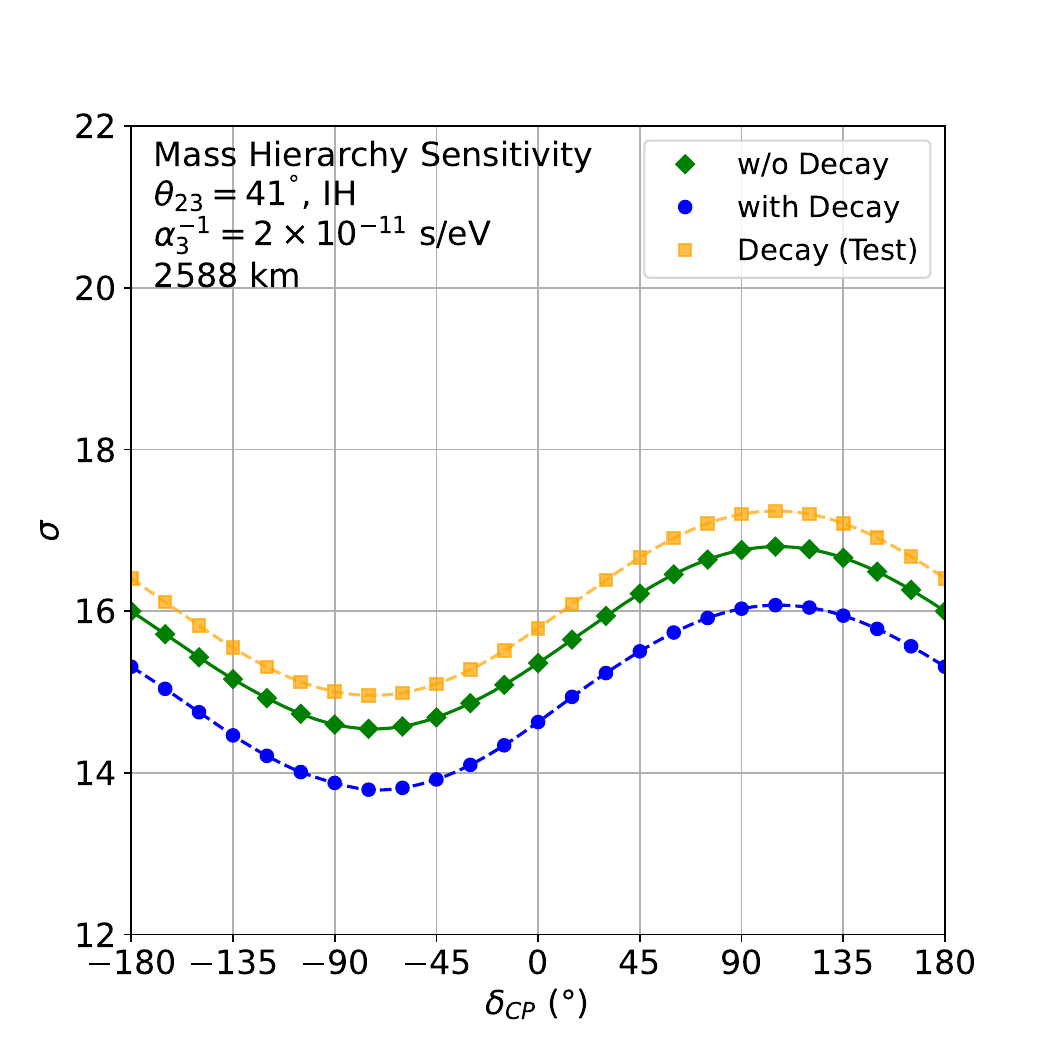}
    \includegraphics[width=.32\textwidth]{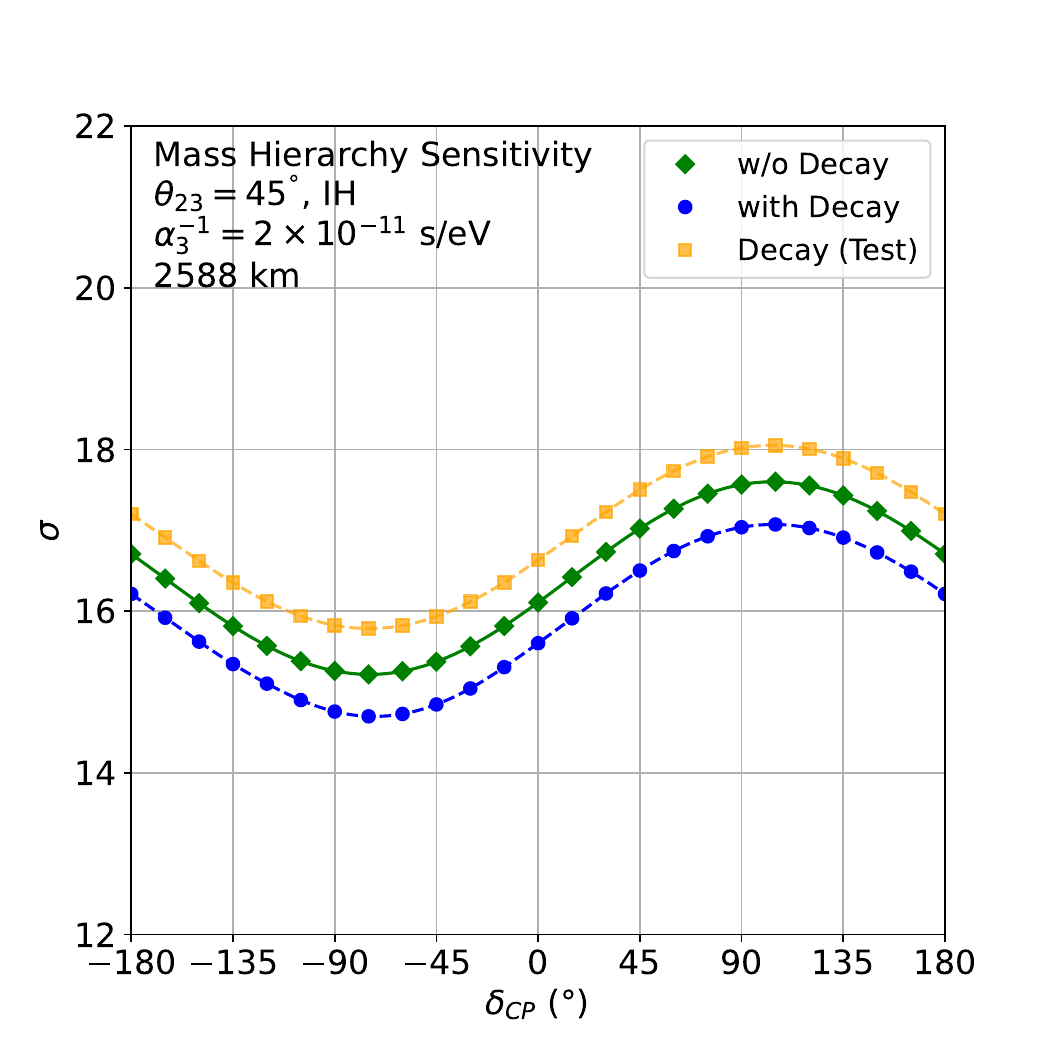}
    \includegraphics[width=.32\textwidth]{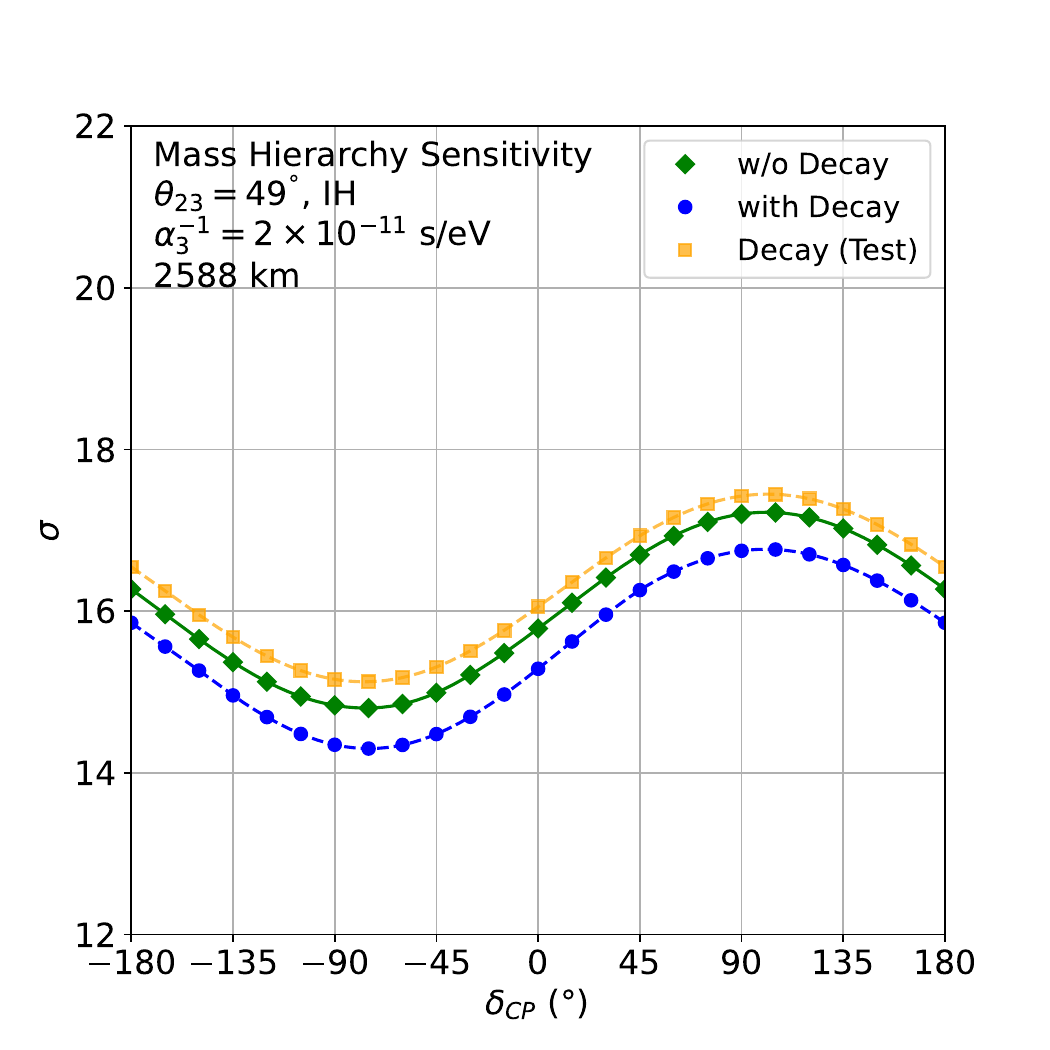}
    \caption{Sensitivity to mass hierarchy as a function of true $\delta_{CP}$ for true values of $\theta_{23}=41^\circ$ (top), $45^\circ$ (middle), $49^\circ$ (bottom) considering $ \alpha_3^{-1} =2.0\times 10^{-11}$ s/eV at 2588 km for NH (top), IH (bottom). The green curves correspond to the standard without decay scenario. The blue and orange curves are for decay present in both true and test, and only in the test, respectively.}
    \label{fig:chi-dcp-hr-p2o}
\end{figure}

In the figure \ref{fig:chi-dcp-hr-p2o}, the sensitivity to the mass hierarchy has been depicted considering $\theta_{23}=41^\circ$ (left), $45^\circ$ (middle) and $49^\circ$(right) for true hierarchy taken as normal (top) and inverted (bottom). The green curve shows the mass hierarchy sensitivity for a standard scenario without decay. We consider two different decay scenarios: the blue curve shows the same in the presence of decay in both the true and test, while the orange curve shows sensitivity in the presence of decay only in the test, i.e., in the opposite hierarchy. The observations from the plots in figure \ref{fig:chi-dcp-hr-p2o}, are as follows,

\begin{itemize}
    \item In the presence of decay (decay is considered in true and test), there is a decrease in hierarchy sensitivity for both NH and IH compared to the standard scenario. However, when decay is considered only in the test, for true IH, the hierarchy sensitivity is more than the no decay case. This is because the decay affects NH more as seen in the probability plots in figure \ref{fig:pme-pmm-dec-nodec-p2o}.

    \item If we compare the plots in the first row for NH, hierarchy sensitivity is seen to rise with the true value of $\theta_{23}$ for all the cases. But, while considering IH as the true hierarchy, the highest sensitivity is observed for the maximal value of $\theta_{23}=45^\circ$. 
\end{itemize}

\begin{figure}[h!]
\centering 
\includegraphics[width=.32\textwidth]{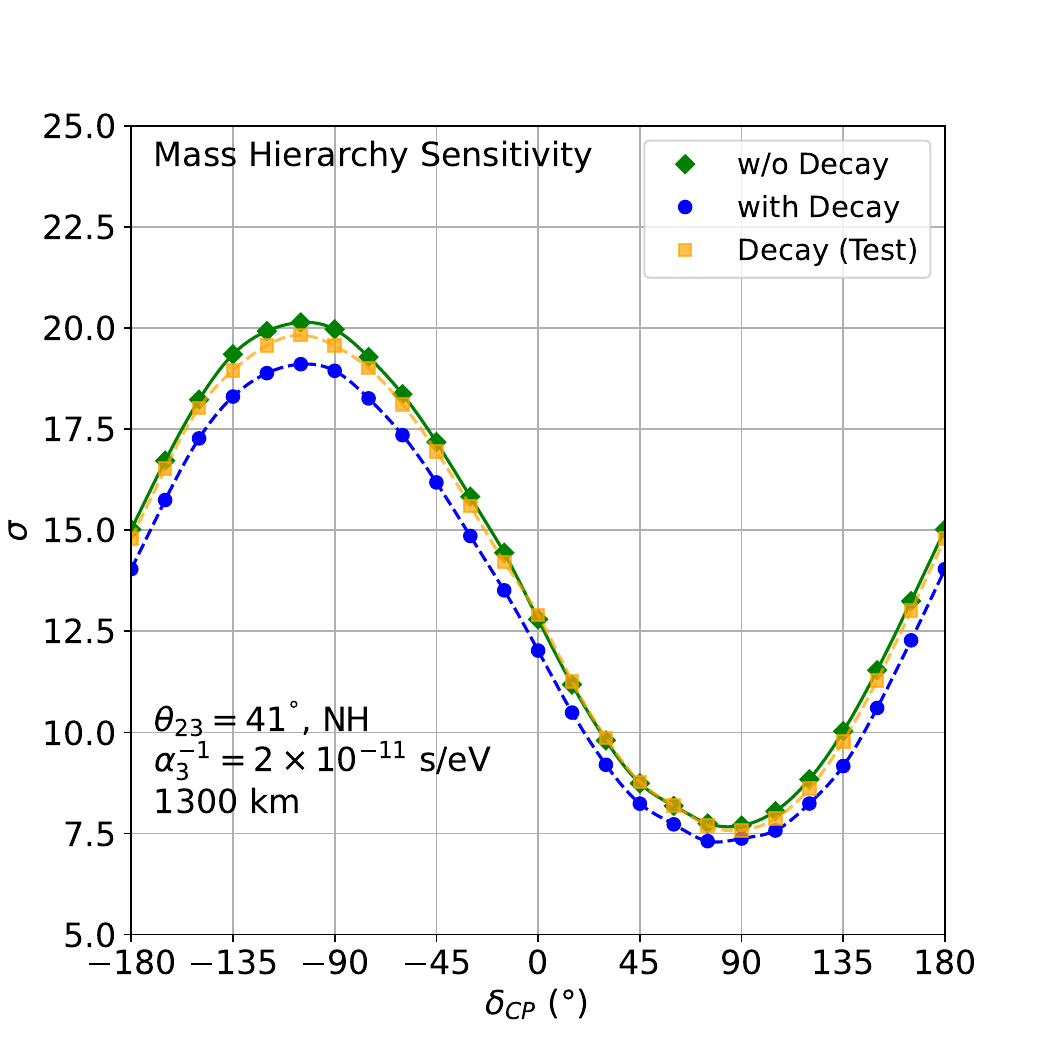}
\includegraphics[width=.32\textwidth]{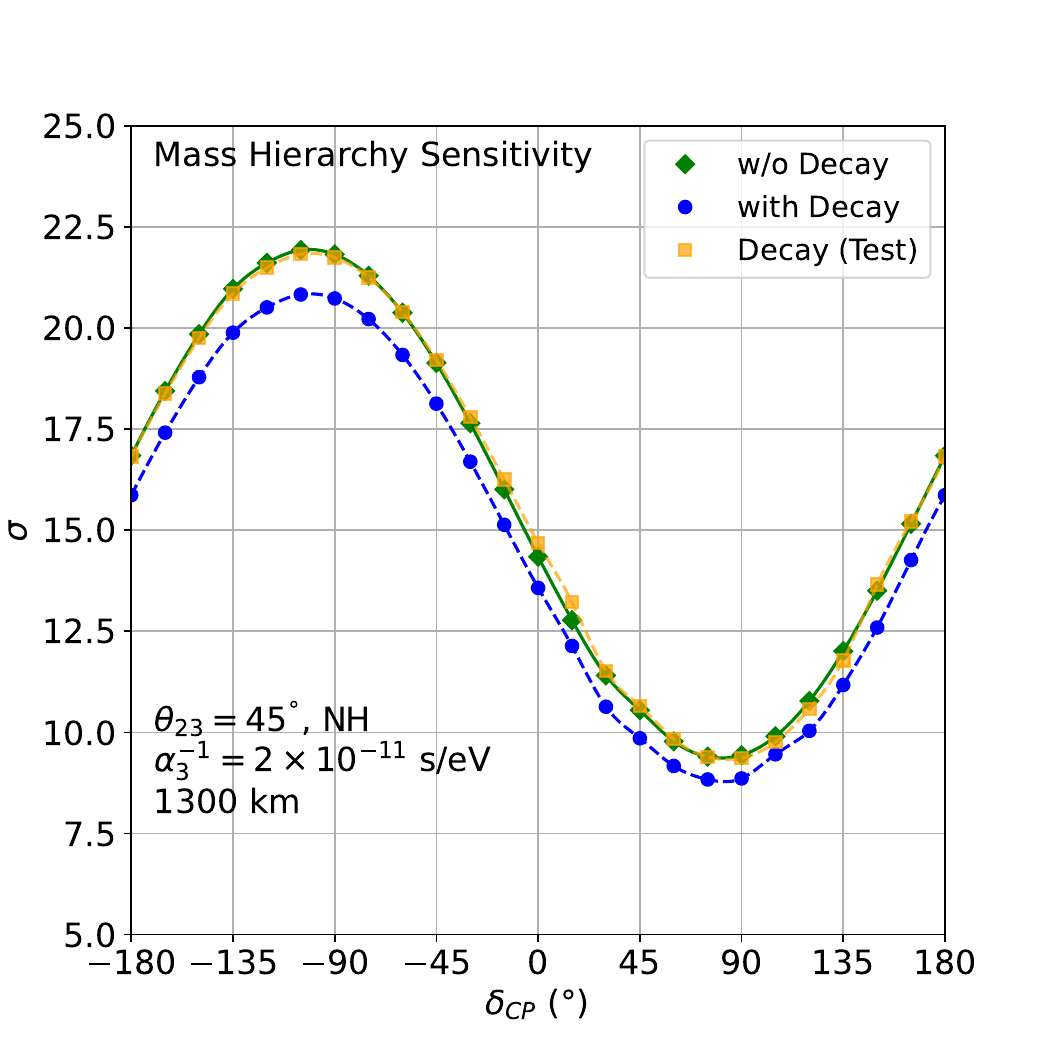}
\includegraphics[width=.32\textwidth]{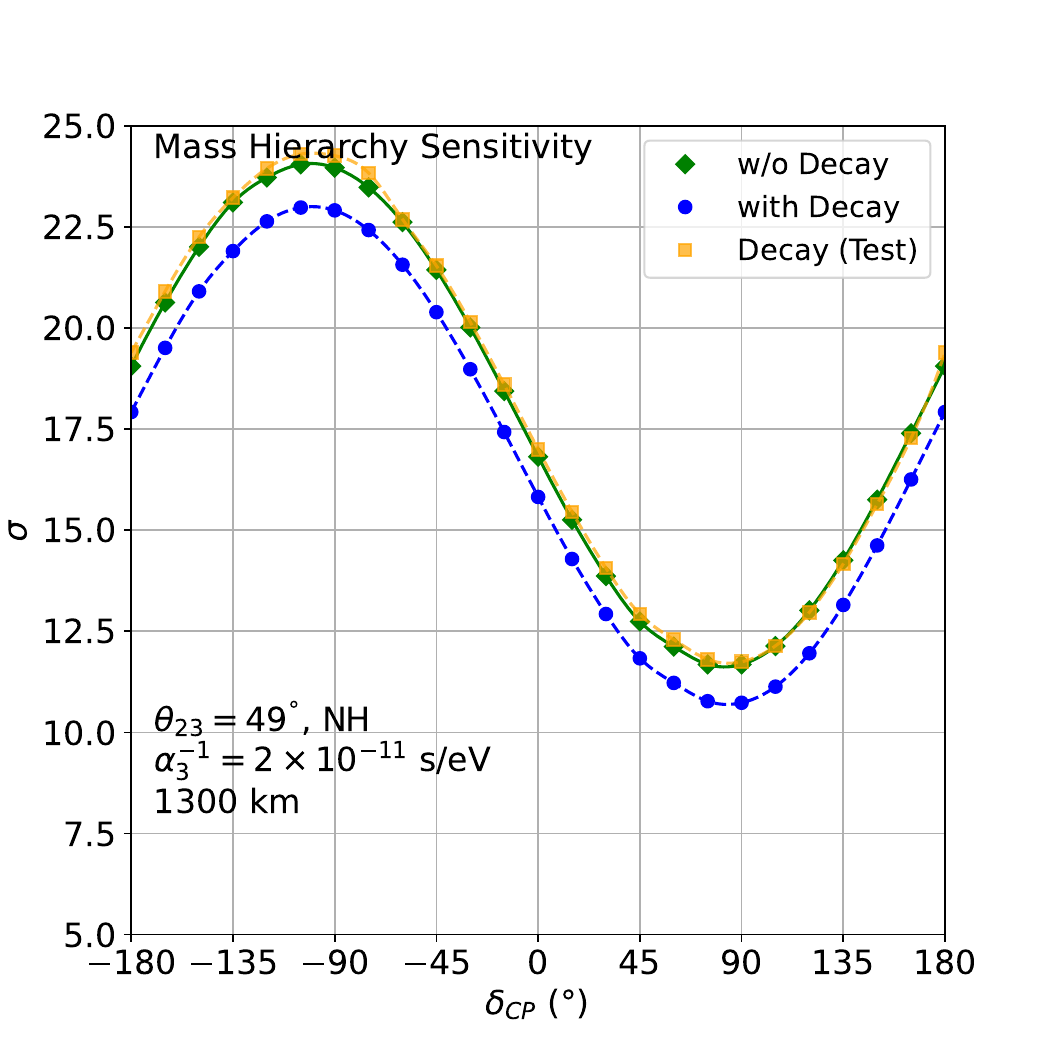}
\includegraphics[width=.32\textwidth]{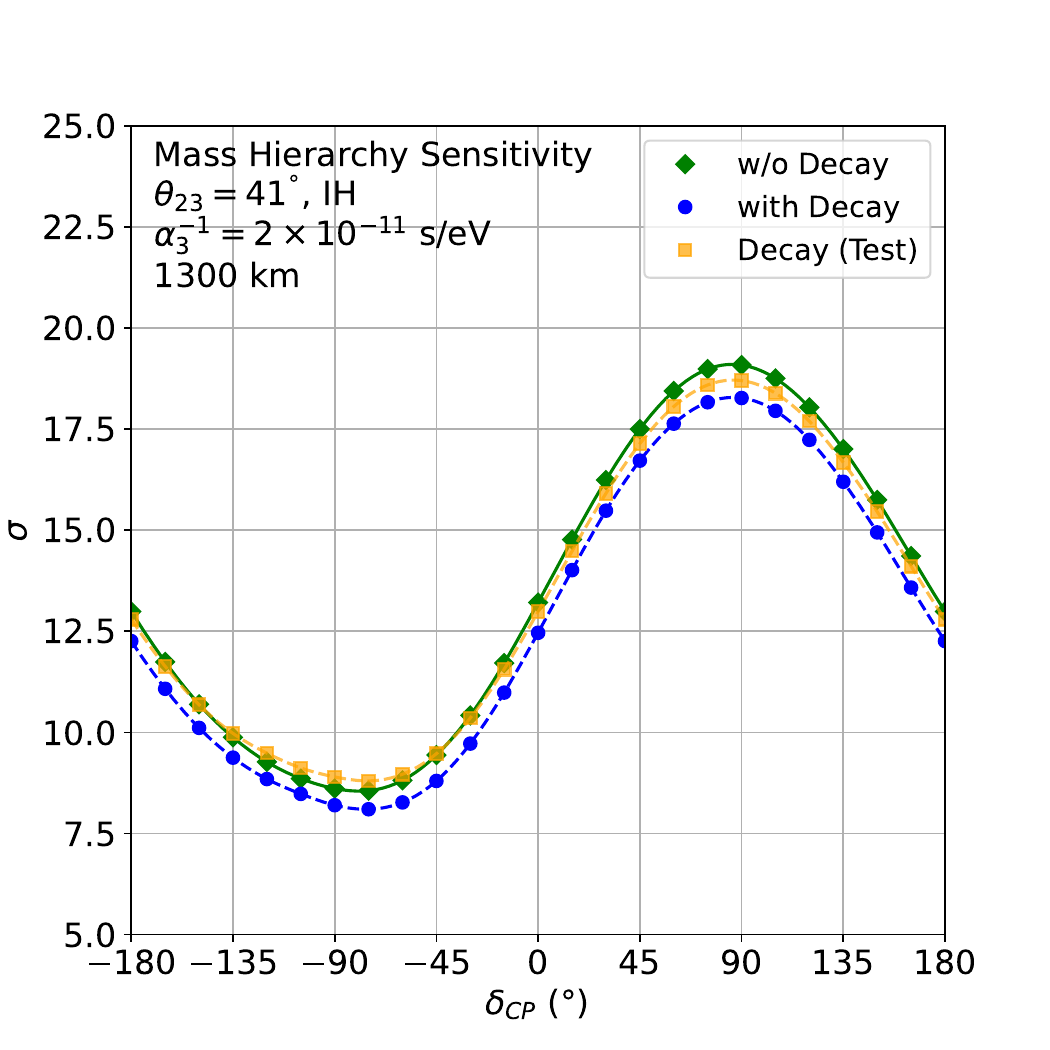}
\includegraphics[width=.32\textwidth]{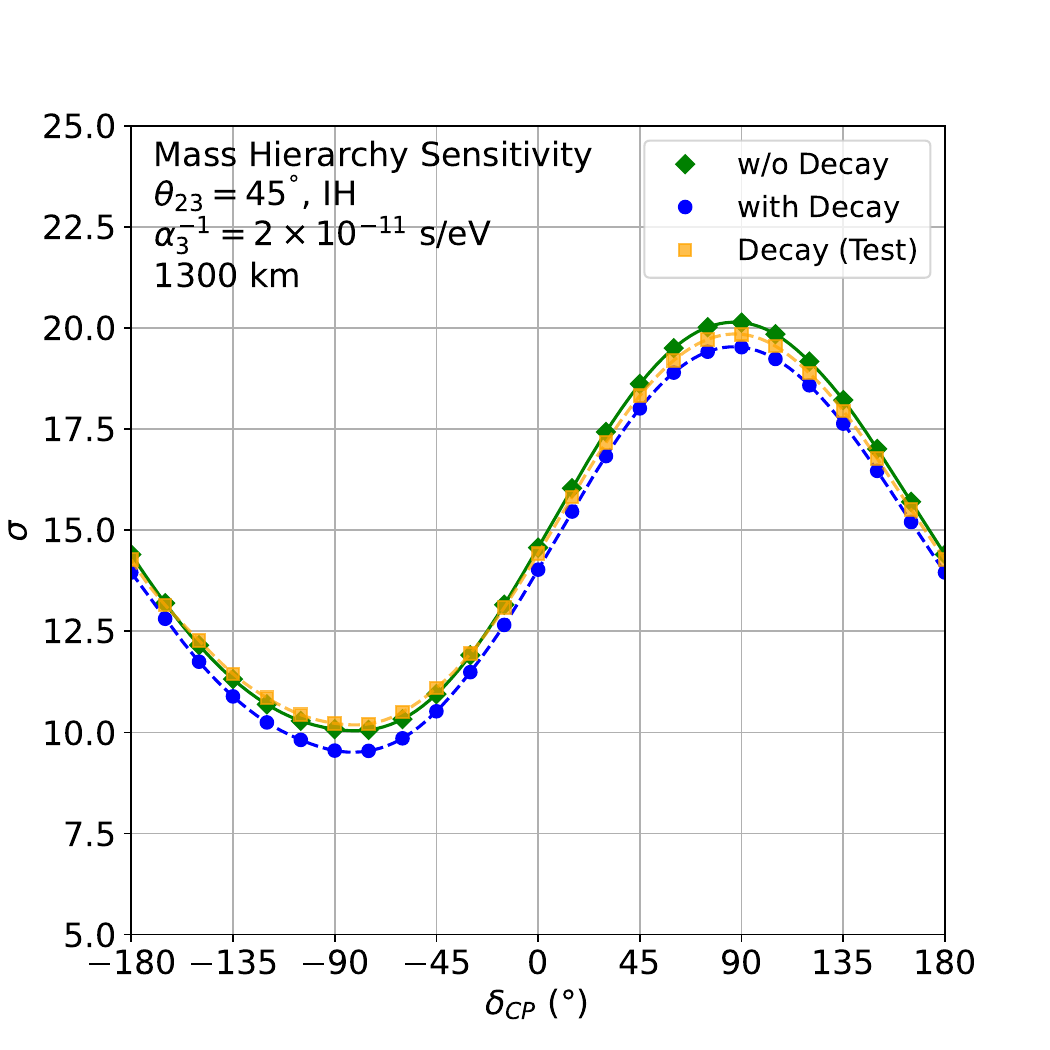}
\includegraphics[width=.32\textwidth]{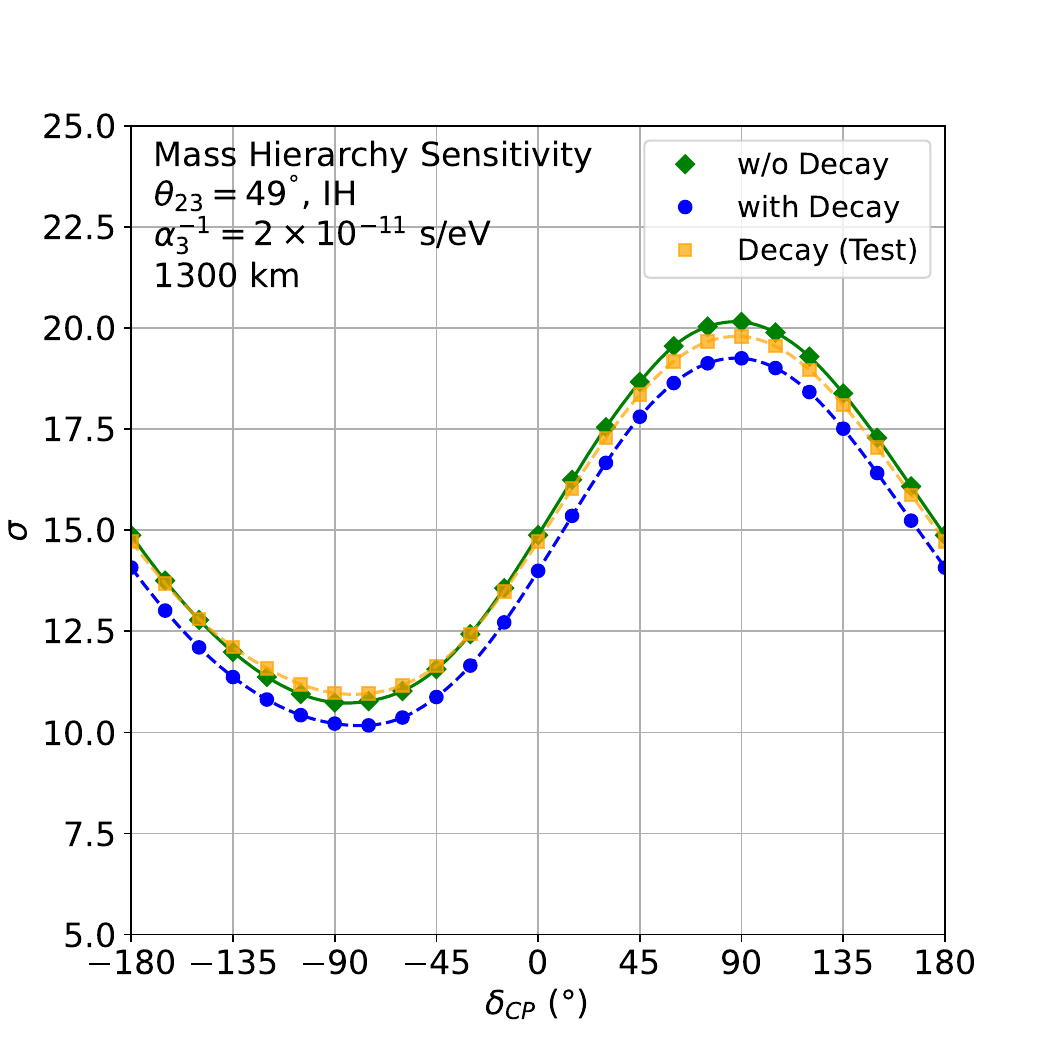}
\caption{Sensitivity to mass hierarchy as a function of true $\delta_{CP}$ for true values of $\theta_{23}=41^\circ$ (top), $45^\circ$ (middle), $49^\circ$ (bottom) considering $ \alpha_3^{-1} =2.0\times 10^{-11}$ s/eV at 1300 km for NH (top), IH (bottom). The green curves correspond to the standard without decay scenario. The blue and orange curves are for decay present in both true and test, and only in the test, respectively.}
\label{fig:chi-dcp-hr-dune} 
\end{figure}

\begin{figure}[H]
    \centering
    \includegraphics[width=0.32\linewidth]{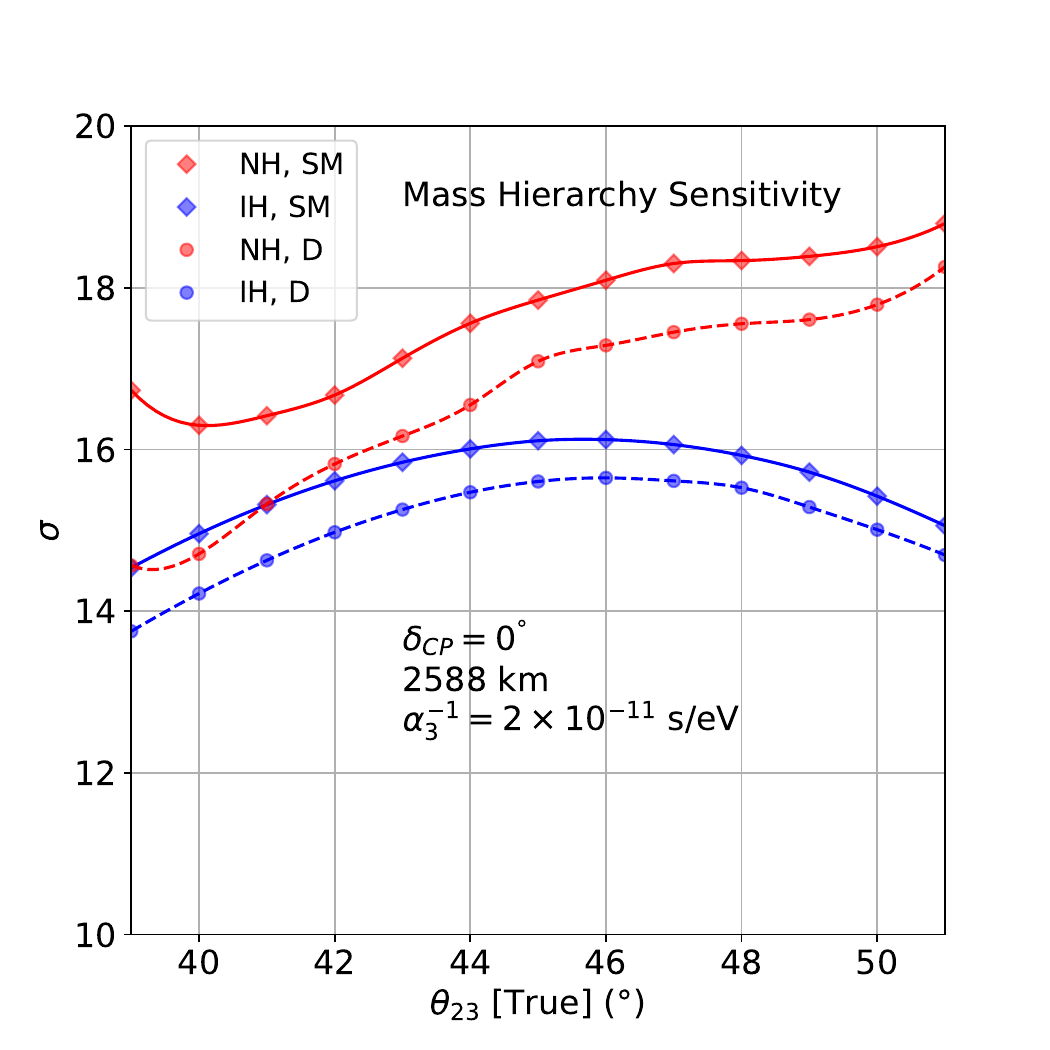}
    \includegraphics[width=0.32\linewidth]{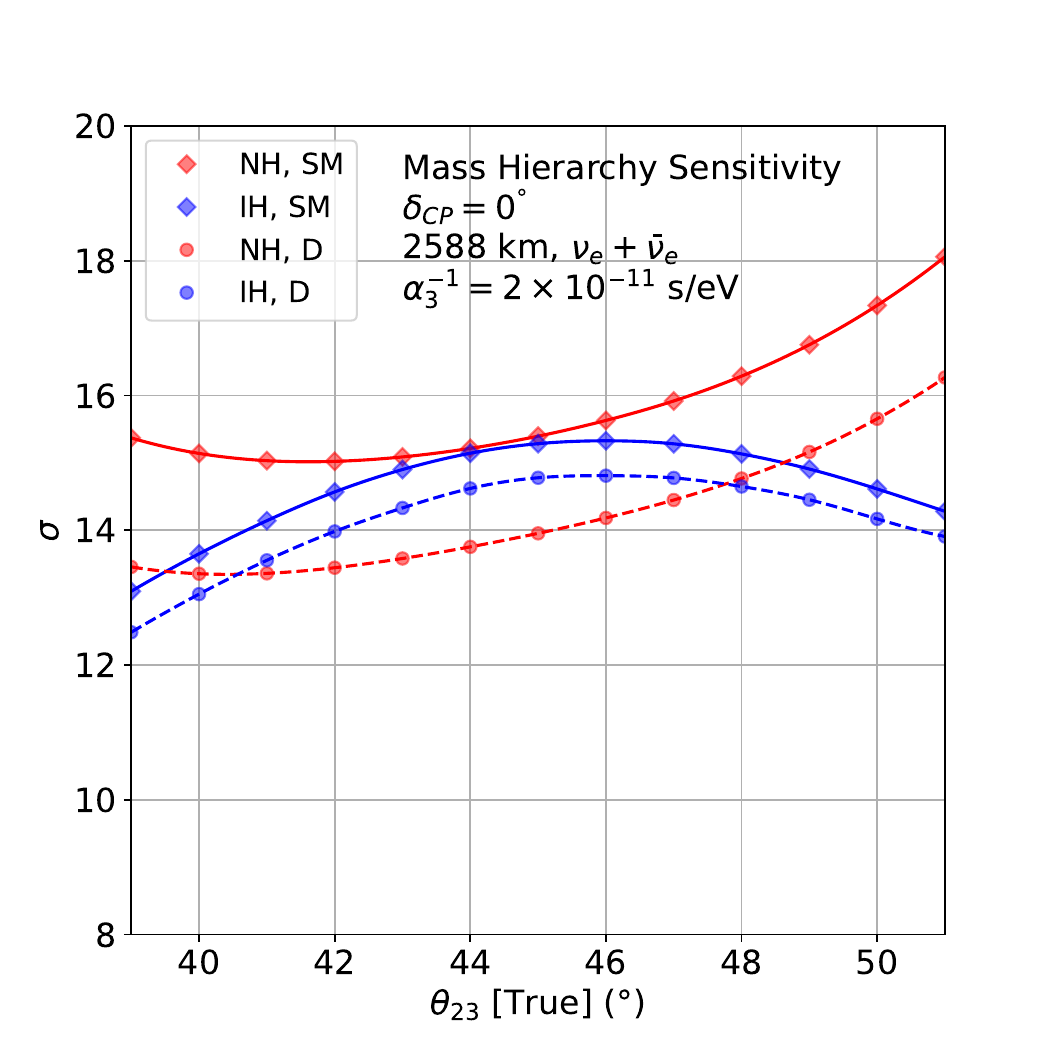}
    \includegraphics[width=0.32\linewidth]{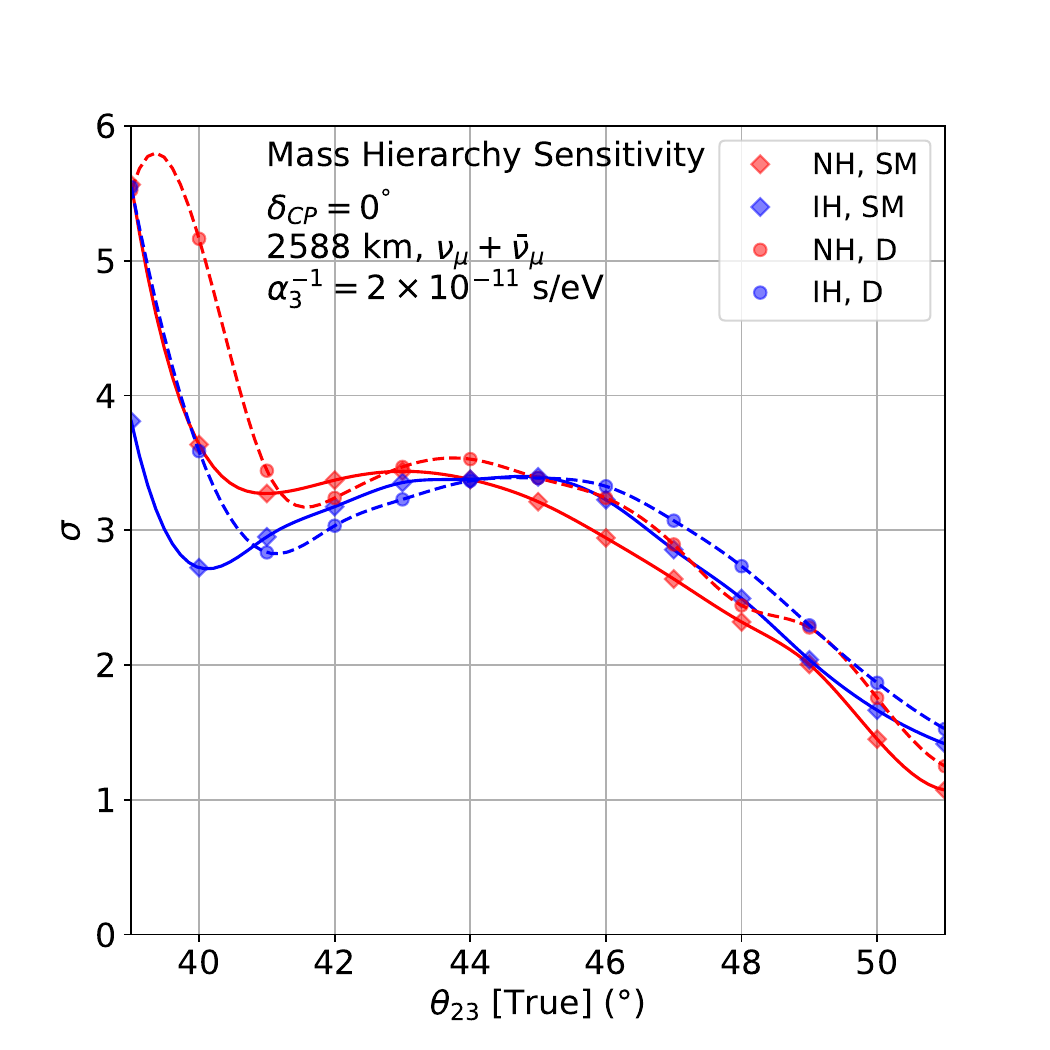}
    \includegraphics[width=0.32\linewidth]{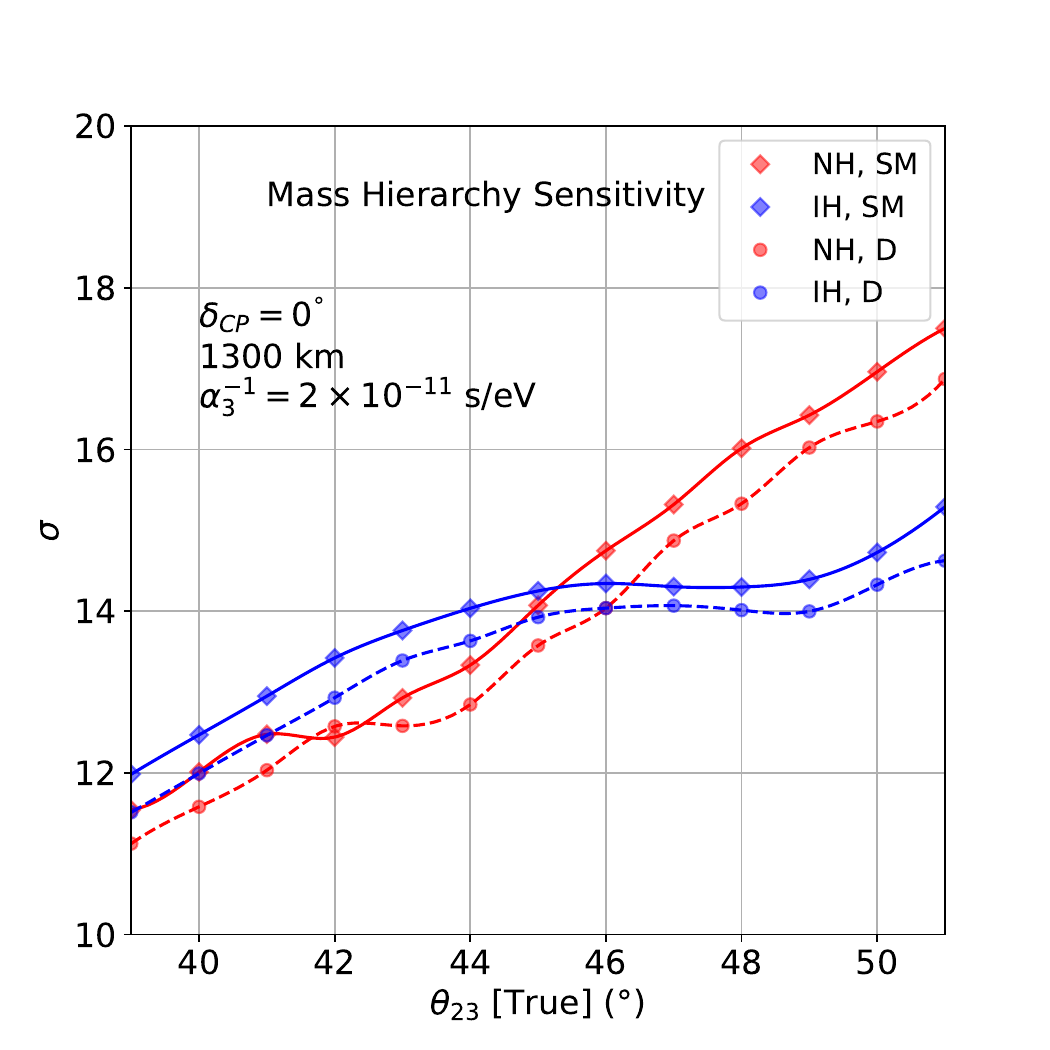}
    \includegraphics[width=0.32\linewidth]{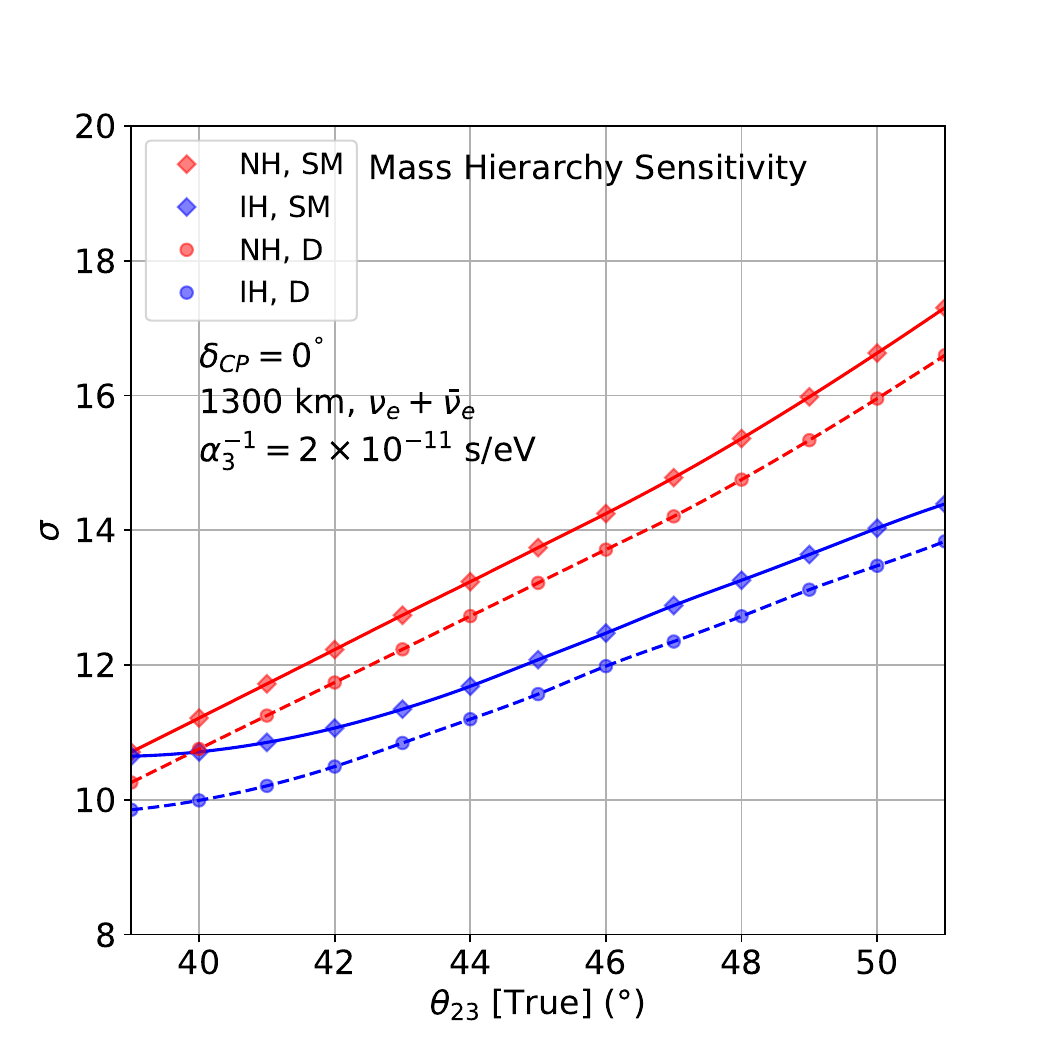}
    \includegraphics[width=0.32\linewidth]{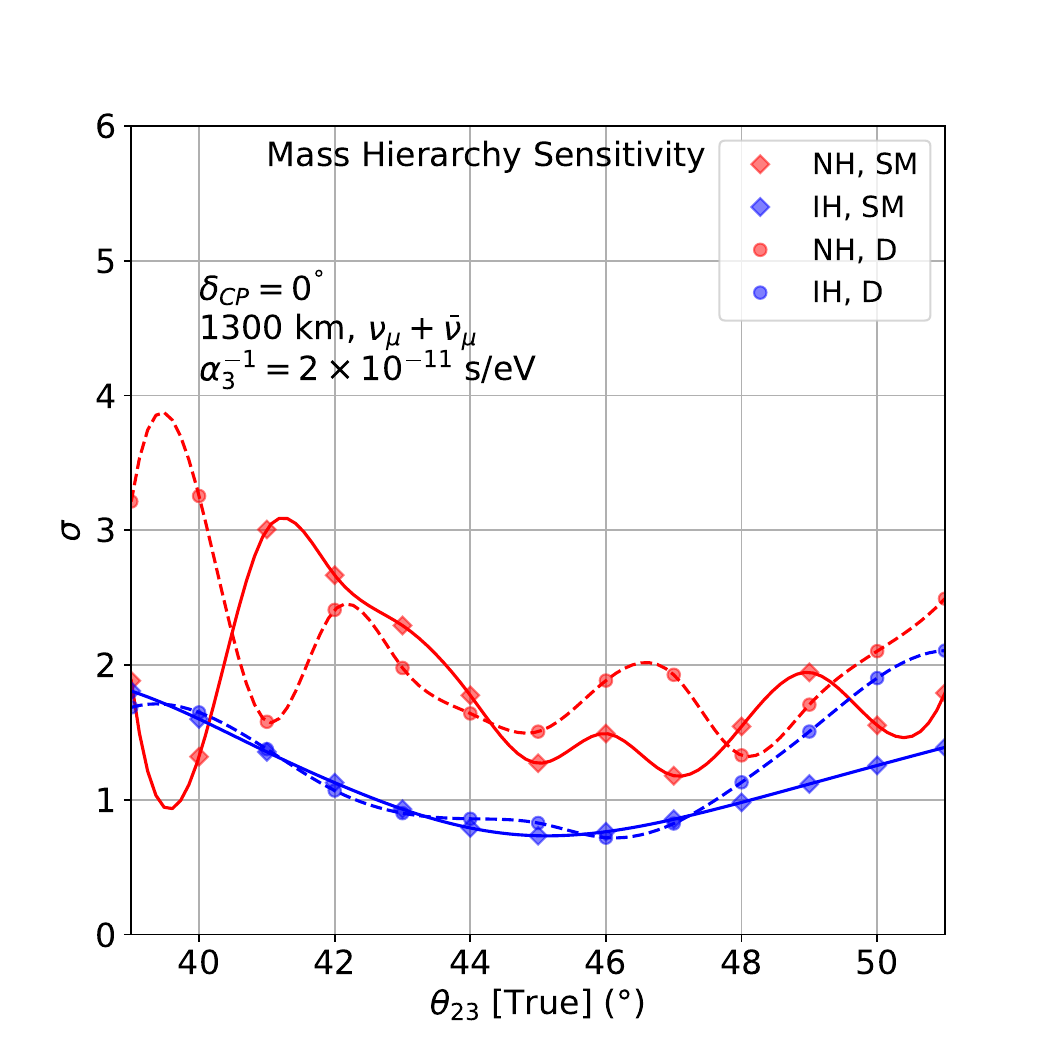}
    \caption{Sensitivity to mass hierarchy as a function of $\theta_{23}$ for without decay scenario (solid) and with decay (dotted) in 2588 km (top) and 1300 km (bottom) for NH (red), IH (blue).}
    \label{fig:chi-hr-th23}
\end{figure}

\begin{figure}[H]
    \centering
    
    \includegraphics[width=0.4\linewidth]{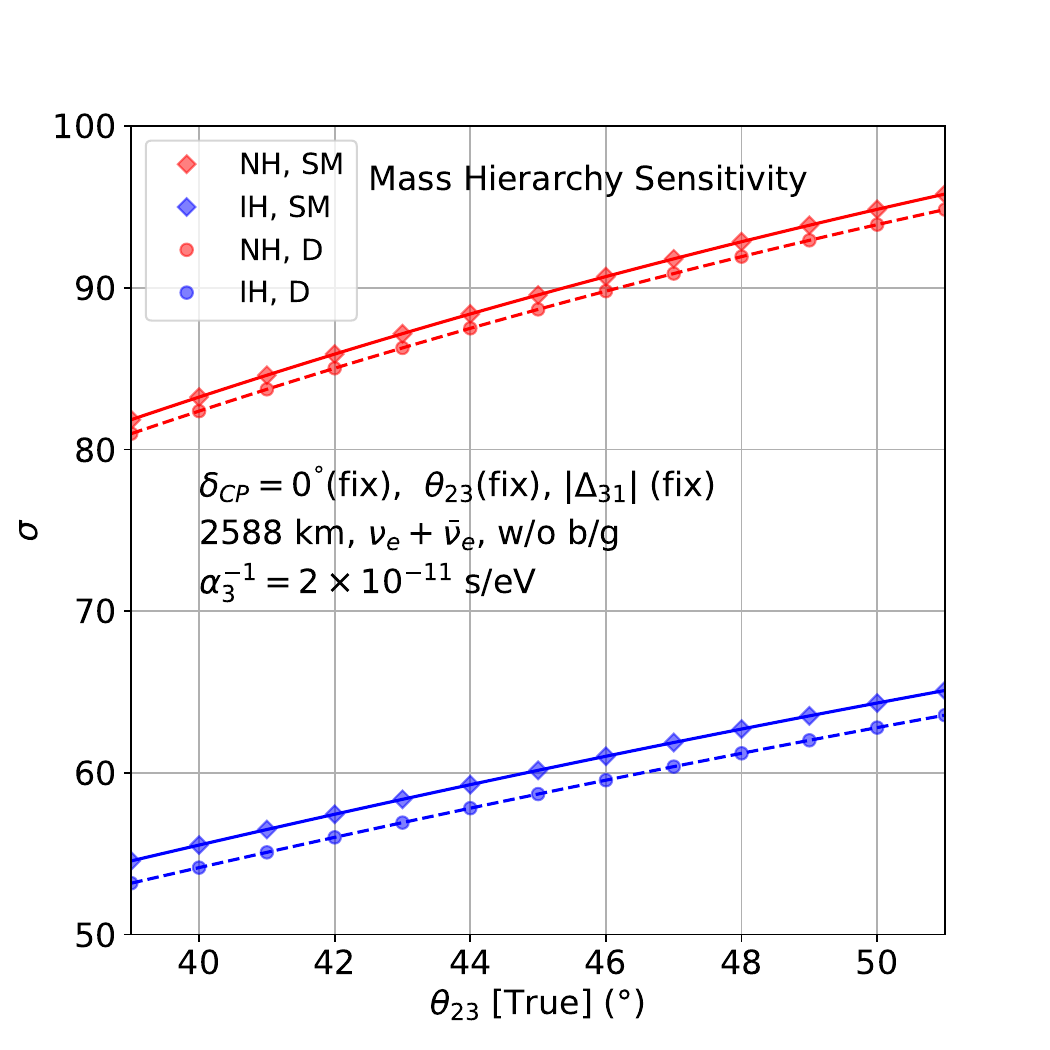}
    \includegraphics[width=0.4\linewidth]{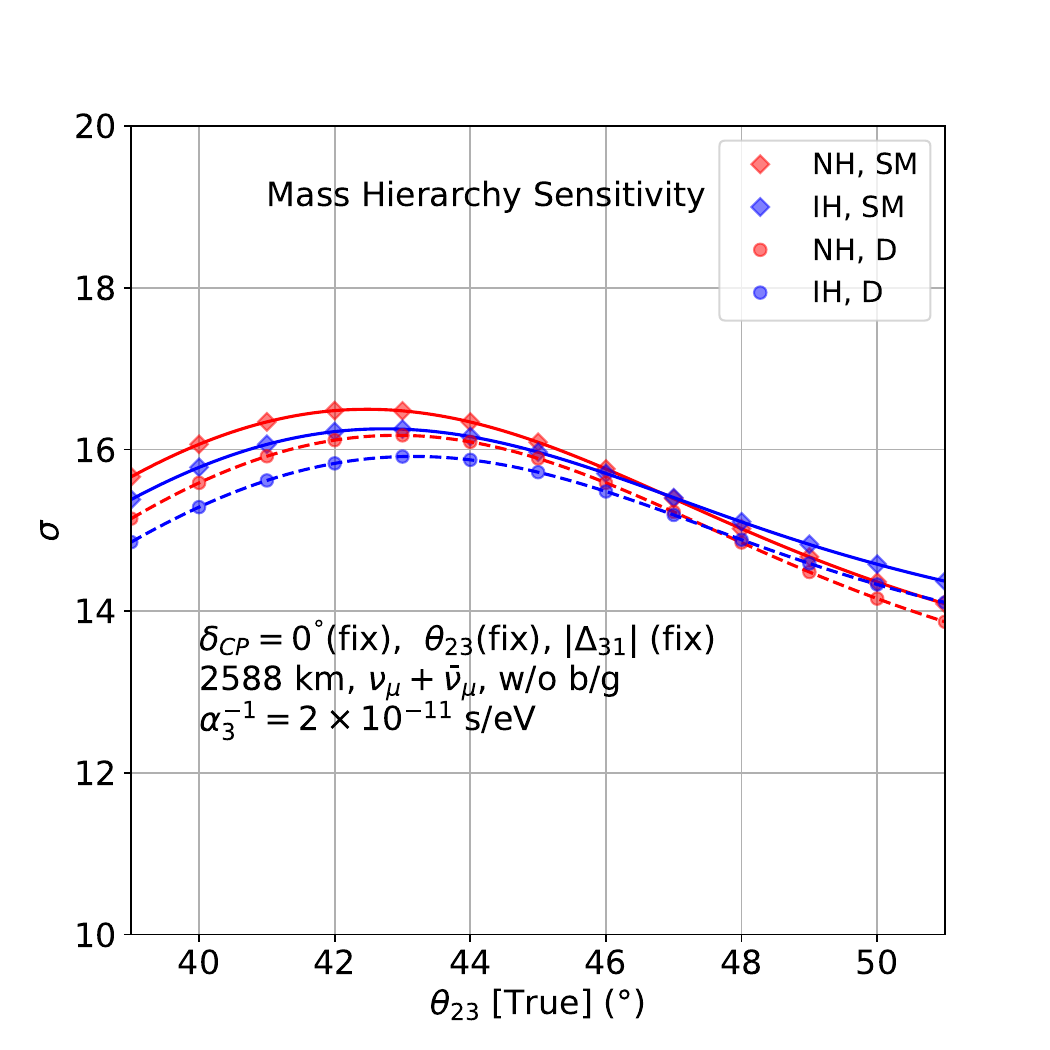}
    \includegraphics[width=0.4\linewidth]{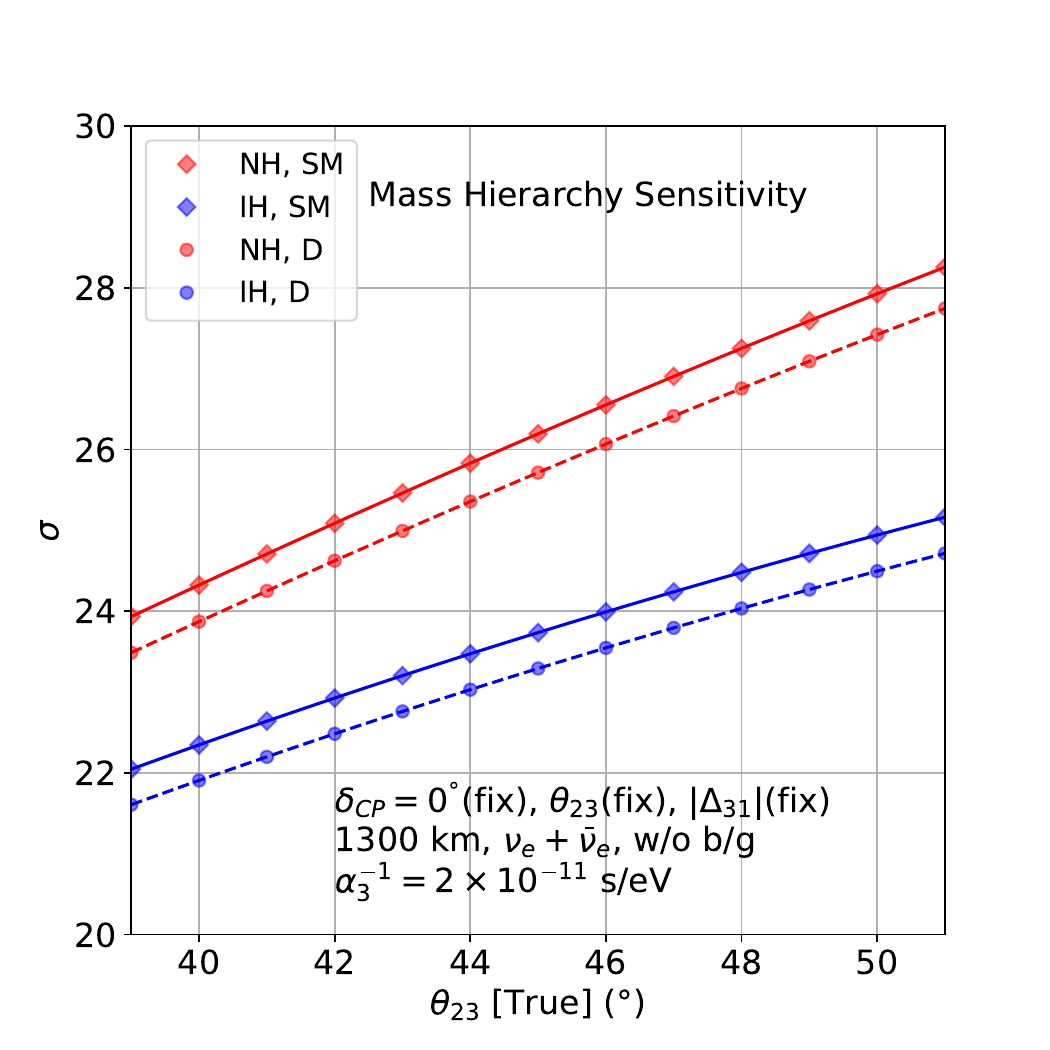}
    \includegraphics[width=0.4\linewidth]{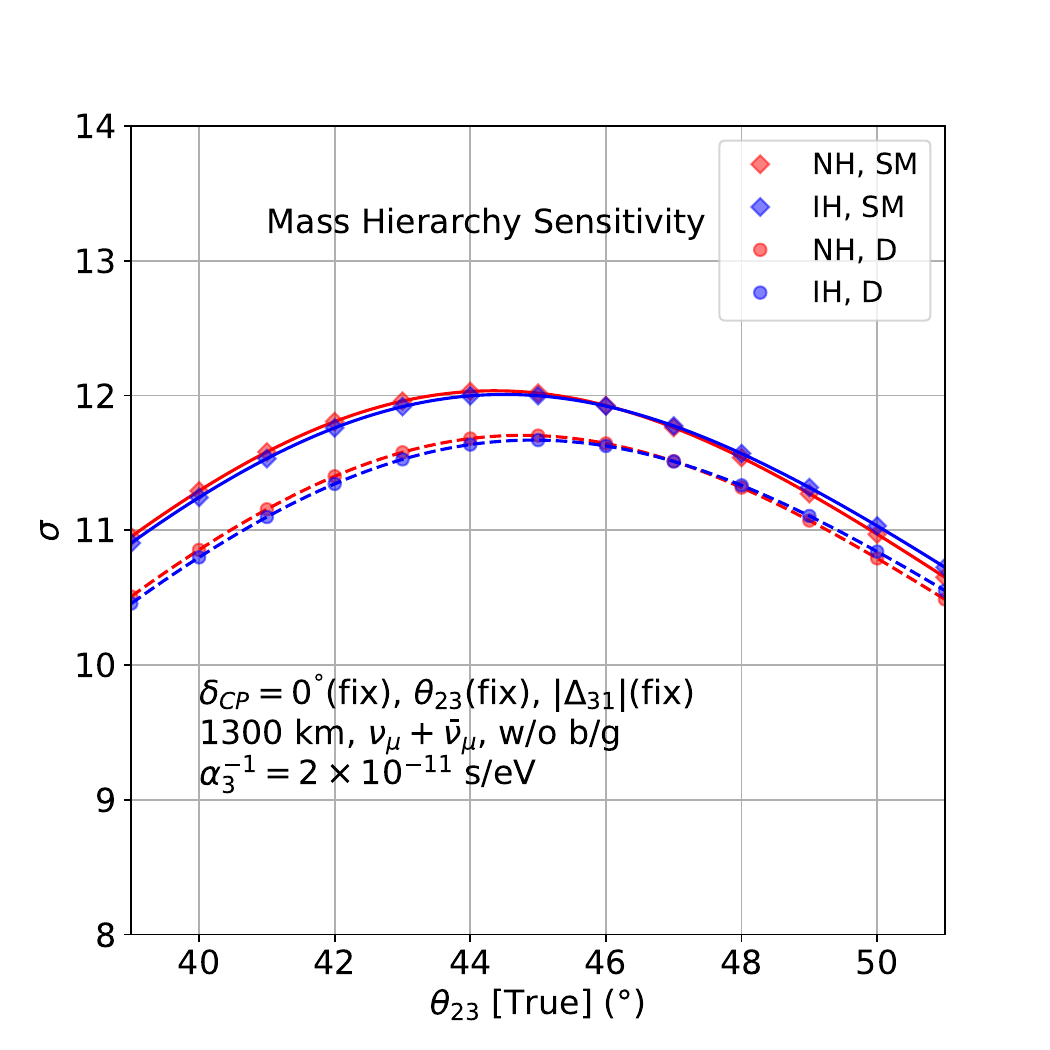}
    \caption{Sensitivity to mass hierarchy as a function of $\theta_{23}$ for the electron (left) and muon (right) channels without background and marginalization at 2588 km (top) and 1300 km (bottom).}
    \label{fig:chi-hr-th23-ch}
\end{figure}

In figure \ref{fig:chi-dcp-hr-dune}, the sensitivity to mass hierarchy as a function of $\delta_{CP}$ is depicted for the LArTPC detector setup for true values of $\theta_{23}=41^\circ$ (left), $45^\circ$ (middle) and $49^\circ$ (right). In this case, $\delta_{CP}$ affects the sensitivity to mass hierarchy more than it does in the case of P2O. The variation of sensitivity is around $10 \sigma$ between the maxima and minima of the plots in LArTPC compared to around $4 \sigma$ variation in P2O. The sensitivity with decay (blue curves) in true and test cases is lowest, while the w/o decay scenario and Decay[Test] scenarios have higher sensitivity. The sensitivity for all three scenarios increases with $\theta_{23}$, i.e., from the left panel to the right panel in NH. In the case of IH, the sensitivity is higher for $\theta_{23}=45^\circ$ than $41^\circ$ but is similar for $45^\circ$ and $49^\circ$. We probe further into dependence on true value of $\theta_{23}$ in figures \ref{fig:chi-hr-th23},\ref{fig:chi-hr-th23-ch}.

In figure \ref{fig:chi-hr-th23}, the sensitivity to MH in the presence of decay (dashed) in both true and test along with standard no decay scenario (solid) are plotted as a function of the true value of $\theta_{23}$ with the red (blue) curves corresponding to sensitivity for the true hierarchy of NH (IH) at P2O (top) and LArTPC (bottom) setups. We see that both with and without decay scenarios, the sensitivity gradually increases with $\theta_{23}$ for NH, whereas, for IH, the sensitivity falls on both sides of some $\theta_{23}$ value for P2O and remains almost constant for $\theta_{23}=(44^\circ,48^\circ)$ for LArTPC setup. To understand this behavior, we have plotted the sensitivity from $\nu_e$ ($\nu_\mu$) channels in second (third) column in figure \ref{fig:chi-hr-th23}. The sensitivity of $\nu_e$ channel for P2O is mostly similar to total sensitivity in IH, but in NH, sensitivity at first decreases till $\theta_{23}=42^\circ$ and then increases slowly. The contribution from $\nu_\mu$ channel is similar in both NH and IH, with a sudden decrease at first and then slightly rising before further decreasing with $\theta_{23}$. In the case of the LArTPC detector, sensitivity from the $\nu_e$ channel rises linearly with $\theta_{23}$ in both NH and IH. However, the sensitivity in the $\nu_\mu$ channel shows an oscillatory nature for NH, while in IH, the sensitivity attains a minima around $\theta_{23}=45^\circ$.

To inspect this different kind of behavior of sensitivity for IH, we have further plotted the contribution from electron and muon channels in figure \ref{fig:chi-hr-th23-ch} without background (b/g) contributions mentioned in table \ref{tab:globes-rules-p2o}, \ref{tab:globes-rules-dune} and no marginalization in the test scenario. We see from figure \ref{fig:chi-hr-th23-ch} that without the b/g contributions, in the electron channel, the sensitivity increases with $\theta_{23}$ that can be explained by the leading term in $P_{\mu e}$ varies with $\sin^2\theta_{23}$ in eq. \ref{eq:pme-dec}. While, in the muon channel, the sensitivity decreases with $\theta_{23}$ after initially increasing, and this nature can be understood from the leading terms in $P_{\mu\mu}$ in eq. \ref{eq:pmm-dec} depending on $\sin^2 2\theta_{23}$. It shows that the muon background present in the electron channel is the reason why the sensitivity in IH is only rising till $\theta_{23}\sim 46^\circ$ and then decreasing in 2588 km and is constant for $\theta_{23} = [44^\circ,48^\circ]$ in 1300 km. This background effect is also present in the true NH scenario; however, in that case, sensitivity w/o background in the electron channel is smaller compared to the muon background.

\subsection{Octant Sensitivity}\label{subsec:octant}
In this section, we discuss the results of octant sensitivity in the presence of invisible decay.
\begin{figure}[H]
    \centering 
    \includegraphics[width=.4\textwidth]{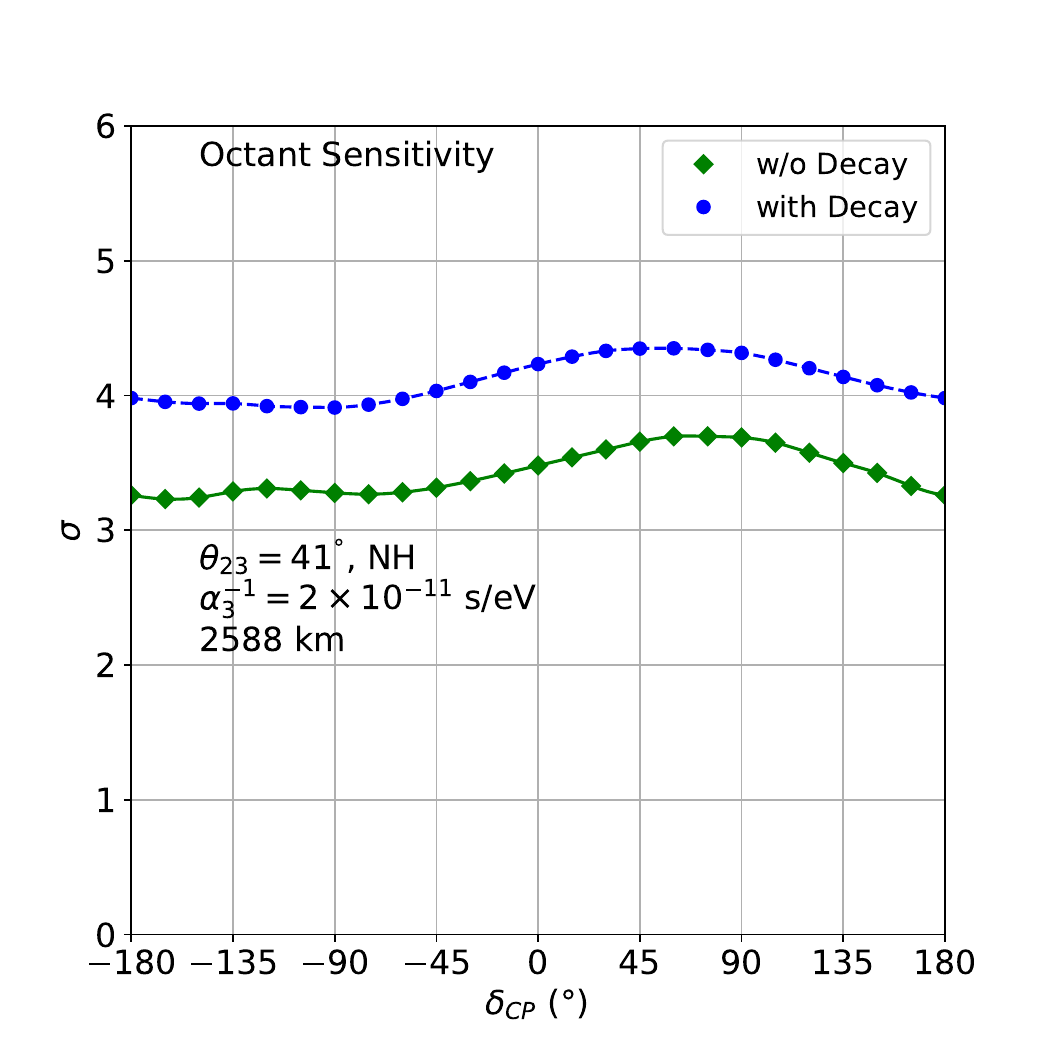}
    \includegraphics[width=.4\textwidth]{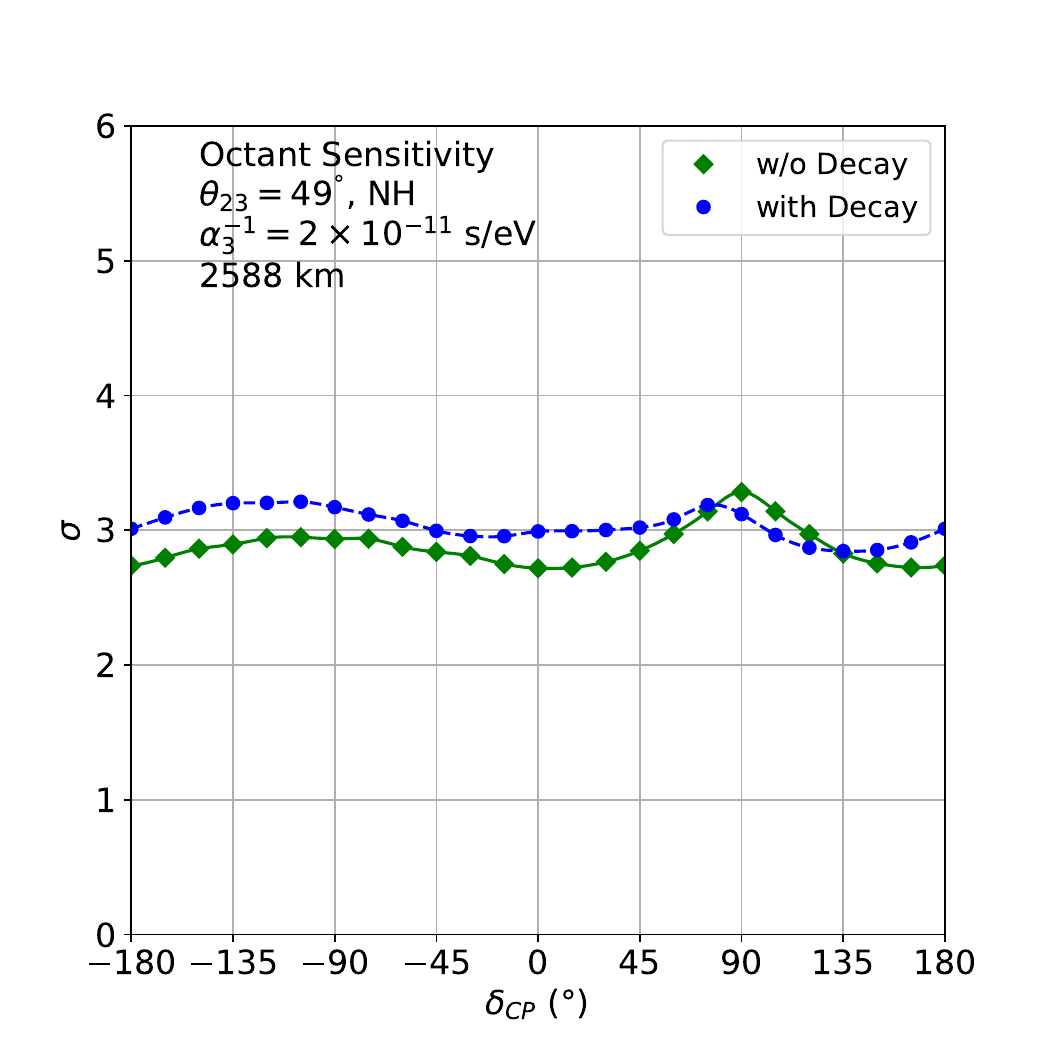}
    \includegraphics[width=.4\textwidth]{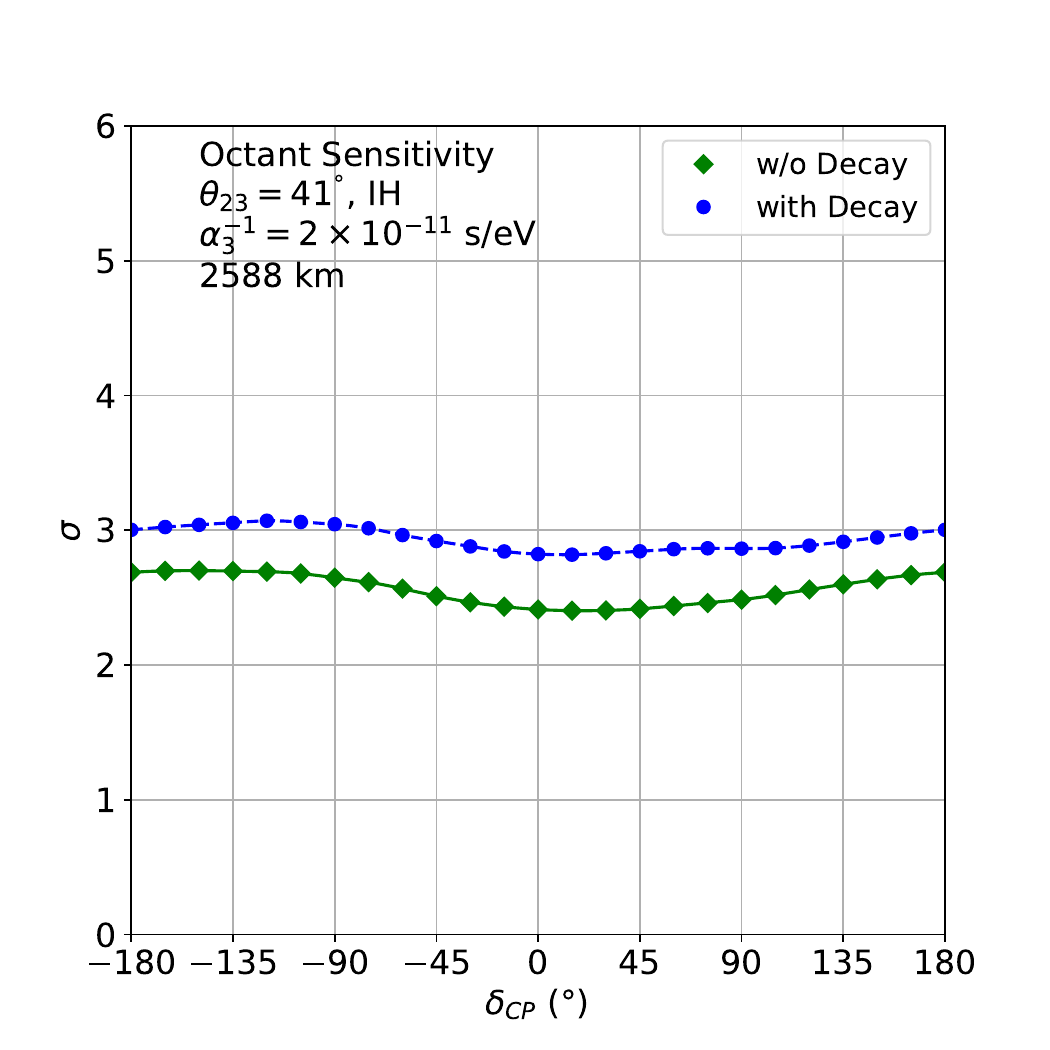}
    \includegraphics[width=.4\textwidth]{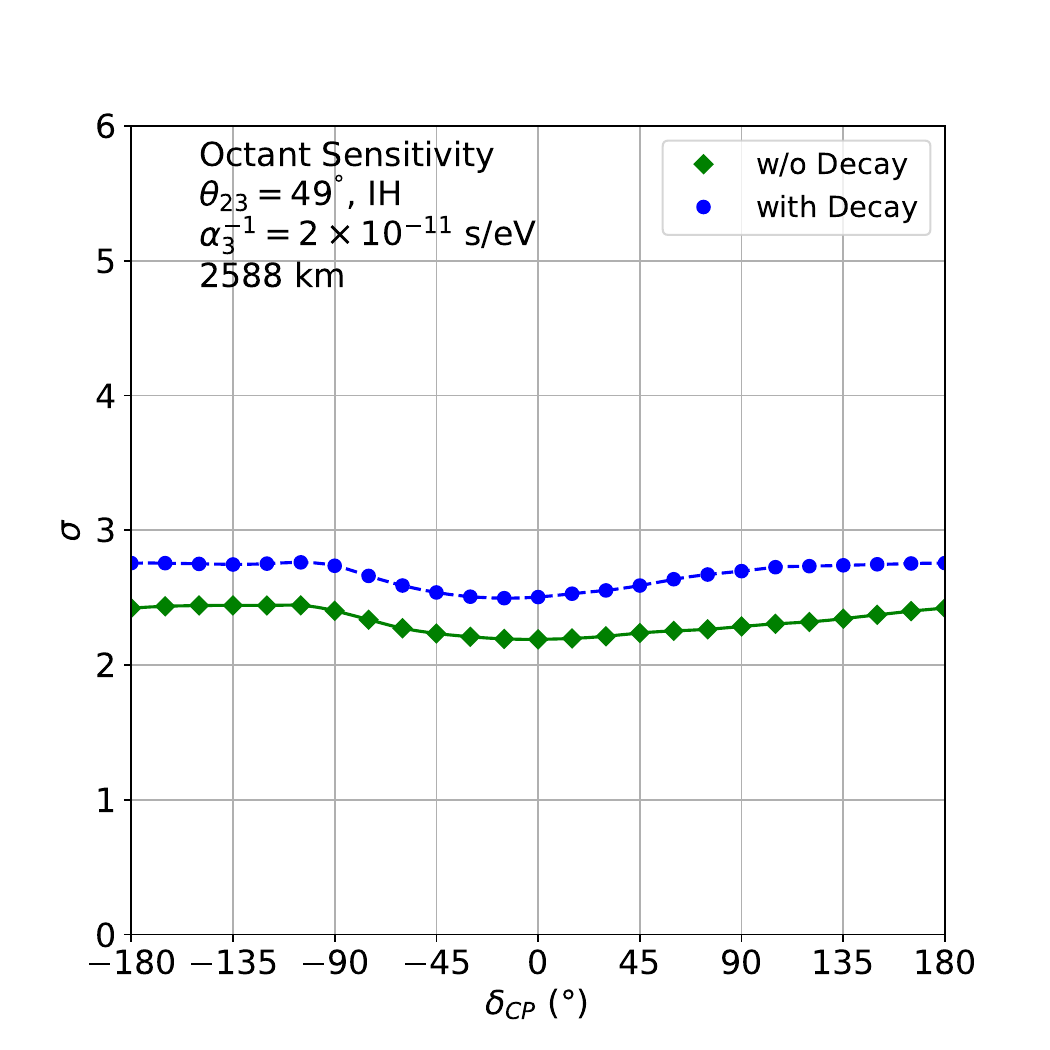}
    \caption{Sensitivity to octant of $\theta_{23}$ as a function of true $\delta_{CP}$ considering true $\theta_{23}=41^\circ$ (left), $49^\circ$ (right) for without decay (green) and with decay (blue) considering true $\alpha_3^{-1} = 2.0\times 10^{-11}$ s/eV for NH (top) and IH (bottom) at 2588 km baseline.}
    \label{fig:oct-dcp-p2o}
\end{figure}

\begin{figure}
\centering 
\includegraphics[width=.4\textwidth]{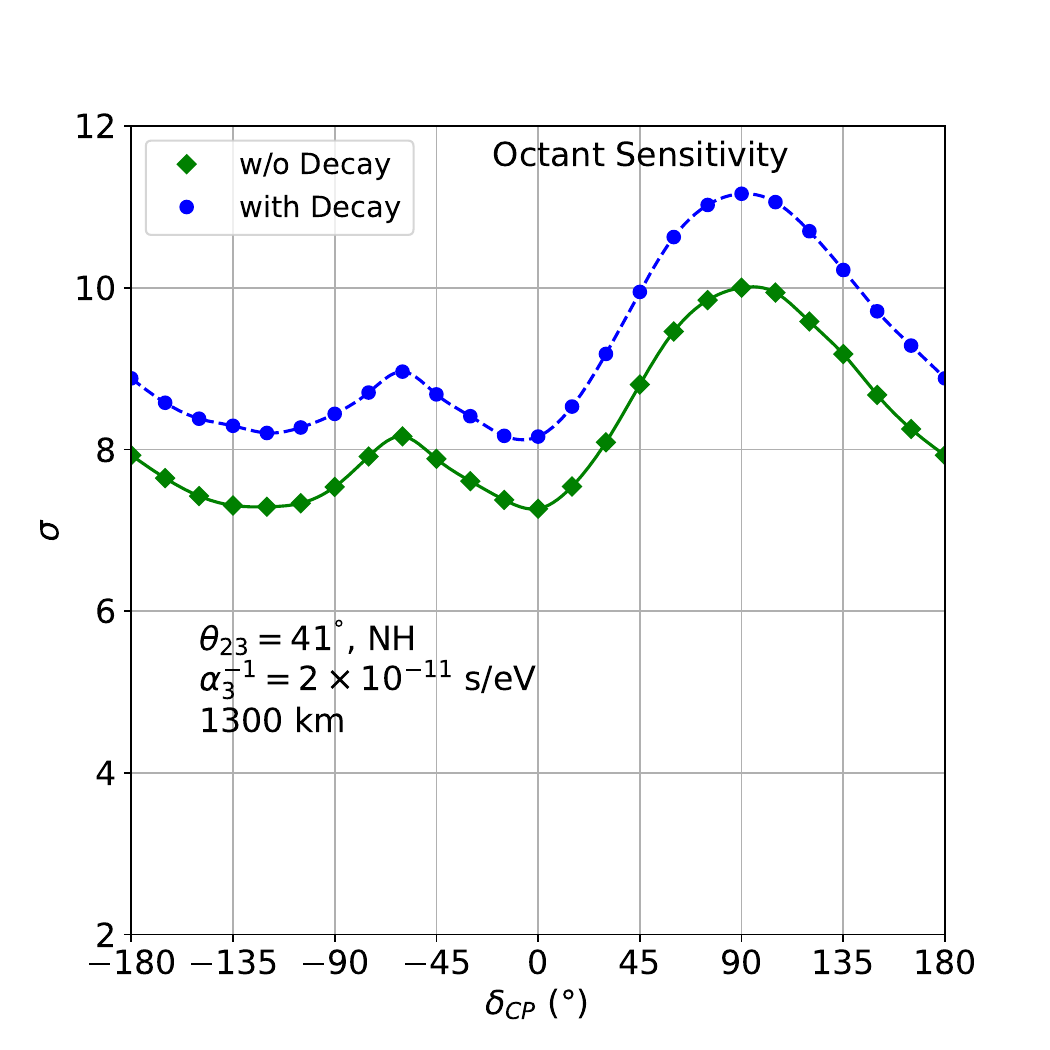}
\includegraphics[width=.4\textwidth]{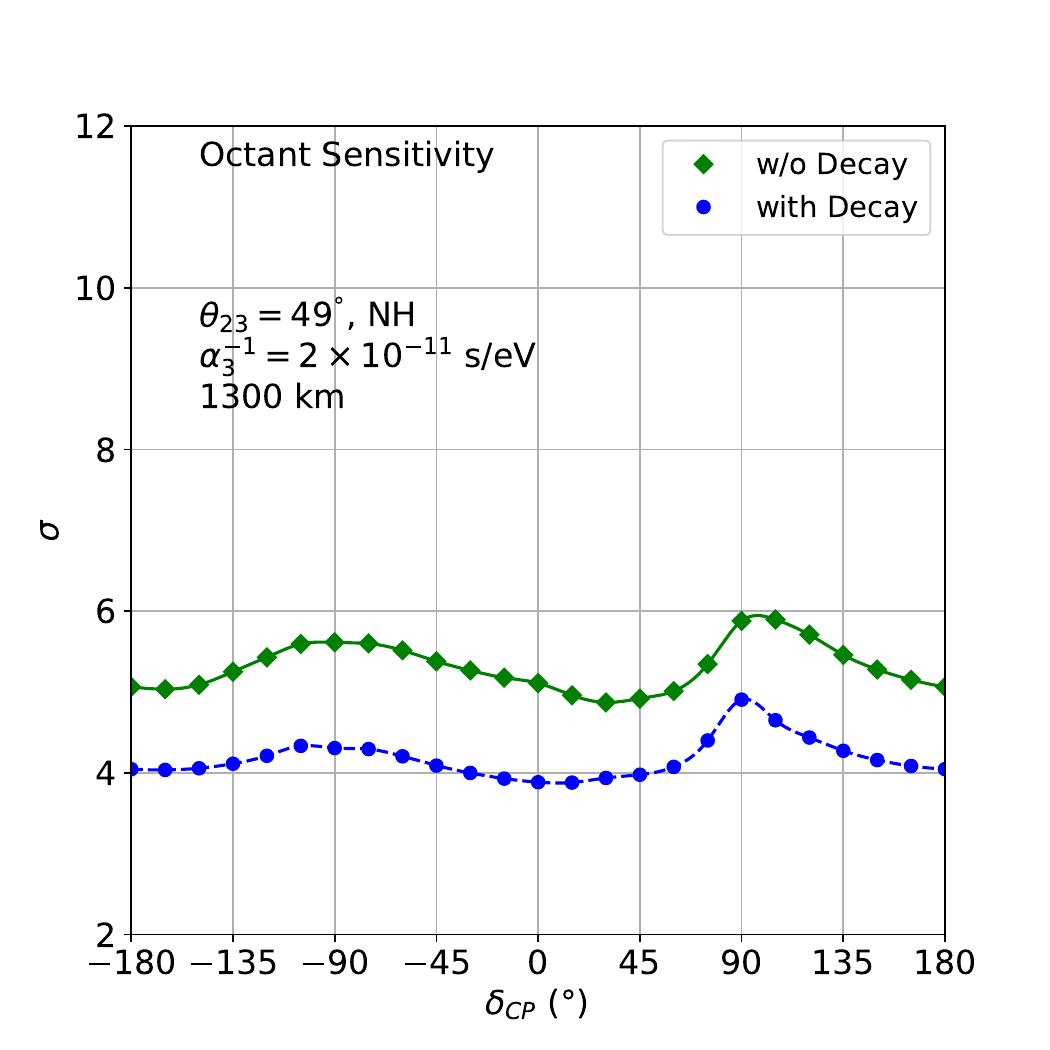}
\includegraphics[width=.4\textwidth]{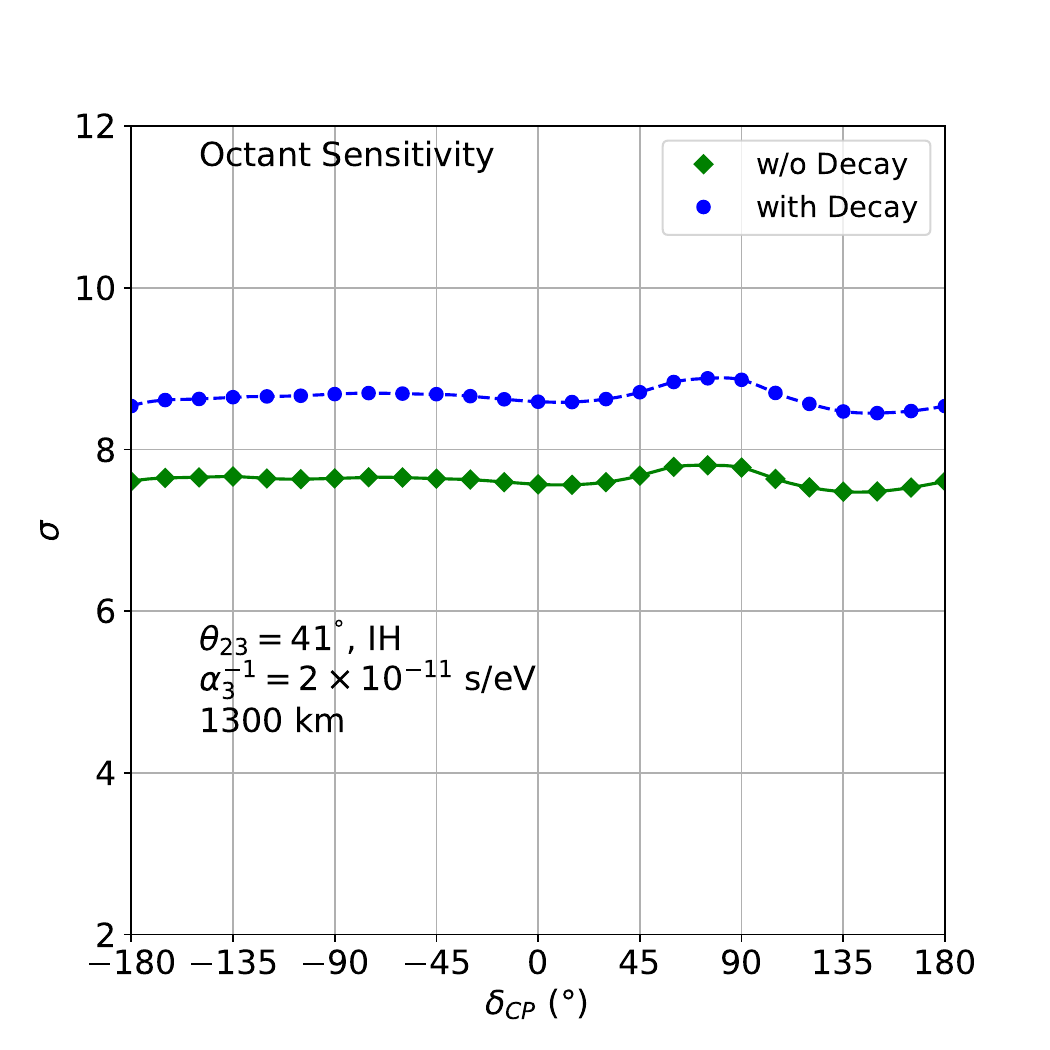}
\includegraphics[width=.4\textwidth]{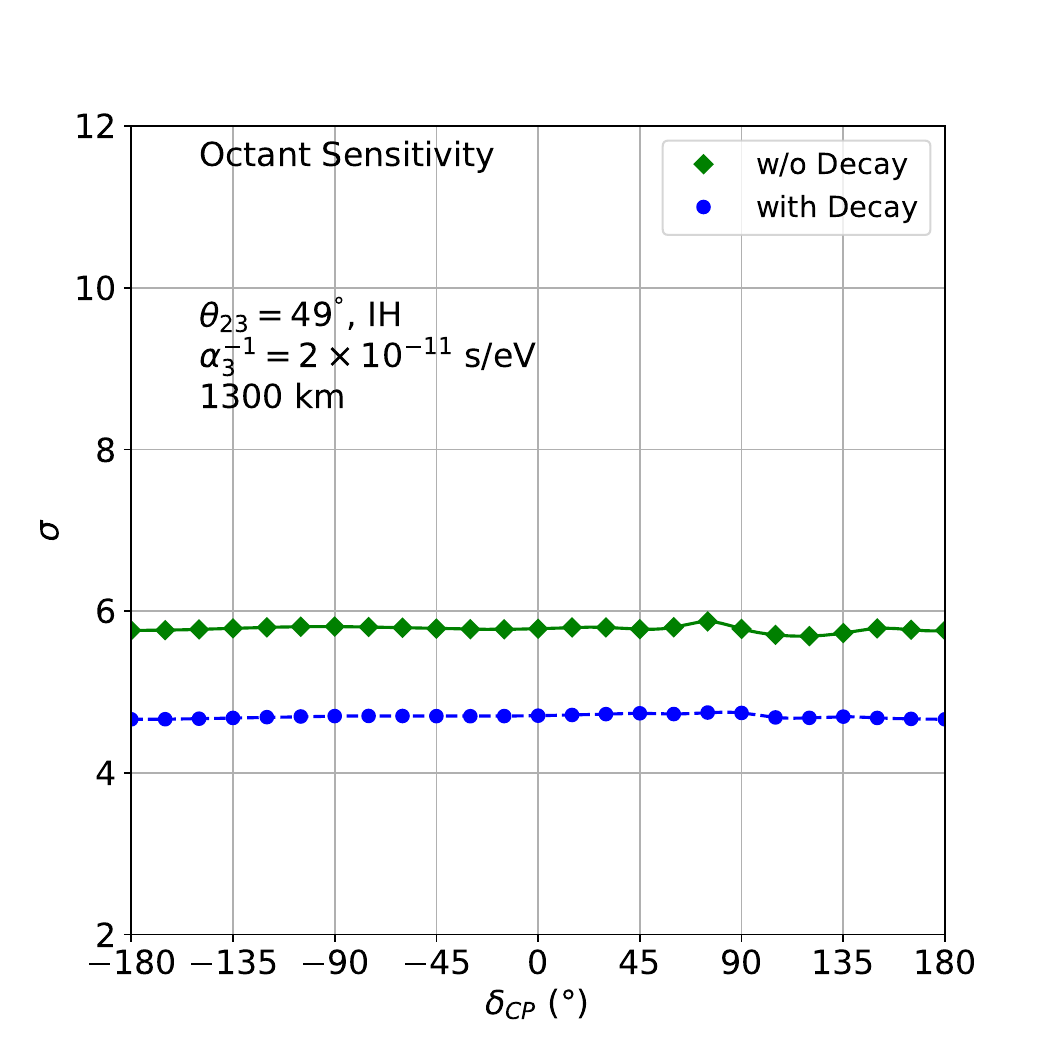}
\caption{Sensitivity to octant of $\theta_{23}$ as a function of true $\delta_{CP}$ considering true $\theta_{23}=41^\circ$ (left), $49^\circ$ (right) for without decay (green) and with decay (blue) considering true $\alpha_3^{-1} = 2.0\times 10^{-11}$ s/eV for NH (top) and IH (bottom) at 1300 km baseline.}
\label{fig:oct-dcp-dune}
\end{figure}

In figure \ref{fig:oct-dcp-p2o} and \ref{fig:oct-dcp-dune}, the octant sensitivity with and without decay scenarios have been shown as a function of true $\delta_{CP}$ for true $\theta_{23}=41^\circ$ (left), $49^\circ$ (right) for NH (top) and IH (bottom) for the detector setups at 2588 km and 1300 km respectively. We present two scenarios: one where decay is present in the true and test scenario (blue dashed), and in another case, the decay is absent in both true and test, i.e., the standard scenario (green solid). In all the cases, the marginalization has been done in $\theta_{23}$ in opposite octant, $|\Delta_{31}|$, $\delta_{CP}$ and $\alpha_3$ (in the presence of decay). The main observed features are;

\begin{itemize}
    \item For 2588 km (fig. \ref{fig:oct-dcp-p2o}), in the presence of decay (blue), the sensitivity is higher than that of the standard no decay scenarios (green). The increase in sensitivity is higher for $\theta_{23}$ in LO ($41^\circ$) as compared to HO $(49^\circ$).

    \item In the case of 1300 km (fig. \ref{fig:oct-dcp-dune}), the octant sensitivity in the presence of decay (blue) is lower for $\theta_{23}^{true}$ in HO and higher for $\theta_{23}^{true}$ in LO relative to the no-decay scenario.
\end{itemize}

\begin{figure}
    \centering
    \includegraphics[width=0.4\linewidth]{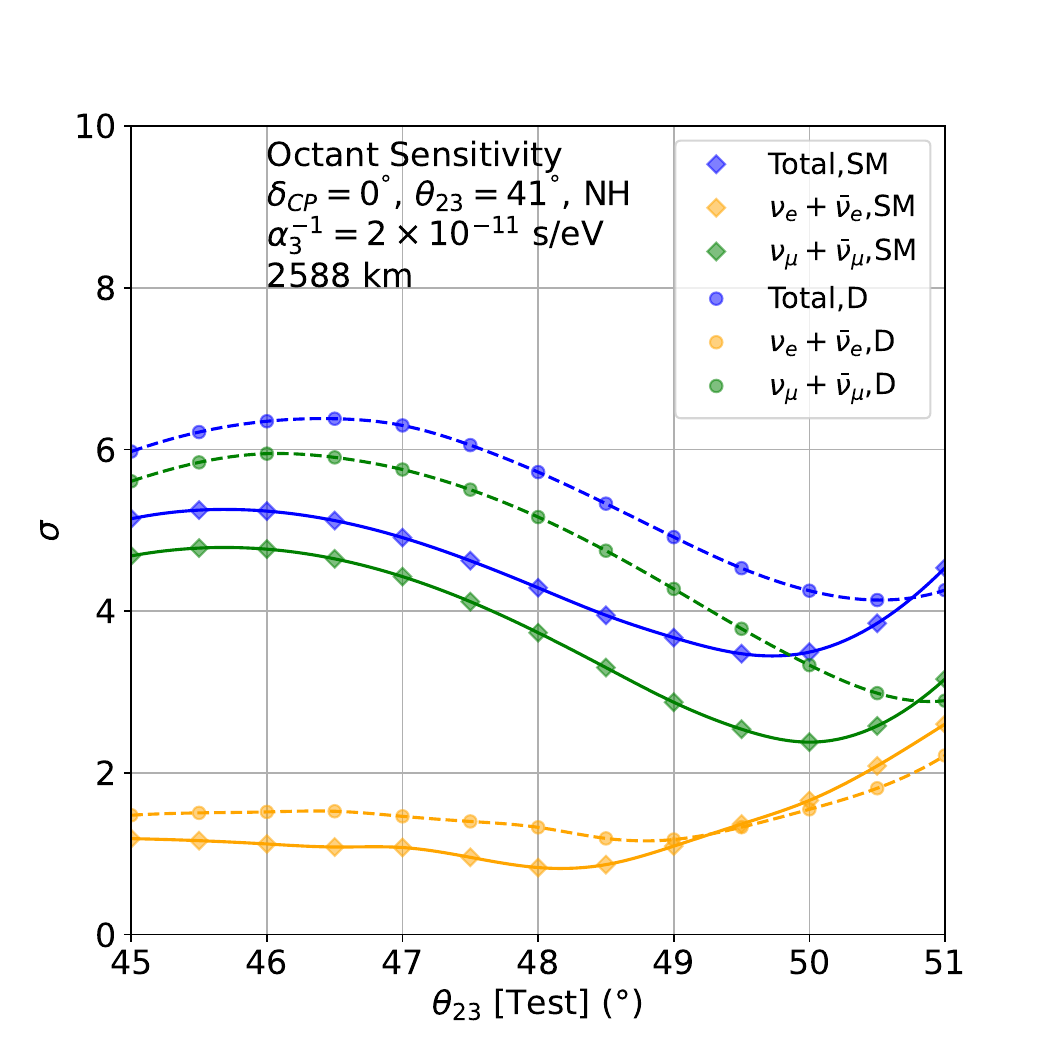}
    \includegraphics[width=0.4\linewidth]{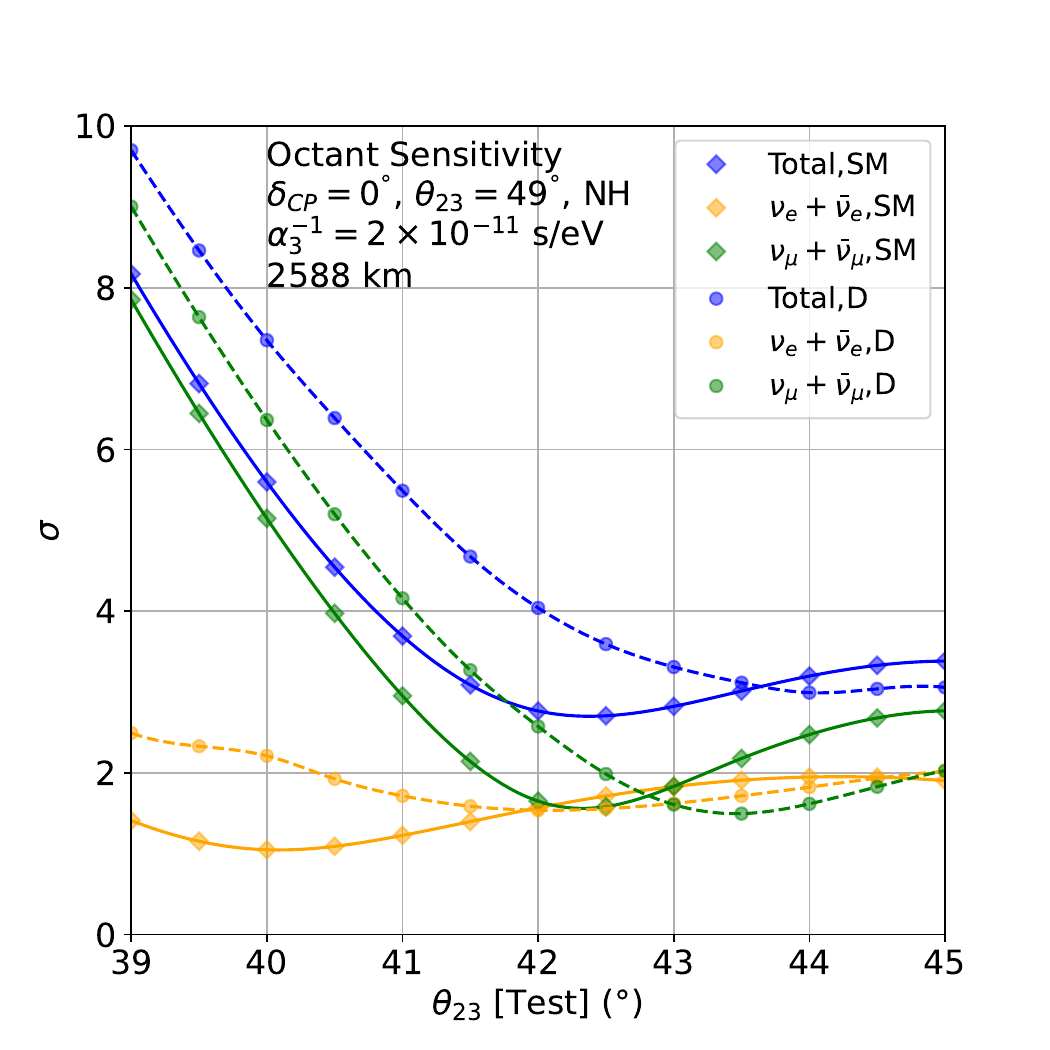}
    \includegraphics[width=0.4\linewidth]{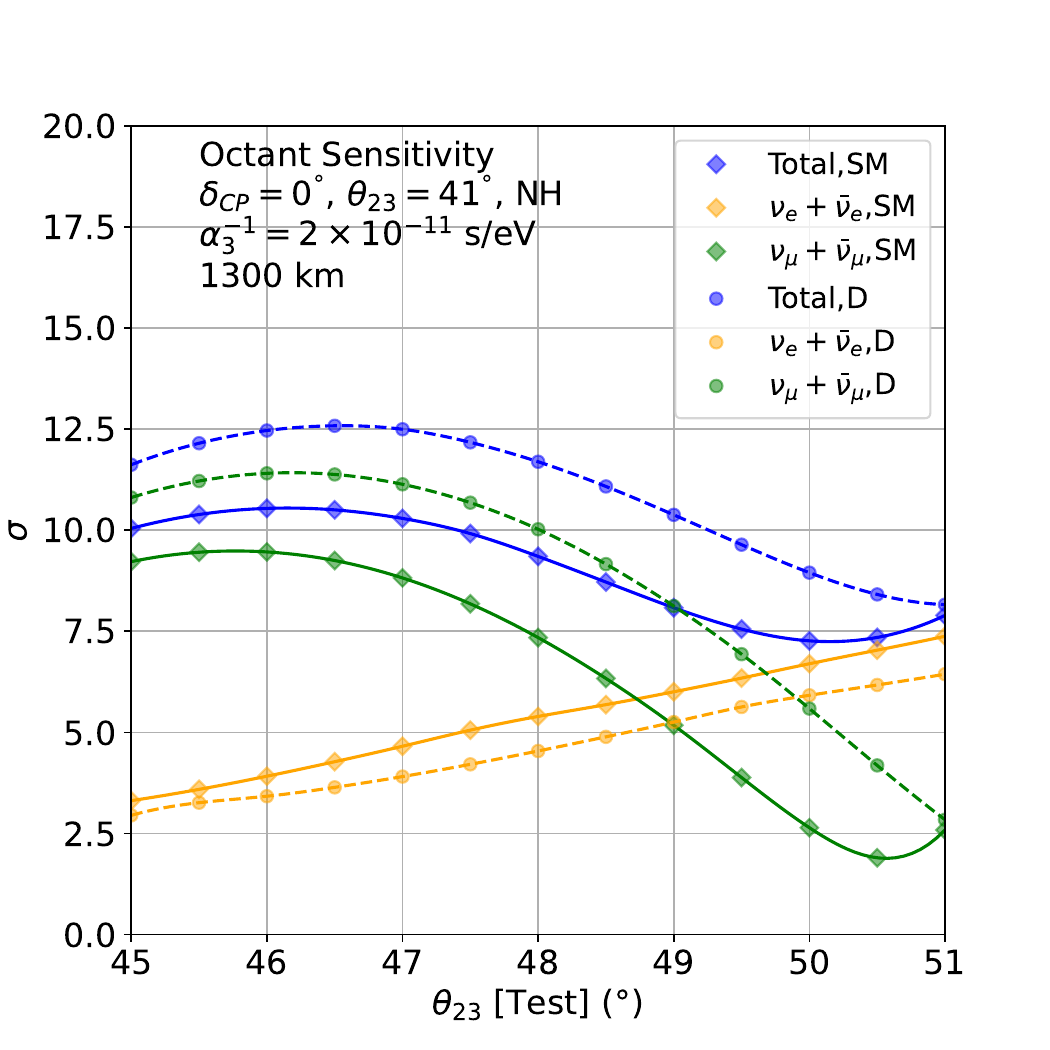}
    \includegraphics[width=0.4\linewidth]{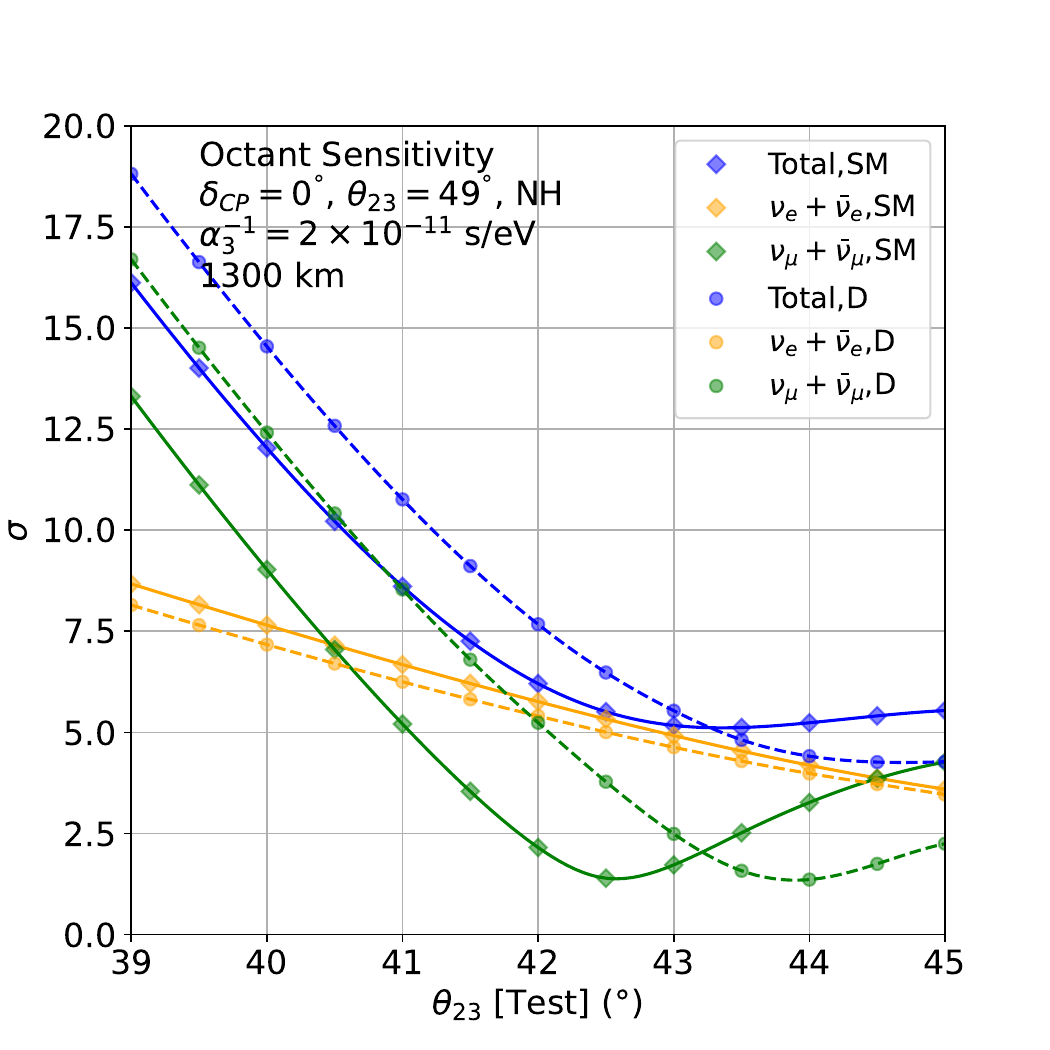}
    \caption{Octant sensitivity as a function of test $\theta_{23}$ showing contribution from $\nu_e$ (orange) and $\nu_\mu$ (green) and total (blue) events for $\theta_{23}$ in LO (left) and HO (right) in 2588 km (top) and 1300 km (bottom). The solid (dashed) curves signify no decay (decay) scenarios.}
    \label{fig:chi-oct-th23-p2o-dune}
\end{figure}

Since the synergy between electron and muon channel is the main contributor in enhancing the octant sensitivity, in order to understand the nature of curves in figures \ref{fig:oct-dcp-p2o}, \ref{fig:oct-dcp-dune}, we have plotted the contributions from these two channels along with total sensitivity in the test $\theta_{23}$ plane in figure \ref{fig:chi-oct-th23-p2o-dune} for 2588 km (top) and 1300 km (bottom). In figure \ref{fig:chi-oct-th23-p2o-dune}, the sensitivity without decay and with decay is shown by solid and dashed curves, respectively. 
For the $P_{\mu e}$ channel (c.f. eq.\ref{eq:pme-sm},\ref{eq:pme-dec}), the leading contribution goes as $\sin^2 \theta_{23}$, and the chi-square in the opposite octant increases with $\theta_{23}$ as can be seen from the bottom-left panel in figure \ref{fig:chi-oct-th23-p2o-dune} for the LArTPC setup. However, for the P2O setup (c.f. the top-left panel), due to the presence of $45\%$ muon background (c.f. Table \ref{tab:globes-rules-p2o}), the nature of the chi-square curve in the electron channel gets modified, and it's not a monotonic function of $\sin^2\theta_{23}$ any longer. For the muon channel (c.f. eq.\ref{eq:pmm-sm},\ref{eq:pmm-dec}), the leading term in the probability proportional to $\sin^2 2\theta_{23}$ doesn't offer octant sensitivity. However, due to the matter effect at large baselines like 1300 km and 2588 km, the contribution from the term proportional to $\sin^2\theta_{23}$ becomes significant, and some octant sensitivity comes from this channel as well. The position of the minima of the total sensitivity curve is governed mainly by the $P_{\mu\mu}$ channel, but there is an octant sensitive contribution coming from $P_{\mu e}$ channel, enhancing the chi-square. Since the chi-square from $P_{\mu e}$ channel at P2O is flatter, this octant sensitive contribution is less, and thus, the overall sensitivity is more for the DUNE-like setup, as can be seen from figure \ref{fig:chi-oct-th23-p2o-dune}. In addition, from the top-left panel, it is seen that in the presence of decay, the minima of the total chi-square shifts from $49.5^\circ$ to $50.5^\circ$ where the value of chi-square is higher in both $\nu_\mu, \nu_e$ channels as compared to the no decay case, leading to an elevated total sensitivity. Similarly, from the top-right panel ($\theta_{23}^{\rm true}$ in HO), the minima of the total sensitivity shifts from $42.5^\circ$ to $44^\circ$, where the contribution of $\nu_\mu$ is similar but $\nu_e$ is slightly more than no decay scenario, leading to a smaller increase in total sensitivity. In the case of 1300 km also, we can see enhanced sensitivity due to the synergy between $\nu_e$ and $\nu_\mu$ channels for both decay and no decay scenarios. For $\theta_{23}$ in LO (the bottom-left panel), in the presence of decay, the minima of $\nu_\mu$ channel shifts to larger $\theta_{23}$ with the value of chi-square higher than that of no decay, resulting in enhanced total sensitivity. However, in 1300 km for HO (the bottom-right panel), it is observed in the presence of decay that minima of the combined sensitivity shifts to a higher $\theta_{23}$ (towards $45^\circ$) driven by the $\nu_\mu$ channel. For such values, the contribution from the $\nu_e$ channel is lower, resulting in a reduced total sensitivity as compared to the no decay case.

\subsection{Sensitivity in $\theta_{23}-\delta_{CP}$ Plane}\label{subsec:th23-dcp}
\begin{figure}
\centering 
\includegraphics[width=.32\textwidth]{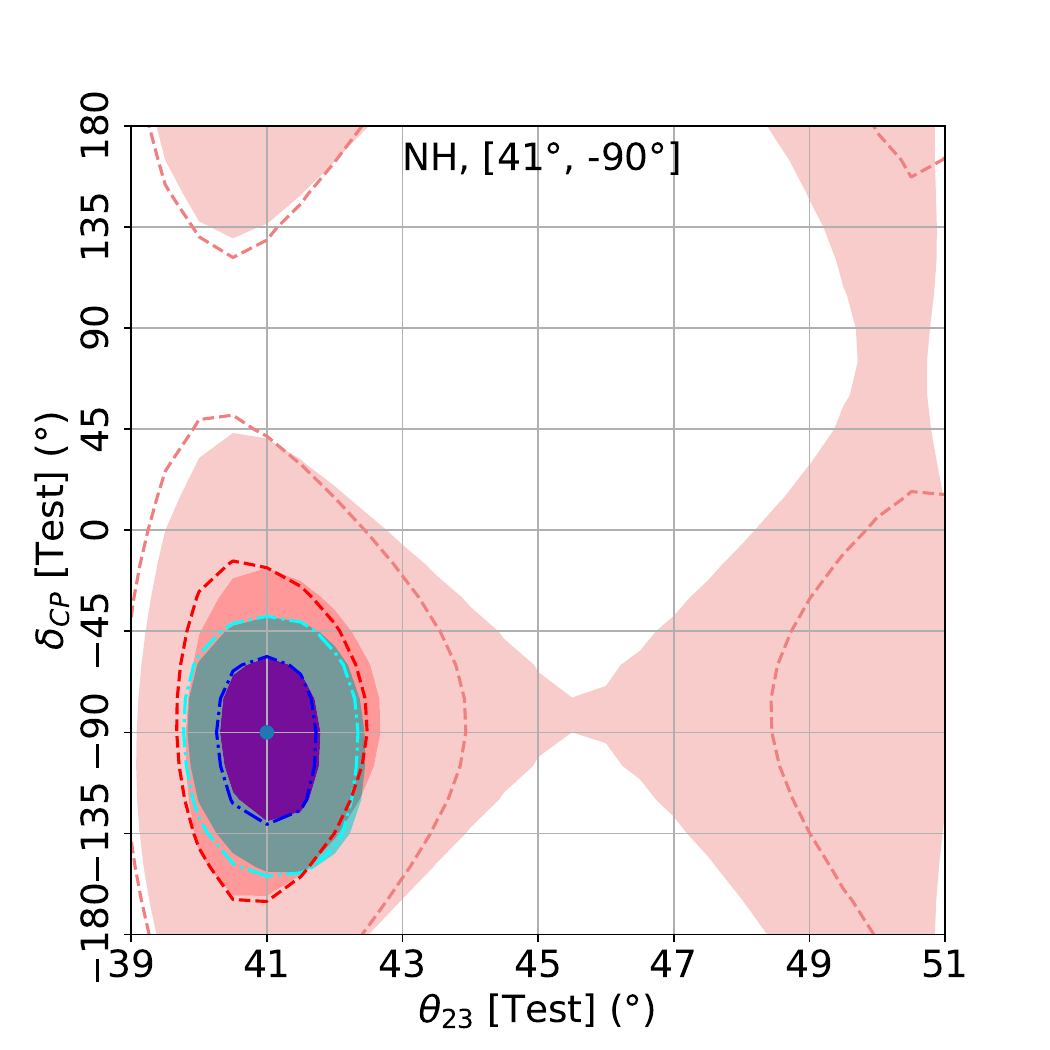}
\includegraphics[width=.32\textwidth]{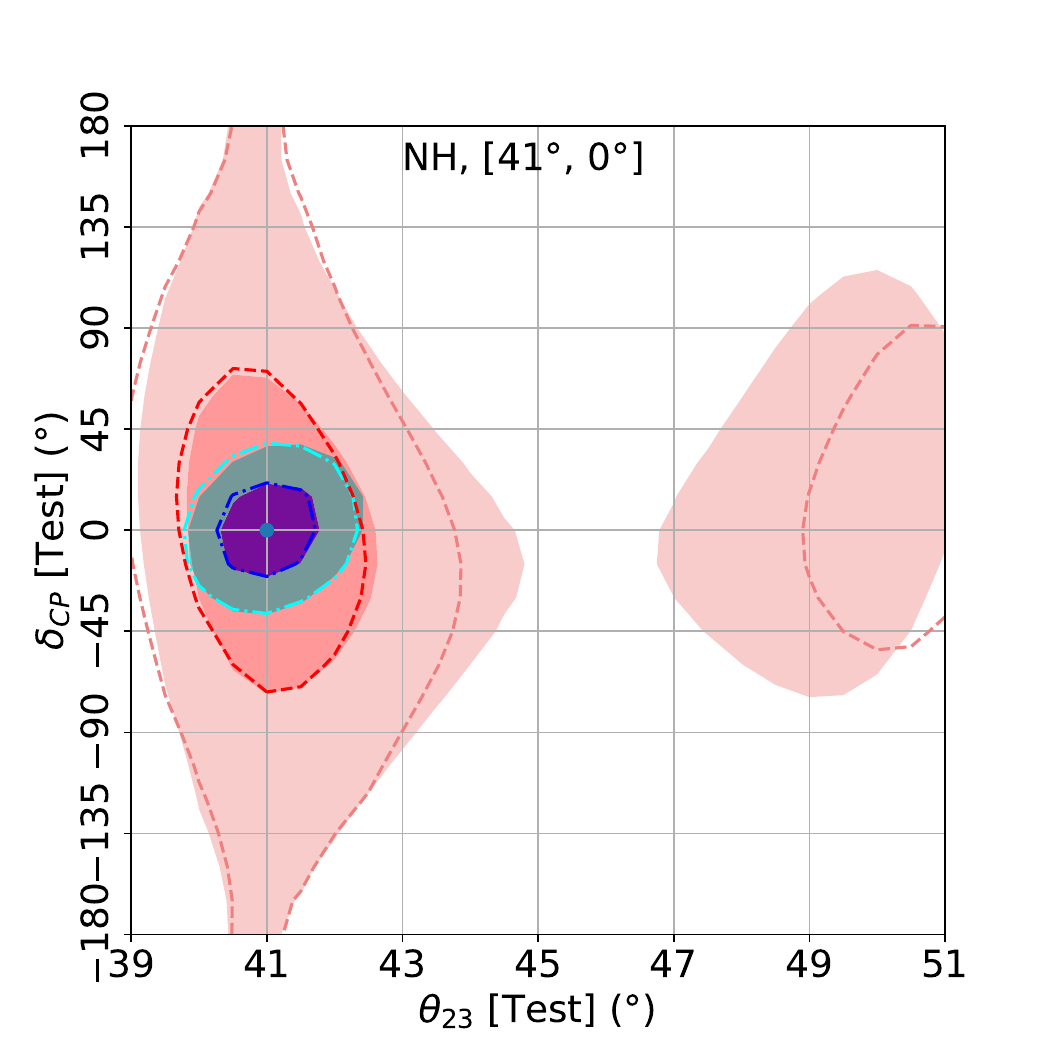}
\includegraphics[width=.32\textwidth]{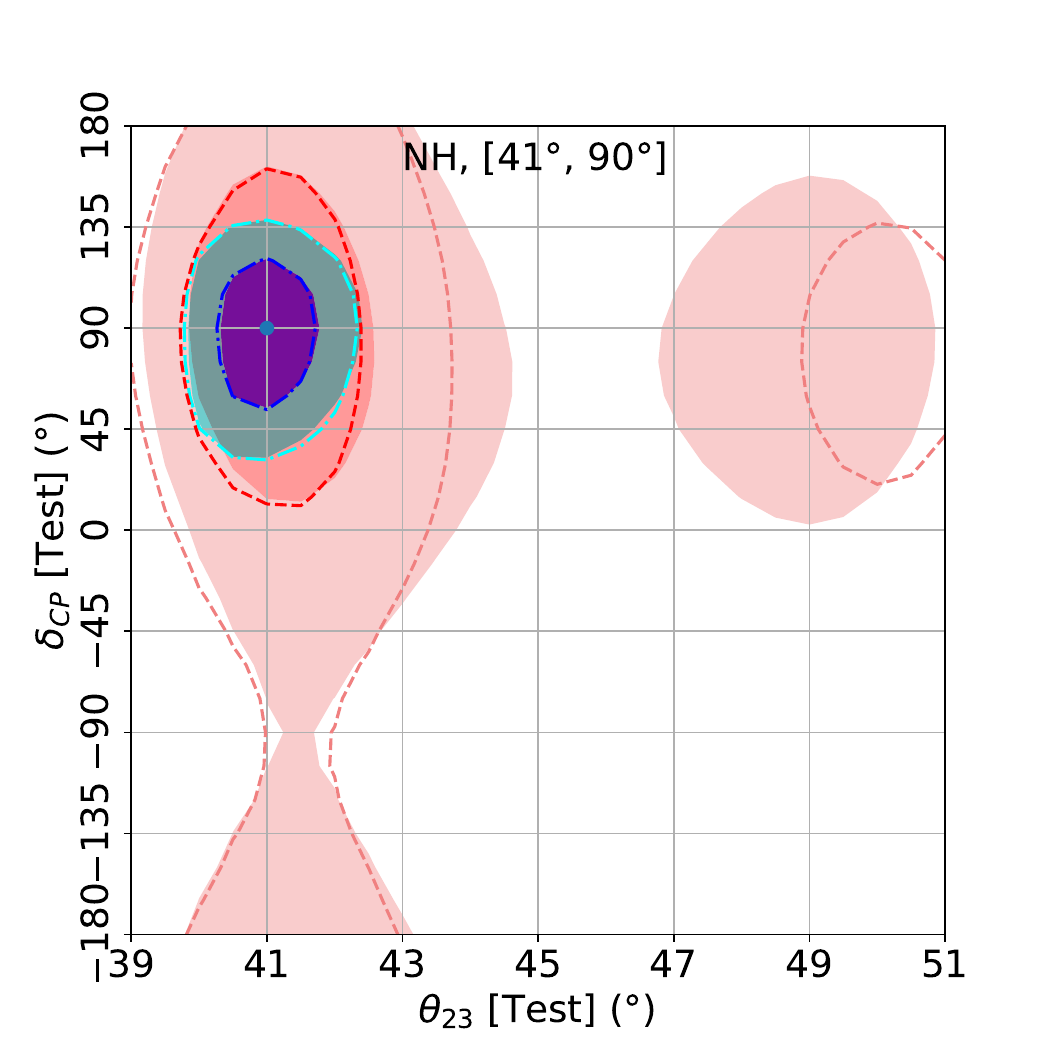}
\includegraphics[width=.32\textwidth]{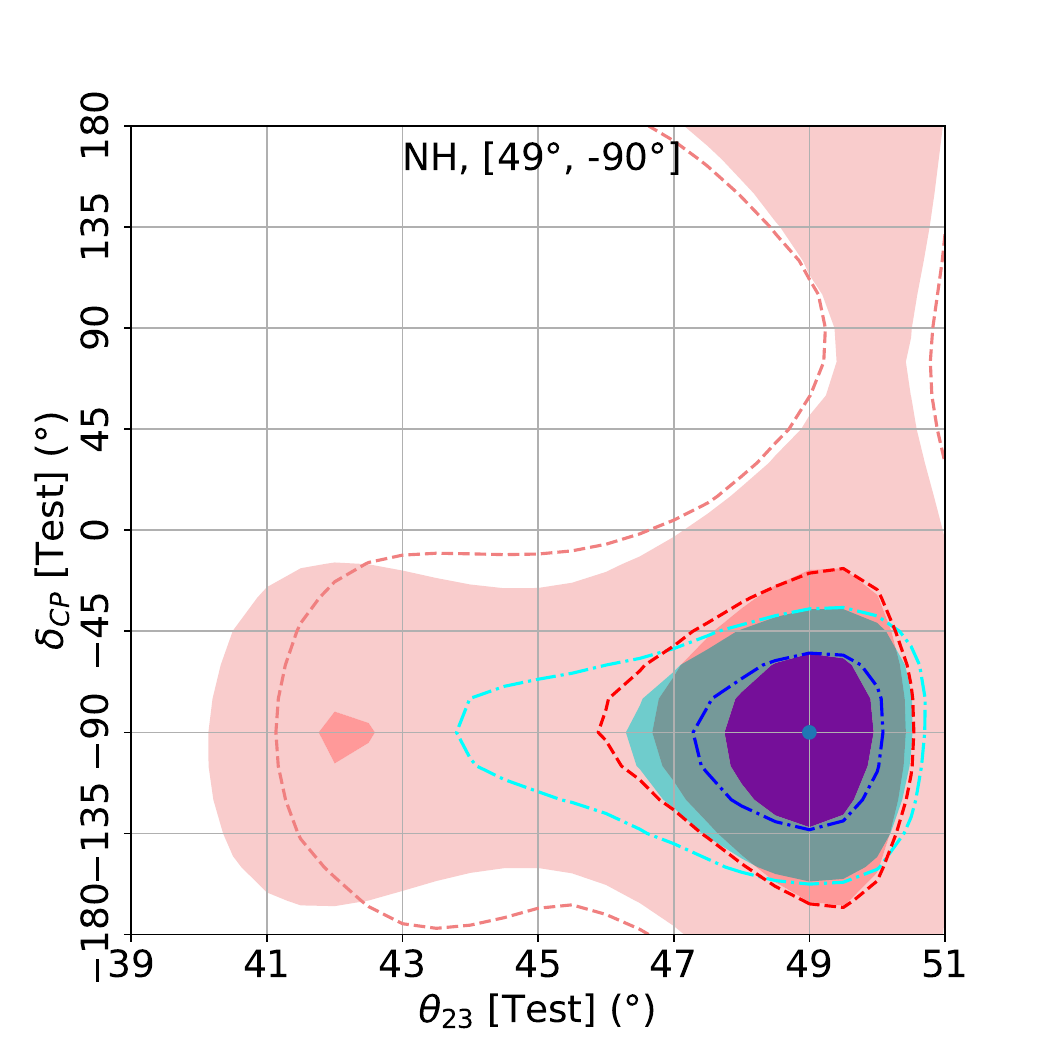}
\includegraphics[width=.32\textwidth]{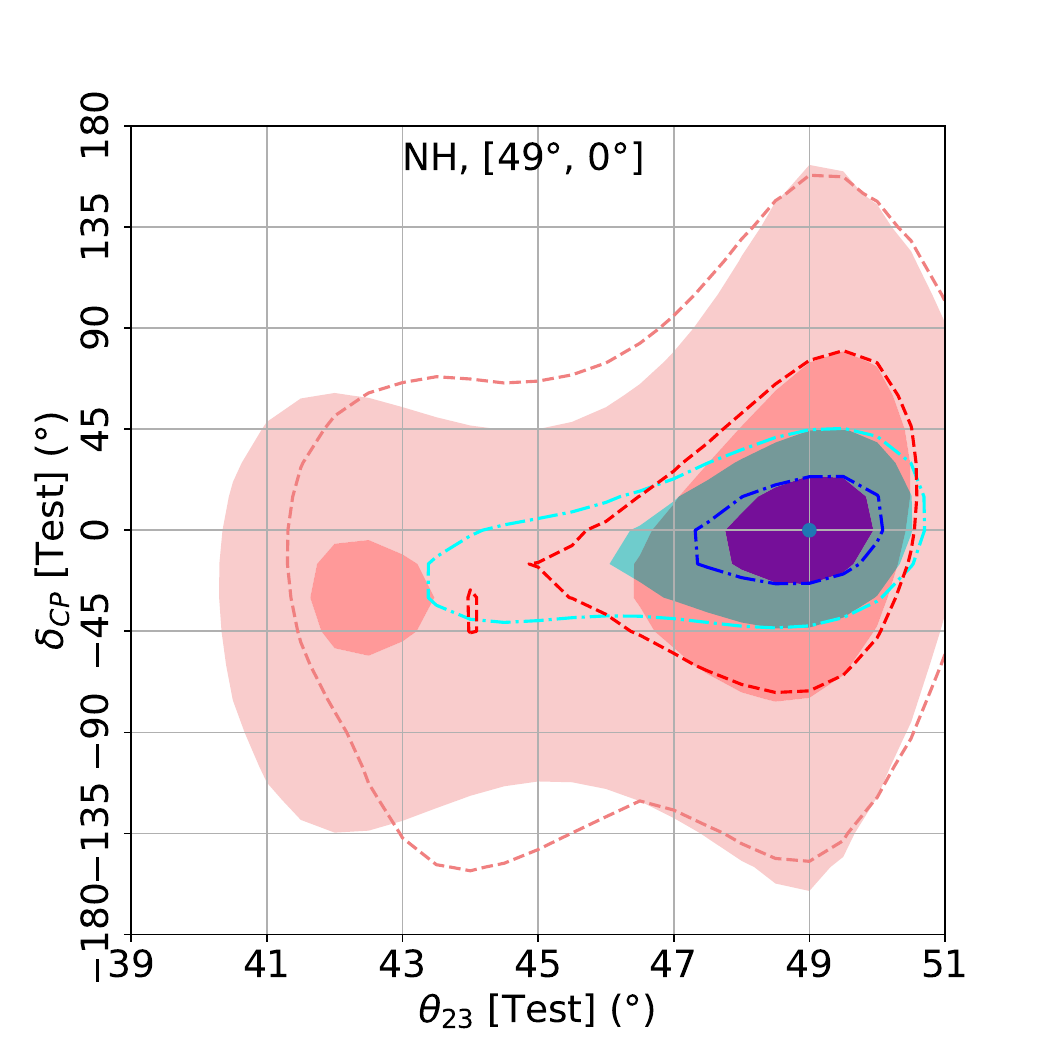}
\includegraphics[width=.32\textwidth]{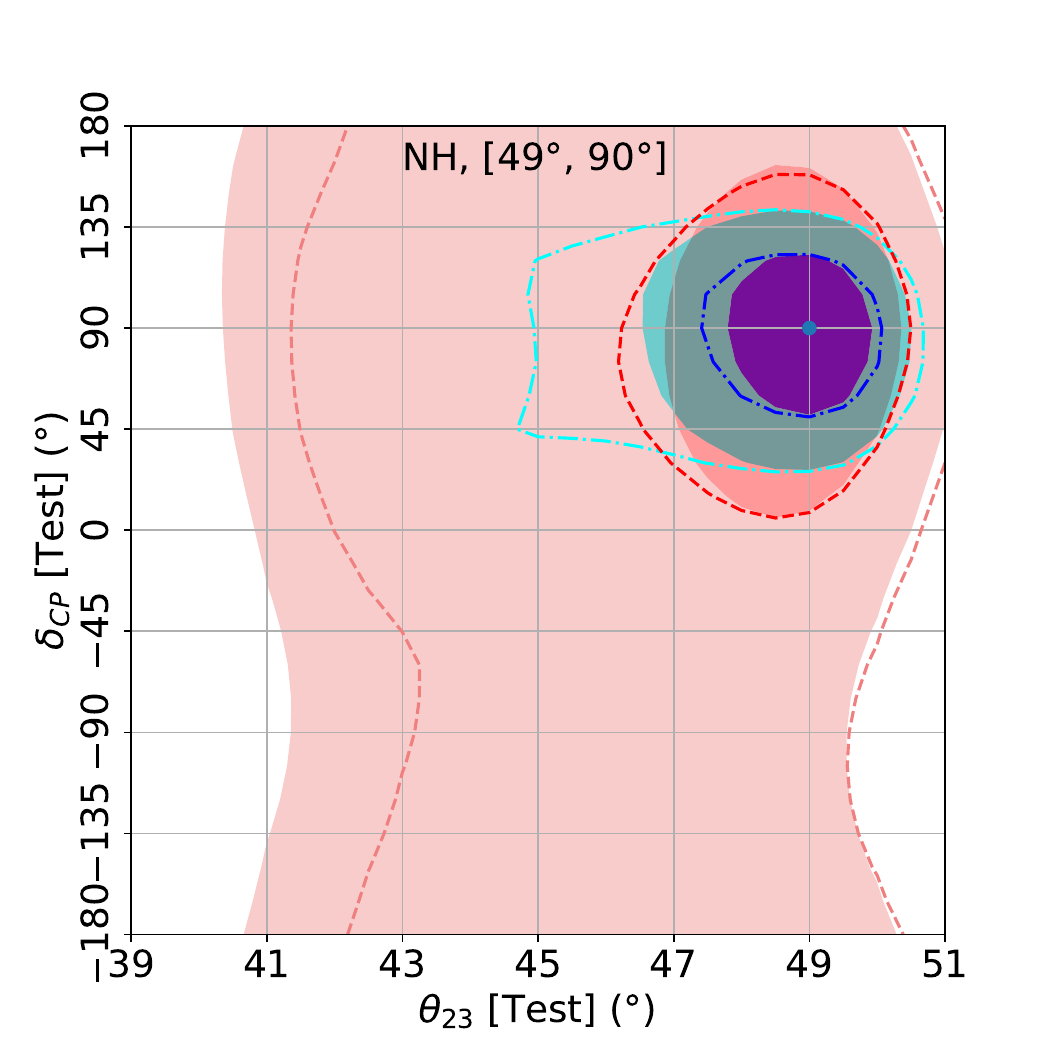}
\includegraphics[width=.32\textwidth]{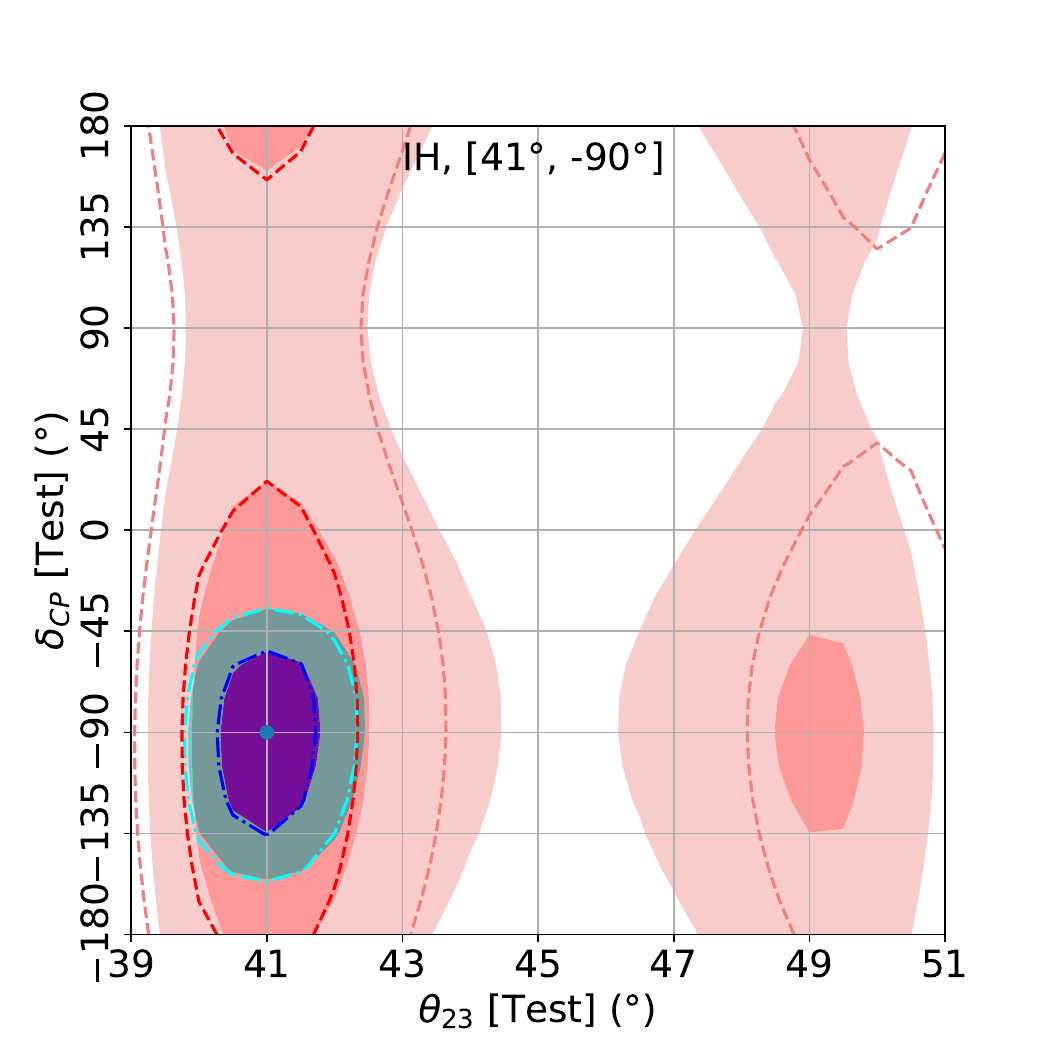}
\includegraphics[width=.32\textwidth]{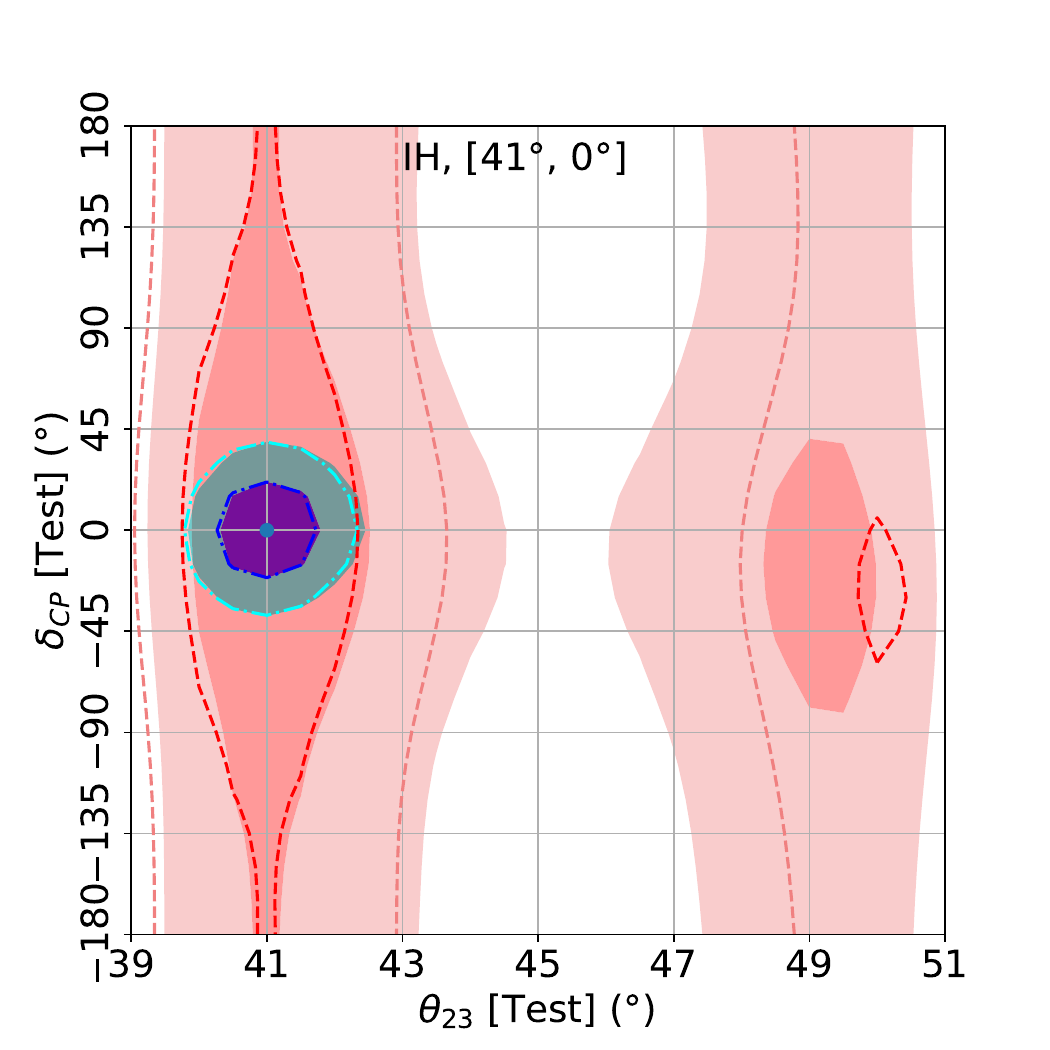}
\includegraphics[width=.32\textwidth]{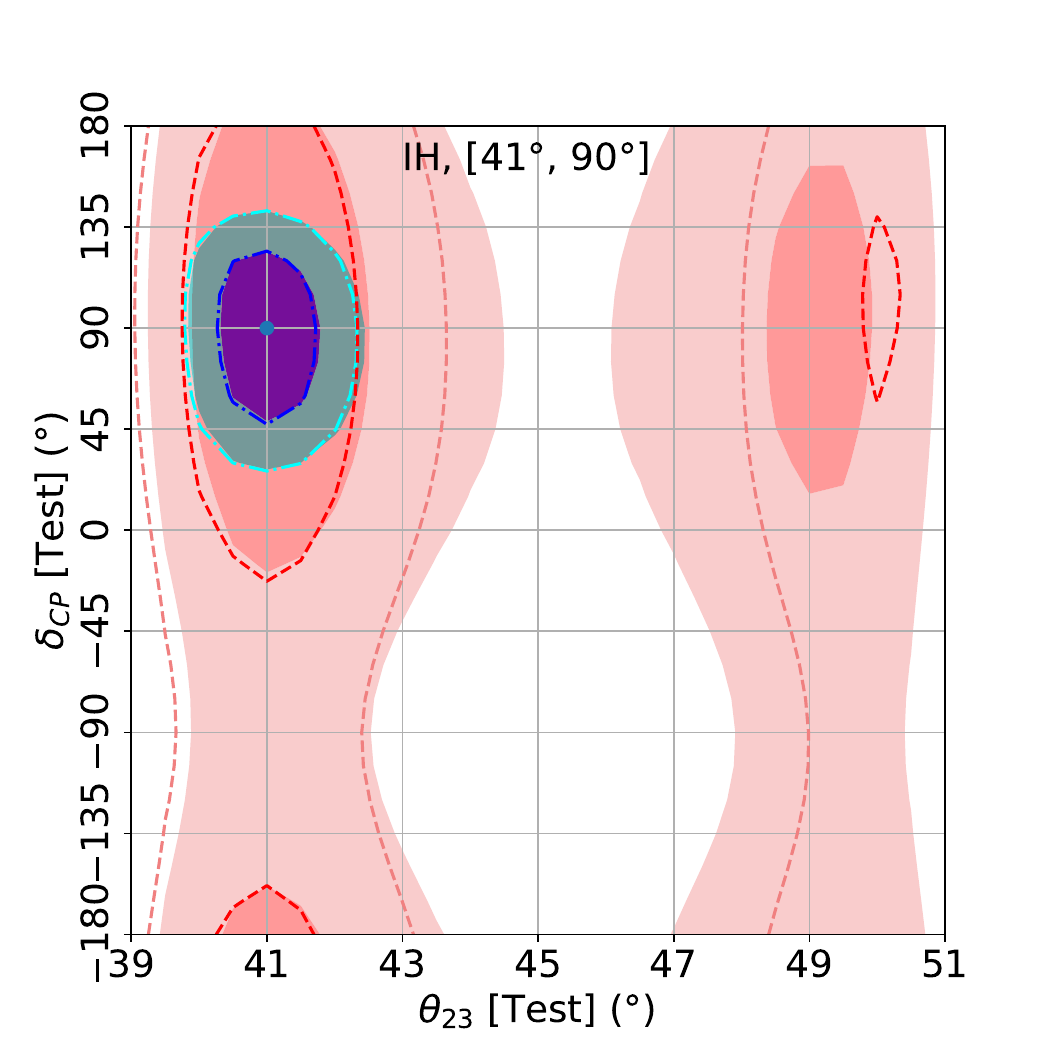}
\includegraphics[width=.32\textwidth]{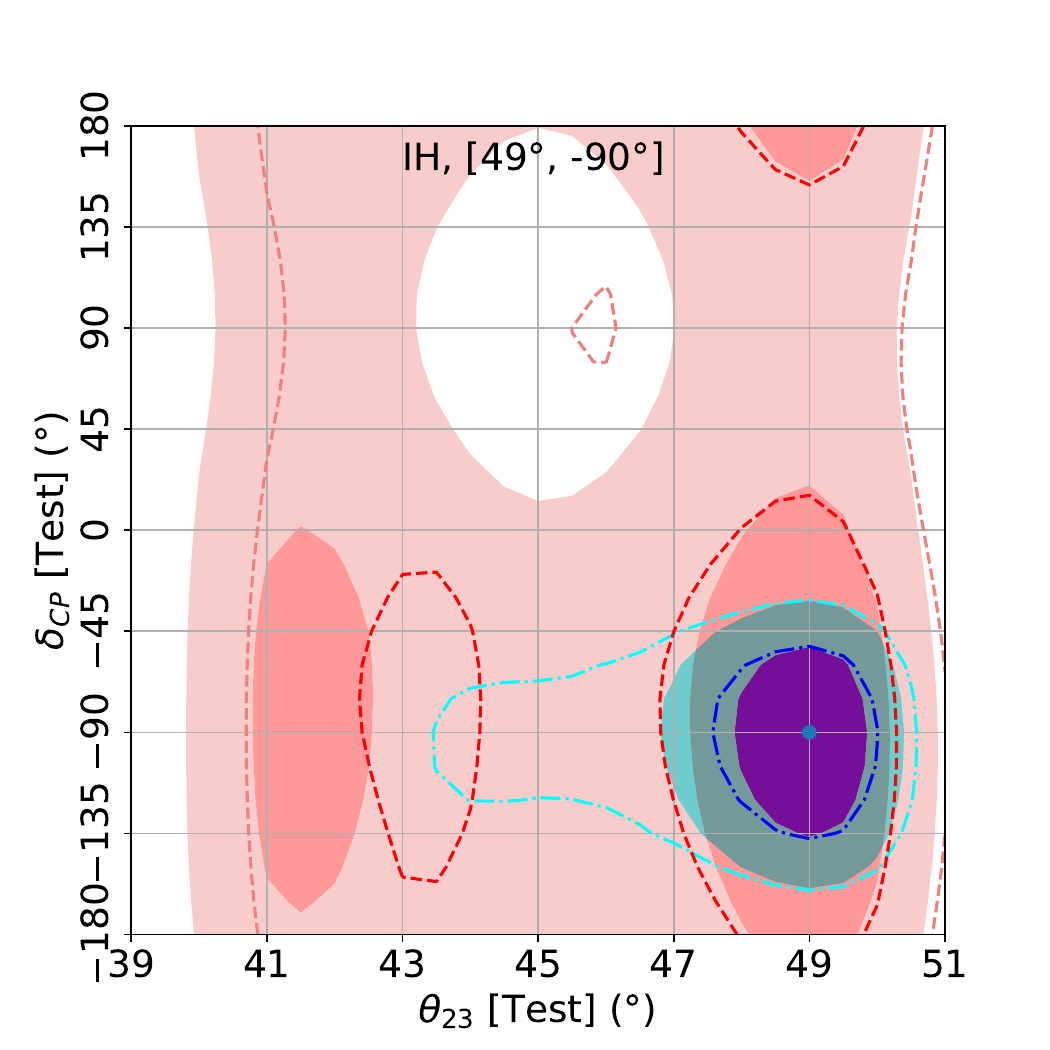}
\includegraphics[width=.32\textwidth]{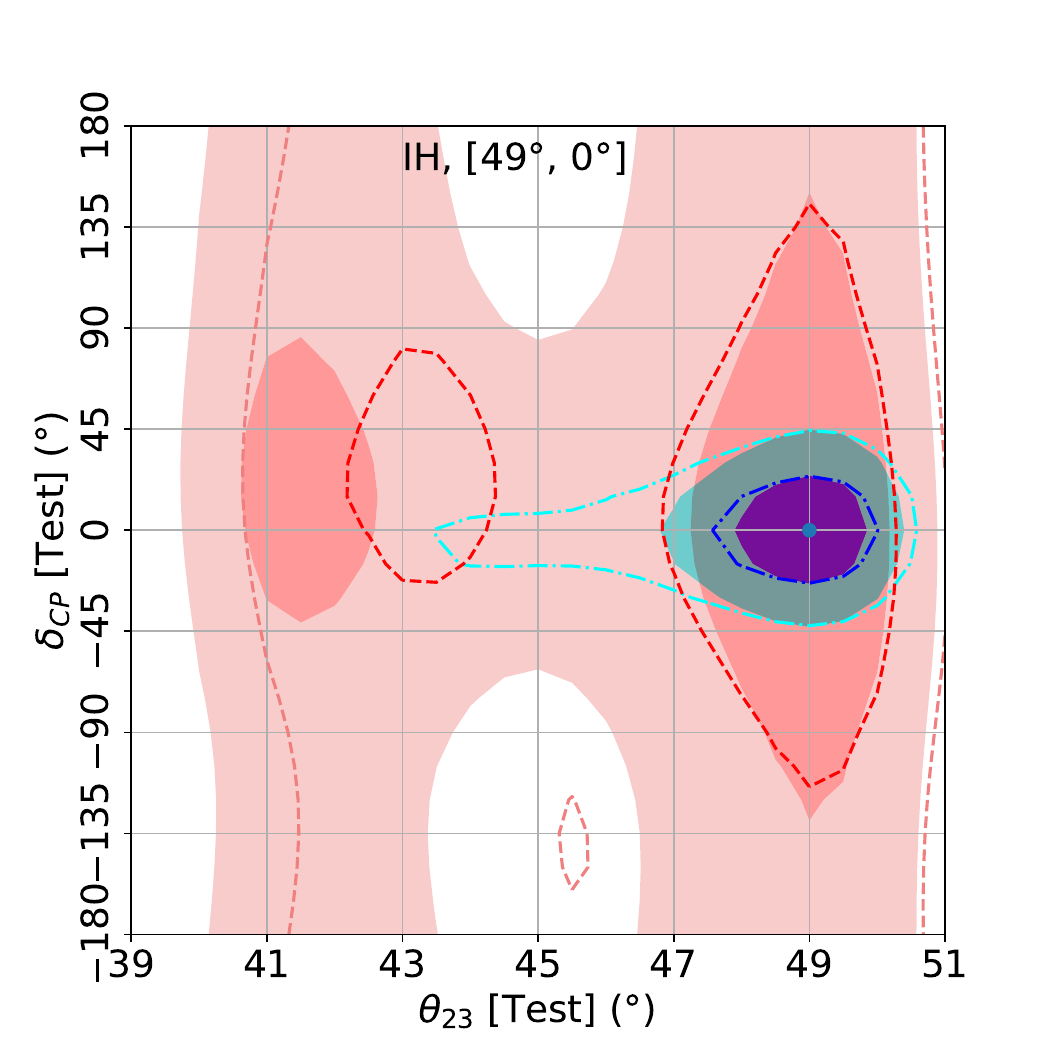}
\includegraphics[width=.32\textwidth]{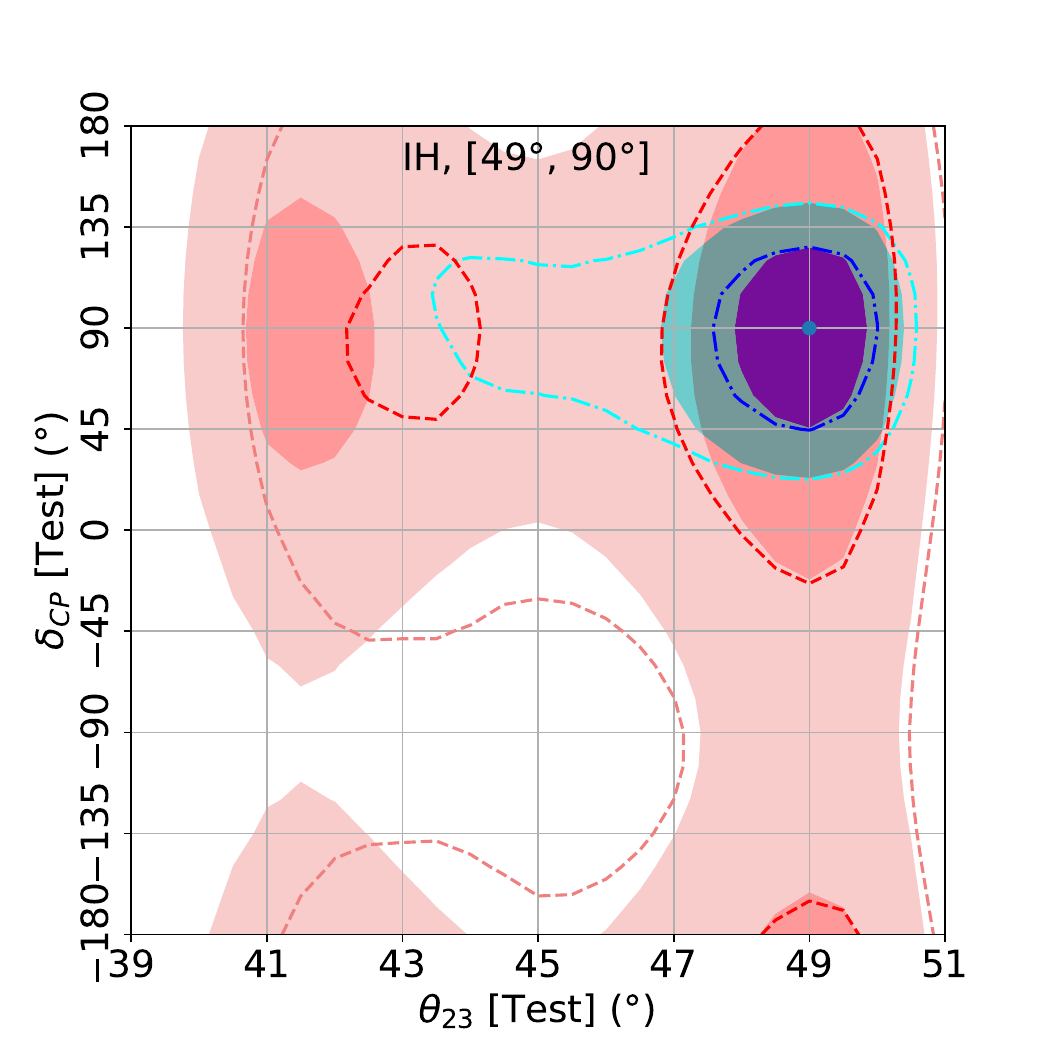}
\caption{Sensitivity in the test $\theta_{23}-\delta_{CP}$ plane for true values of $\theta_{23} = 41^{\circ}, 49^{\circ}$, and $\delta_{CP}=-90^\circ,0^\circ, 90^\circ$ for both normal and inverse hierarchy. The shaded contours are for the decay case, and the contours encompassed by curves are for the no-decay case. The dark red and light red correspond to $3\sigma, 5\sigma$ contours in P2O, while blue and cyan stand for  $3\sigma, 5\sigma$ contours in LArTPC.}
\label{fig:th23-dcp-comp}
\end{figure}

\begin{figure}
\centering 
\includegraphics[width=.32\textwidth]{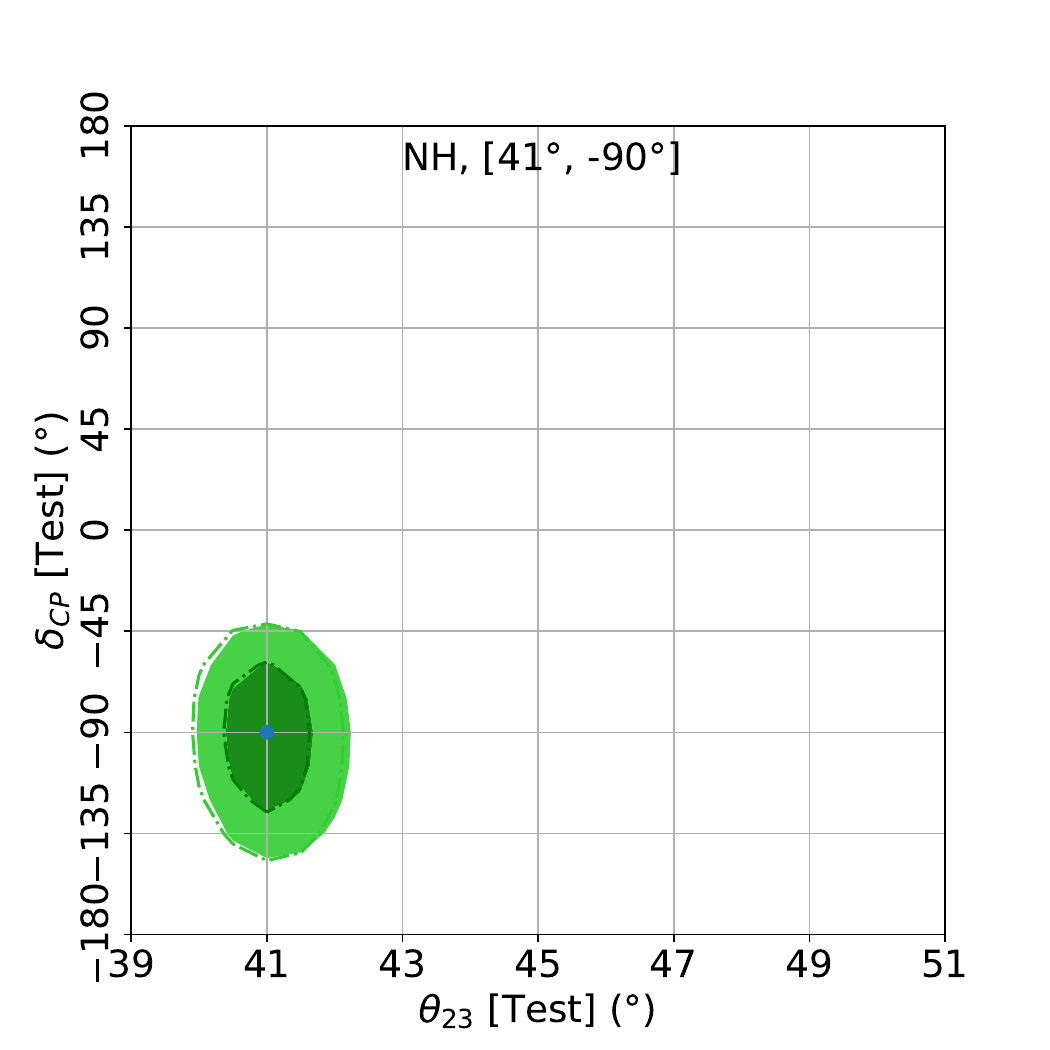}
\includegraphics[width=.32\textwidth]{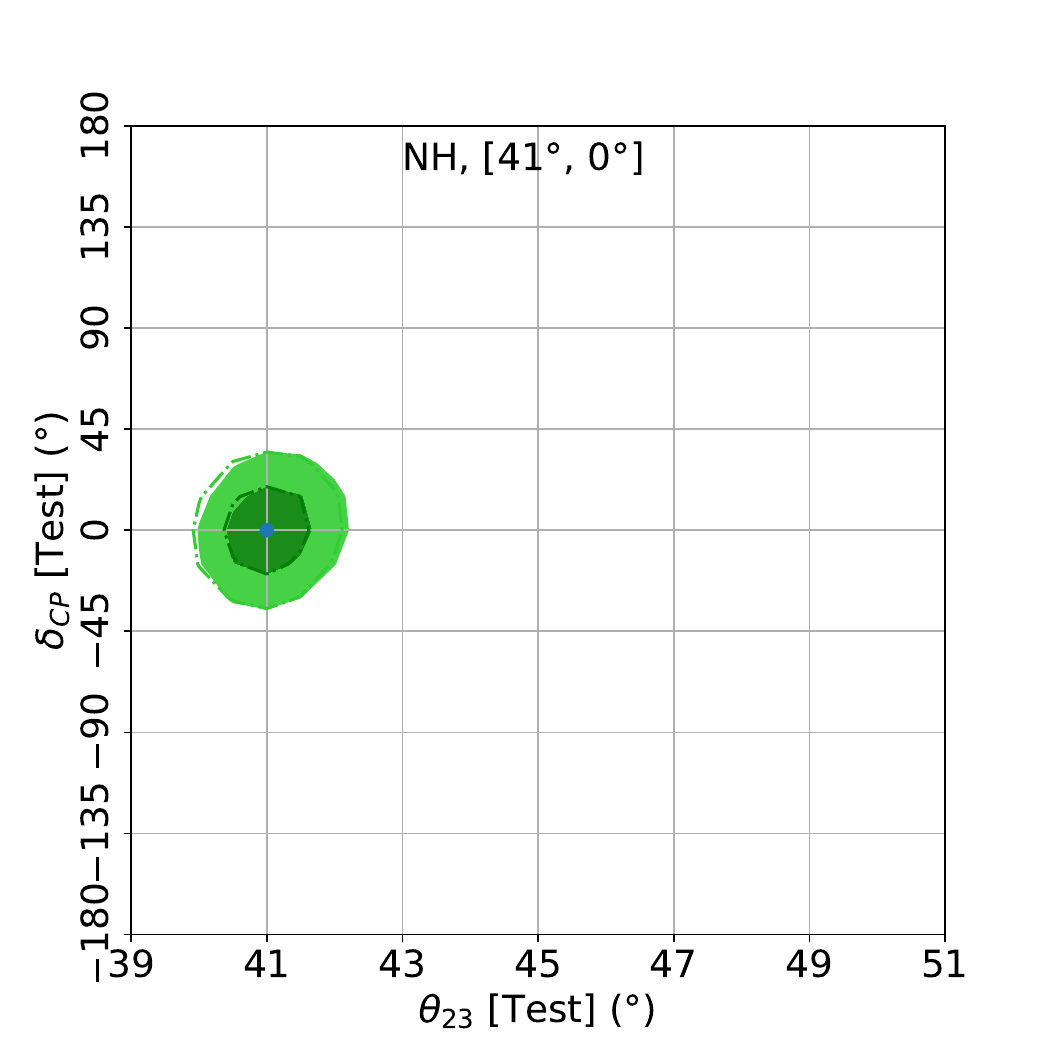}
\includegraphics[width=.32\textwidth]{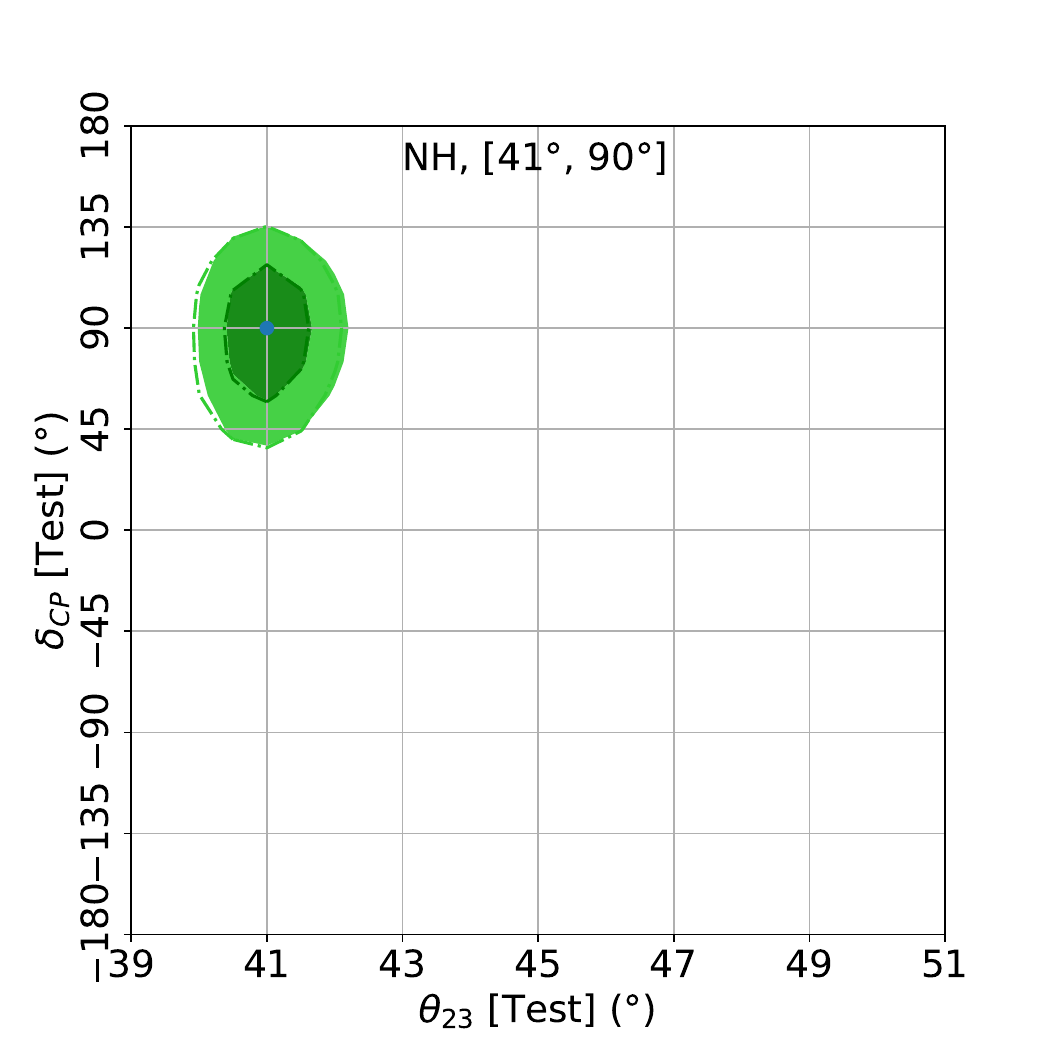}
\includegraphics[width=.32\textwidth]{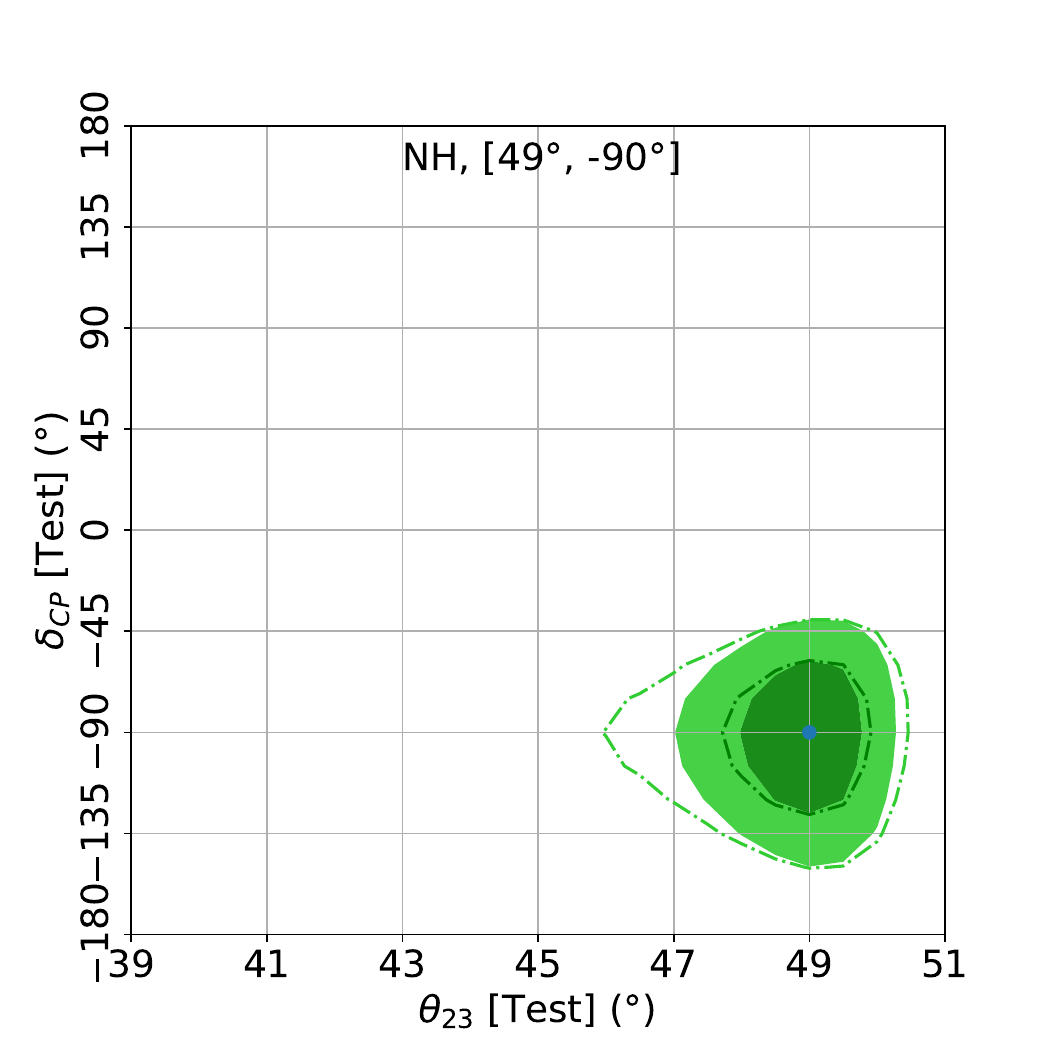}
\includegraphics[width=.32\textwidth]{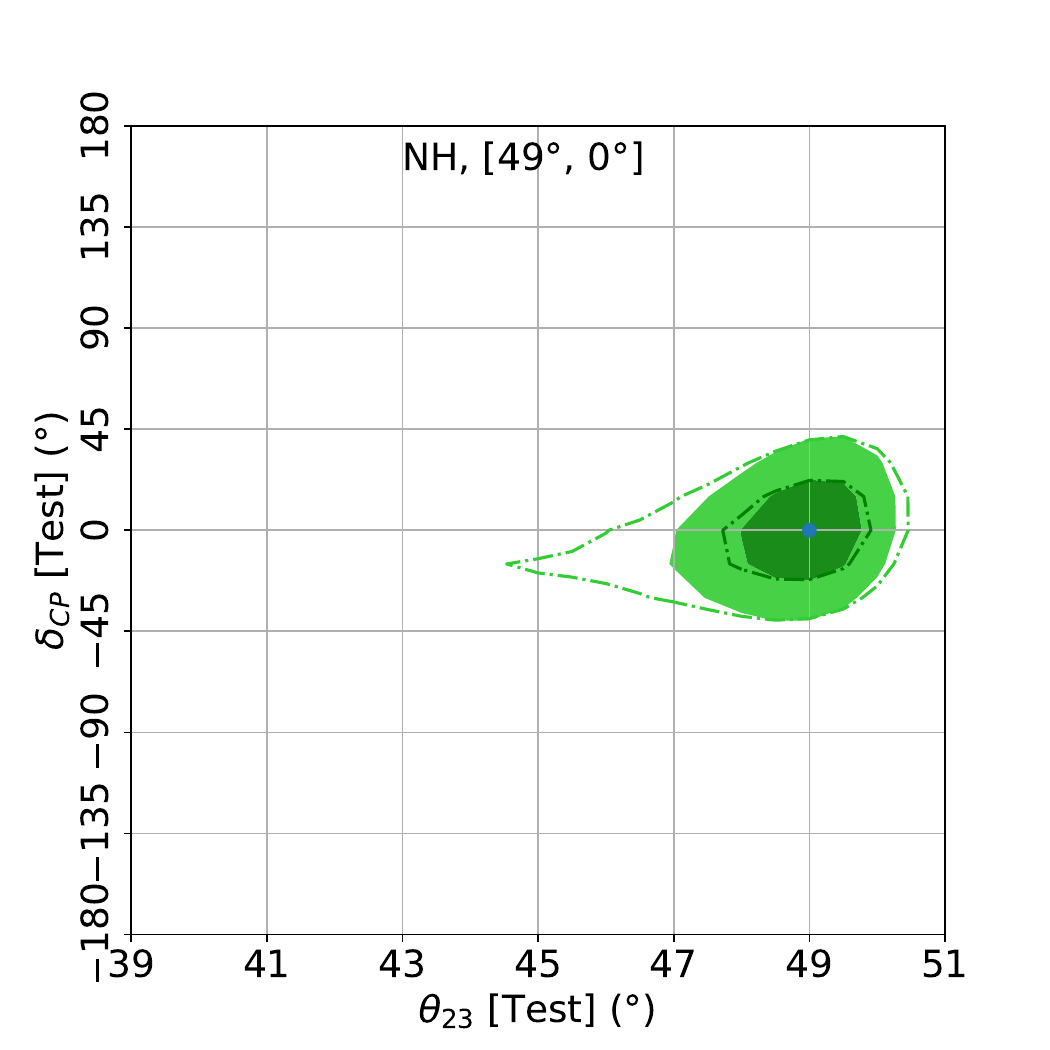}
\includegraphics[width=.32\textwidth]{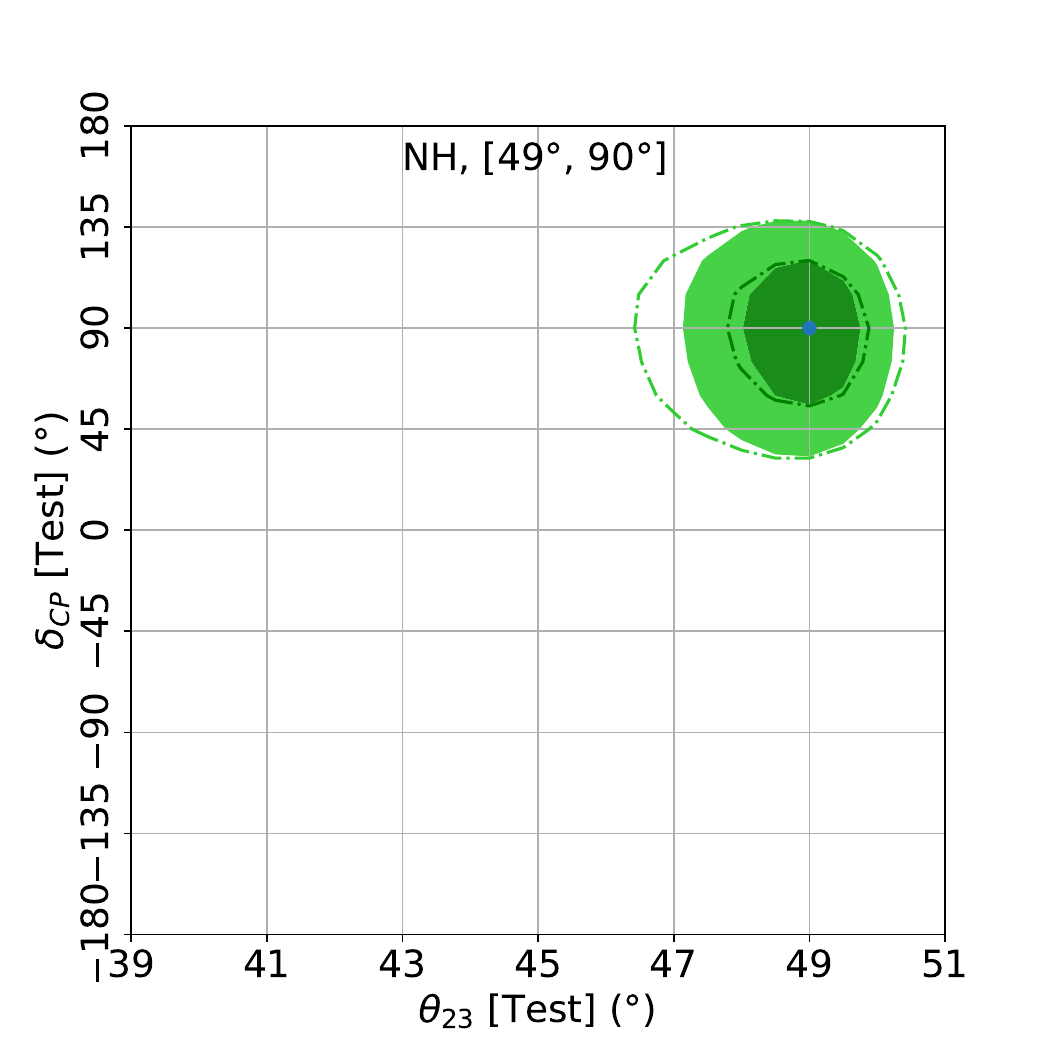}
\includegraphics[width=.32\textwidth]{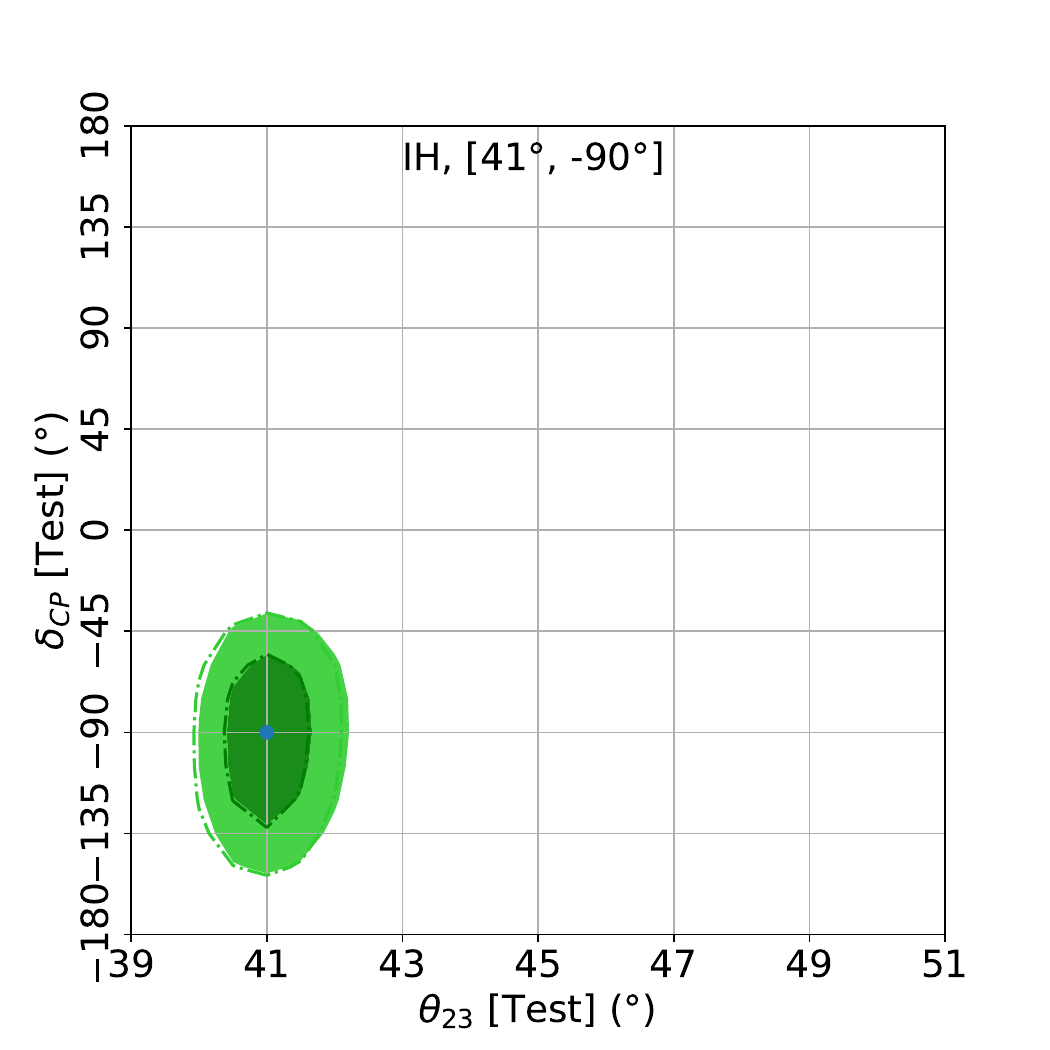}
\includegraphics[width=.32\textwidth]{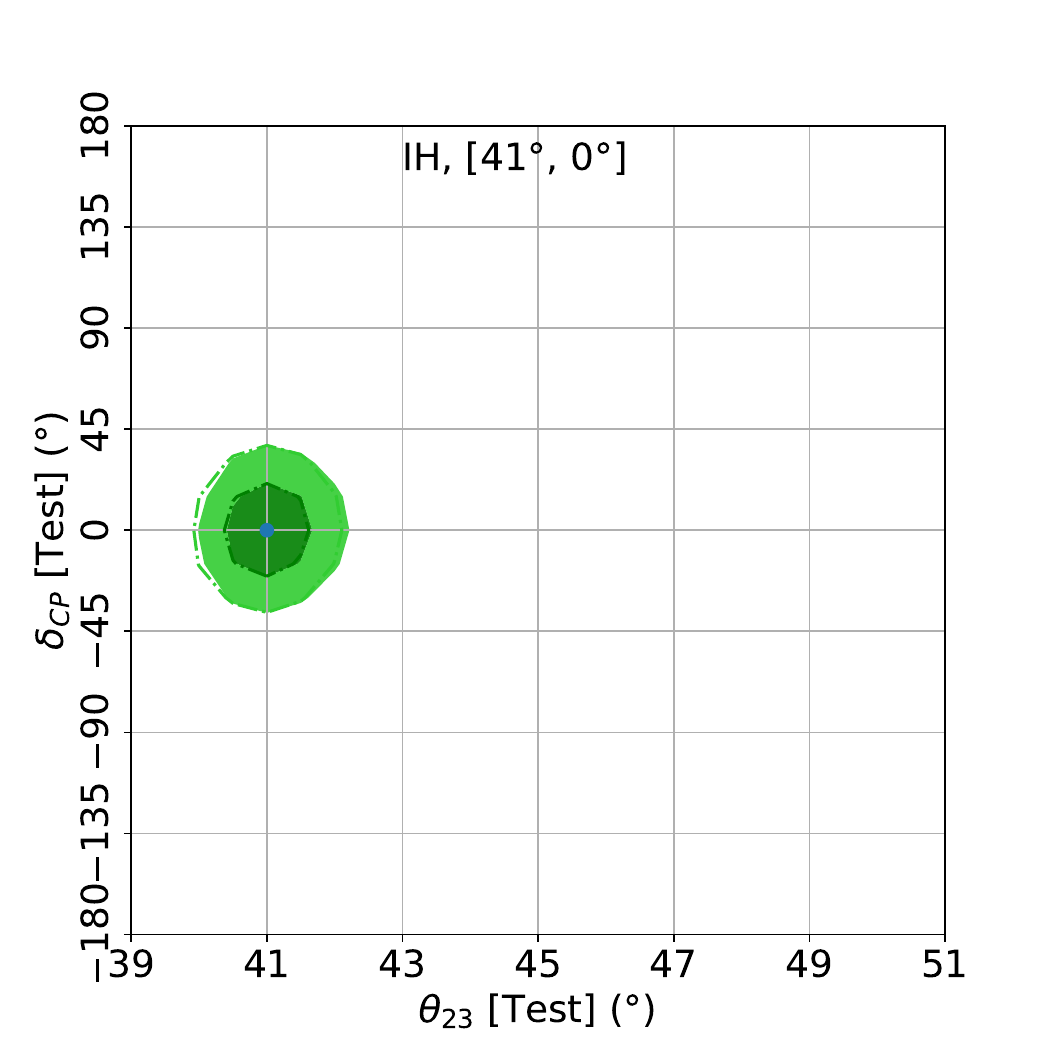}
\includegraphics[width=.32\textwidth]{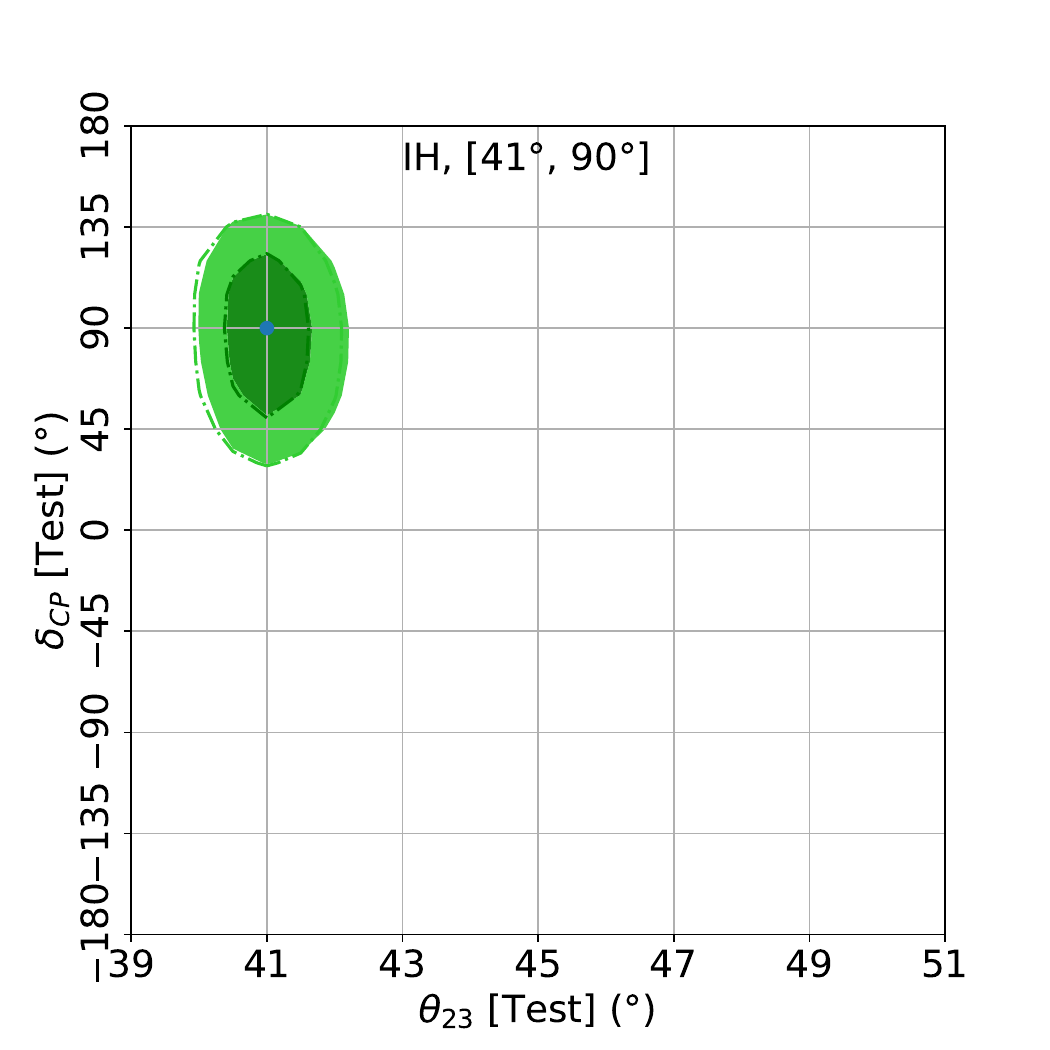}
\includegraphics[width=.32\textwidth]{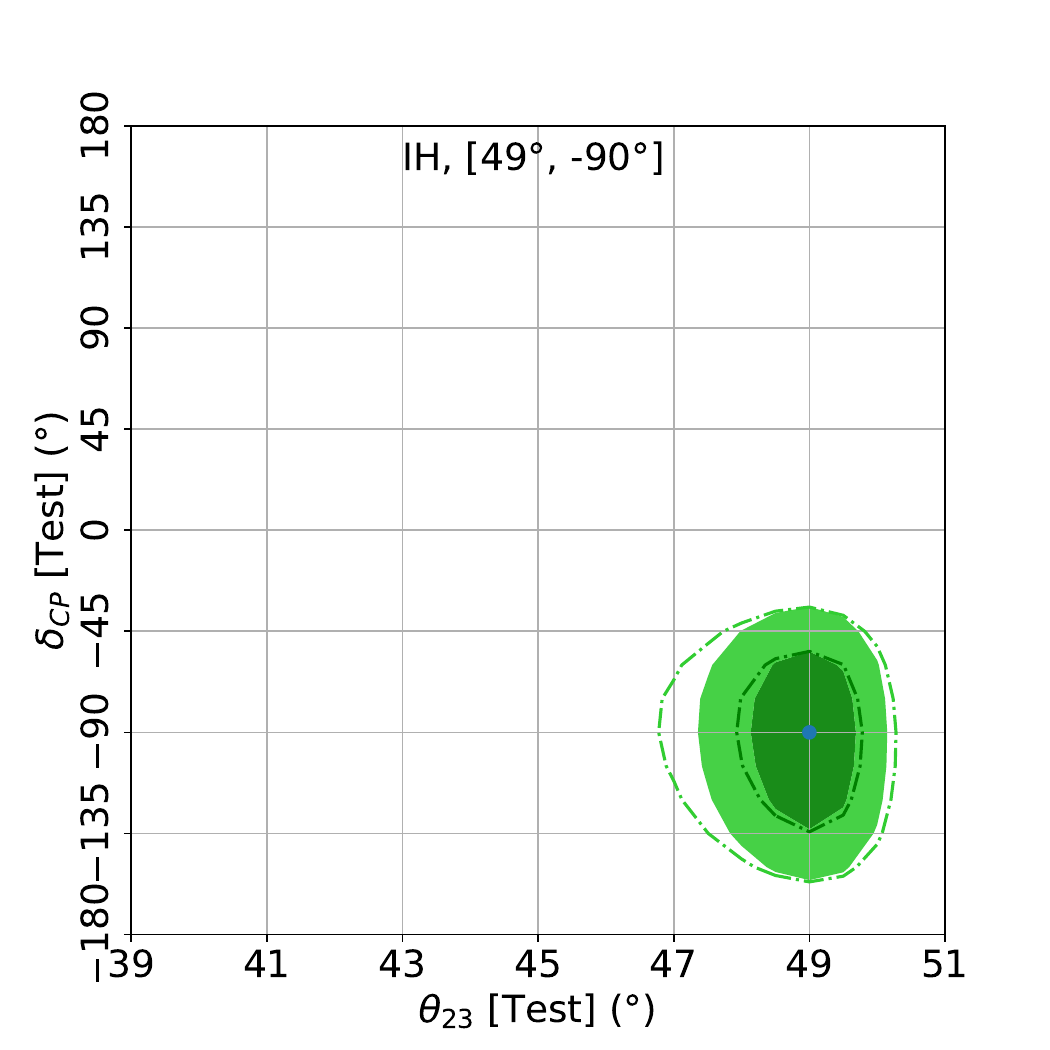}
\includegraphics[width=.32\textwidth]{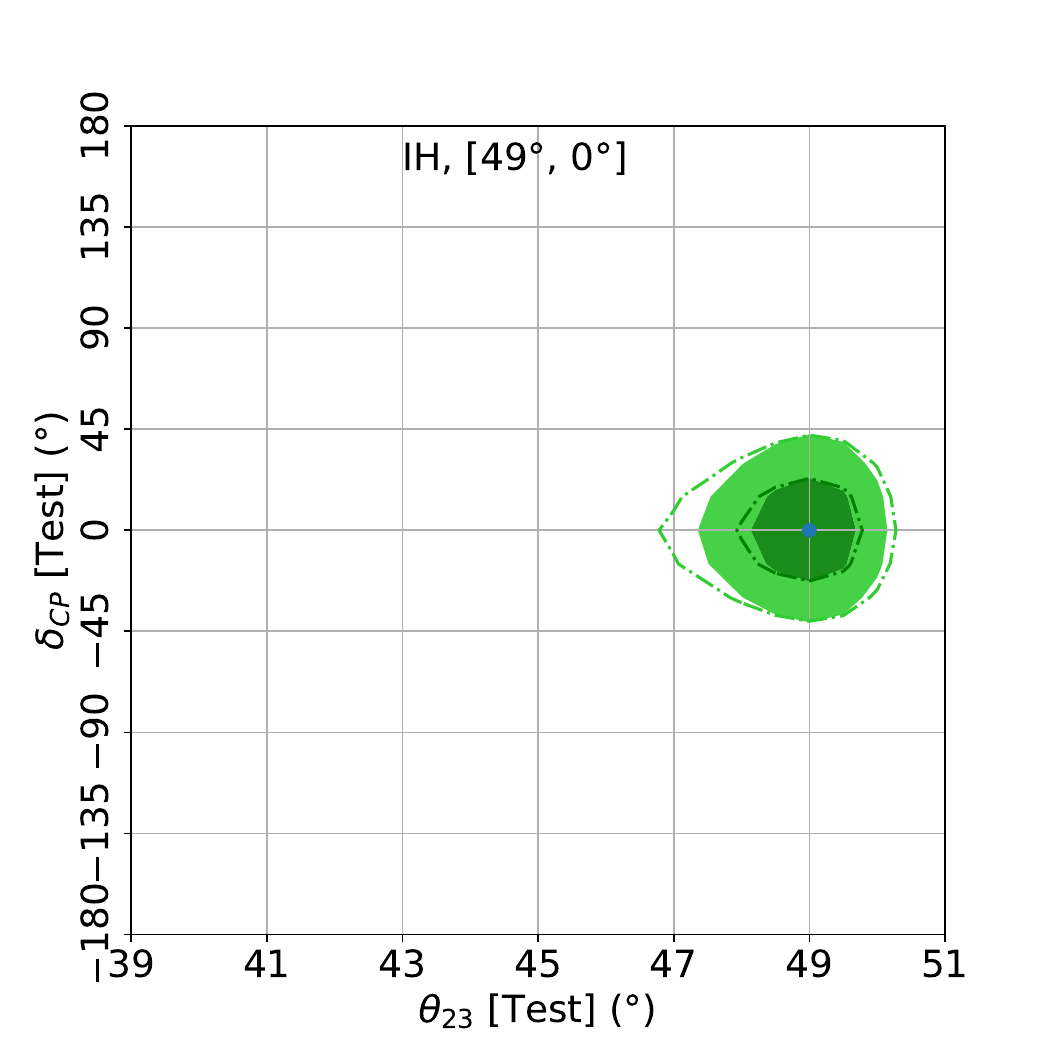}
\includegraphics[width=.32\textwidth]{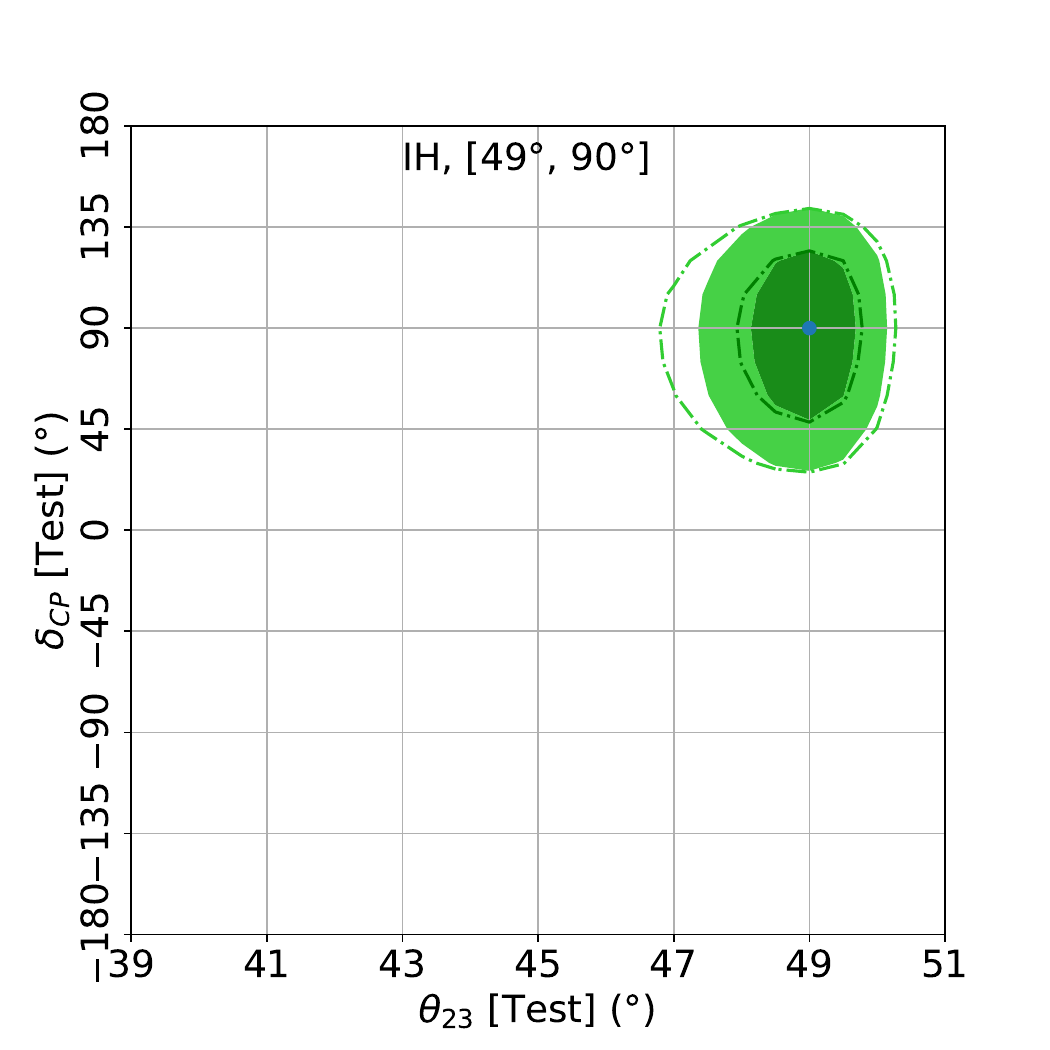}
\caption{Sensitivity in the test $\theta_{23}-\delta_{CP}$ plane for true values of $\theta_{23} = 41^{\circ}, 49^{\circ}$, and $\delta_{CP}=-90^\circ,0^\circ, 90^\circ$ for both normal and inverse hierarchy. The shaded contours are for the decay case, and the contours encompassed by curves are for the no-decay case. The dark green and light green correspond to $3\sigma, 5\sigma$ contours in the combined analysis of P2O+LArTPC setups.}
\label{fig:th23-dcp-comb}
\end{figure}

In this section, we explore the degenerate solutions in the presence of decay in the test $\theta_{23}-\delta_{CP}$ plane for fixed mass hierarchy in the case of the LArTPC detector (DUNE-like) and the water Cherenkov detector (P2O) as well the combination of these setups. In figures \ref{fig:th23-dcp-comp} and \ref{fig:th23-dcp-comb}, we have plotted regions of [$3\sigma$, $5\sigma$] sensitivity in $\theta_{23}-\delta_{CP}$ plane for no-decay (decay) scenarios by shaded contours (contour lines) in P2O [dark red and light red], DUNE-like [blue and cyan] setup and the combined analysis [dark green and light green] of these two setups, respectively. We considered different combinations of true values in the case of both NH (first and second rows) and IH (third and fourth rows) as the true hierarchy. In case of decay, the true value of $\alpha_3^{-1}$ is $2\times 10^{-11}$ s/eV and marginalization is over $[1:3]\times 10^{-11}$ s/eV. The main observations from figures \ref{fig:th23-dcp-comp}, \ref{fig:th23-dcp-comb} are as follows,
\begin{itemize}
    \item The allowed contours are larger for P2O, signifying a lower sensitivity in $\theta_{23}$ and $\delta_{CP}$ than the DUNE-like setup for both decay and no-decay cases.

    \item Comparing the decay and no-decay scenarios, for P2O, allowed regions in $\theta_{23}$ are seen to be more restrictive in the presence of decay. This is consistent with the earlier observation that octant sensitivity is enhanced in the presence of decay in P2O. This also leads to the removal of wrong octant solutions at $3\sigma$ in some cases, e.g., $\theta_{23}=49^\circ, \delta_{CP}=0^\circ$, and $\theta_{23}=41^\circ, \delta_{CP}=-90^\circ$.
    
    \item For the DUNE-like setup, the allowed regions are almost identical for decay and no-decay cases at $3\sigma$, and there are no wrong octant solutions. However, at $5\sigma$, the sensitivity of $\theta_{23}$ decreases for $\theta_{23}=49^\circ$ in the presence of decay, and in some cases, the regions extend to opposite octant. 
    
    %For all the chosen true values in NH, (c.f. the first two rows of Fig. \ref{fig:th23-dcp-comp}) the wrong octant solutions with $3\sigma$ sensitivity are absent in both setups for decay and no-decay cases except for a small region for $\theta_{23}=49^\circ, \delta_{CP}=0^\circ$. However, for IH, (c.f. last two rows of Fig. \ref{fig:th23-dcp-comp}), small regions of wrong octant solutions are present for P2O, e.g., red regions around $42^\circ - 44^\circ$ in the fourth column. For P2O, the wrong octant regions shrink in the presence of decay.

    %\item At $5\sigma$, wrong octant solutions also appear for true values in the higher octant for the DUNE-like setup in NH and IH. For P20, the wrong octant solution arises for both octants, with a lower sensitivity with true values in the higher octant (2nd row).
    
    \item The wrong octant and wrong $\delta_{CP}$ solutions are removed with combined analysis as seen from light green contours in figure \ref{fig:th23-dcp-comb} even at $5\sigma$ in decay and no-decay scenarios. 

    \item The CP sensitivity is more for LArTPC setup as compared to P2O. Moreover, in the case of P2O, the sensitivity for IH is less than that of NH.
    For P2O, the red-dashed regions cover the entire parameter space of $\delta_{CP}$ in most of the panels in figure \ref{fig:th23-dcp-comb}, signifying no sensitivity to $\delta_{CP}$ at $5\sigma$. But for a combined analysis of the two setups, the wrong $\delta_{CP}$ solutions are excluded even at $5\sigma$ significance. 

\end{itemize}

\section{Conclusions}\label{sec:conclusion}
In this work, we have studied the effects of invisible decay of the third neutrino state on the determination of mass hierarchy and octant of $\theta_{23}$ in the context of two different long baseline experimental setups: a LArTPC setup (similar to DUNE) at 1300 km and a water Cherenkov setup (proposed by P2O) at 2588 km. For mass hierarchy sensitivity, we explored two scenarios in the presence of decay: (i) decay in both true and test cases and (ii) no decay in true and decay in the test case for the opposite hierarchy. We find that the mass hierarchy sensitivity is reduced in the first case compared to the no-decay scenario in both setups. However, the sensitivity increases in the second case compared to the first case. In the water Cherenkov setup, the mass hierarchy sensitivity is seen to be increasing steadily with $\theta_{23}$ for true NH but peaks at $\theta_{23}\sim 46^\circ$ for true IH. This nature of sensitivity is due to a higher muon background of $45\%$ in P2O in the electron channel. We also noticed the mass hierarchy sensitivity to be flat between $\theta_{23}=44^\circ-46^\circ$ for true IH in the LArTPC setup in 1300 km, which can be attributed to the presence of muon backgrounds.

In the study of the sensitivity to octant of $\theta_{23}$, we have only considered the presence of decay (in both true and test) and compared it with the no-decay scenario. It is observed that in the presence of decay, the sensitivity is increased for both the setups for true values of $\theta_{23}$ in LO due to enhanced contribution from the $\nu_\mu$ channel. For $\theta_{23}$ in HO, we see reduced sensitivity in the P2O setup as compared to $\theta_{23}$ in LO but still higher than the no-decay scenario. In the LArTPC setup, the sensitivity in presence decay for $\theta_{23}$ in HO is lower than the no decay. These features are explained by the synergy between electron and muon channels.

We also explore the degeneracies present in the test $\theta_{23}-\delta_{CP}$ plane for various combinations of true values in the presence of decay in both true and test and compare with the no decay scenario. In the P2O setup, the wrong octant solutions exist at $3\sigma$ (for certain combinations of true values) and $5\sigma$ in no-decay scenarios. However, the inclusion of decay elevates the octant sensitivity, resulting in the removal or reduction of wrong octant solutions. The $3\sigma$ contours of DUNE-like setups are similar for the decay and no-decay cases, and there is no wrong octant solution in $3\sigma$. In the case of a DUNE-like setup, the sensitivity is lower in the presence of decay for $\theta_{23}$ in HO, leading to enlarged contours extending to opposite octant at $5\sigma$. When we do the combined analysis of these setups, the wrong solutions disappear even at $5\sigma$.
\newpage
\acknowledgments
AC acknowledges the Ramanujan Fellowship (RJF/2021/000157), of the Science and Engineering Research Board of the Department of Science and Technology, Government of India, and CERN EP-Nu group. SG acknowledges the J.C. Bose Fellowship (JCB/2020/000011) of the Science and Engineering Research Board of the Department of Science and Technology, Government of India. It is to be noted that this work has been done solely by the authors and is not representative of the DUNE collaboration. SP appreciates the guidance of Dr. Monojit Ghosh regarding the writing of the glb file for the P2O experiment.
% The bibliography will probably be heavily edited during typesetting.
% We'll parse it and, using the arxiv number or the journal data, will
% query inspire, trying to verify the data (this will probalby spot
% eventual typos) and retrive the document DOI and eventual errata.
% We however suggest to always provide author, title and journal data:
% in short all the informations that clearly identify a document.

\bibliographystyle{unsrt}
\bibliography{reference}
\end{document}